\documentclass[12pt]{article}
\pdfoutput=1

\usepackage{cite}

\usepackage[utf8]{inputenc}
\usepackage[nointlimits,reqno]{amsmath}
\usepackage{amssymb}
\usepackage{slashed}
\usepackage{dsfont}
\usepackage{graphicx}
\usepackage[heightadjust]{marginnote}

\usepackage[pdftex,
 pdfproducer=TeXShop,
 pdfcreator=pdflatex]{hyperref}

\usepackage{xspace}
\usepackage{luty}

\makeatletter
\renewcommand{\subsubsection}{\@startsection{subsubsection}{2}{0pt}%
    {-3.25ex plus -1ex minus -.2ex}%
    {1.5ex plus .2ex}%
    {\centering\normalsize\itshape}}
\makeatother

\usepackage{titlesec}
\usepackage{subfigure}
\usepackage{feynmp-auto}
\usepackage{bbm}

\numberwithin{equation}{section} 
\newcommand{\id}{\mathbbm{1}}
\newcommand{\timeordering}{\text{T}}
\newcommand{\antitimeordering}{\overline{\timeordering}}
\renewcommand{\O}{\scr{O}}
\newcommand{\M}{\scr{M}}
\newcommand{\I}{\mathcal{I}}
\newcommand{\MP}{M_\text{Planck}}
\newcommand{\nS}{n_\text{s}}
\newcommand{\nF}{n_\text{f}}
\newcommand{\nV}{n_\text{v}}

\begin{document}
\unitlength = 1mm


\begin{titlepage}

\vspace*{1cm}

\title{Graviton Scattering and\\
a Sum Rule for the \emph{c} Anomaly\\
in 4D CFT}

\author{Marc Gillioz$^1$,\  \
Xiaochuan Lu$^2$,\  \
and\ \ Markus A.~Luty$^2$}

\address{$^1$Theoretical Particle Physics Laboratory,
Institute of Physics, EPFL
\\ CH--1015 Lausanne, Switzerland}

\address{$^2$Center for Quantum Mathematics and Physics (QMAP)\\
University of California, Davis, California 95616}

\vspace{1cm}

\begin{abstract}
4D CFTs have a scale anomaly characterized by the coefficient $c$, which appears as the coefficient of logarithmic terms in momentum space correlation functions of the energy-momentum tensor. By studying the CFT contribution to 4-point graviton scattering amplitudes in Minkowski space we derive a sum rule for $c$ in terms of $TT\mathcal{O}$ OPE coefficients. The sum rule can be thought of as a version of the optical theorem, and its validity depends on the existence of the massless and forward limits of the $\langle TTTT \rangle$ correlation functions that contribute. The finiteness of these limits is checked explicitly for free scalar, fermion, and vector CFTs. The sum rule gives $c$ as a sum of positive terms, and therefore implies a lower bound on $c$ given any lower bound on $TT\mathcal{O}$ OPE coefficients. We compute the coefficients to the sum rule for arbitrary operators of spin 0 and 2, including the energy-momentum tensor.
\end{abstract}

\end{titlepage}


\section{Introduction}
\scl{intro}

In this paper we consider two related problems in 4D conformal field theory
(CFT).
The first is the computation of physical rates for processes involving
particles coupled to a CFT defined by its operator spectrum and operator
product expansion (OPE) coefficients.
(This has been studied in the phenomenology literature as
``unparticle physics'' \cite{Georgi:2007ek, Georgi:2007si}.)
Here we assume that the coupling of the ordinary particles to the CFT is
sufficiently weak that it does not affect the dynamics of the CFT.
This is similar in spirit to the study of electromagnetism as a probe of QCD,
for example in processes like $e^+ e^- \to \text{hadrons}$ or deep inelastic
scattering.
It is interesting to extend our theoretical understanding of such processes to
general CFTs.
The second problem we consider is the relation of these rates to scale anomalies
in the CFT.

This paper builds on \Ref{Gillioz:2016jnn}, which developed the formalism
needed to address these questions for processes involving probe particles
without spin.
We extend the results to rates involving external gravitons, and
relate these to the $c$ anomaly of 4D CFT.
The final result of this work is \Eq{csumrule},
a sum rule that gives $c$ as a
positive sum over $TT\O$ OPE coefficients, for all primary operators $\O$
other than $\id$ and $T^{\mu\nu}$ itself.

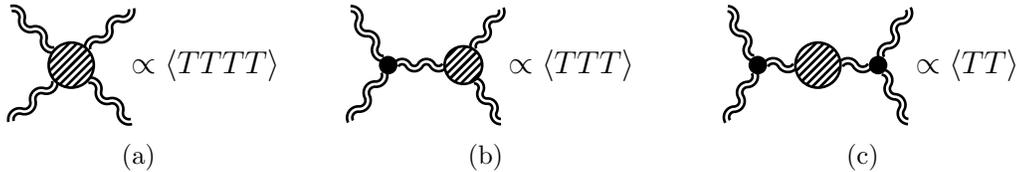
\begin{figure}[t]
\centering
\vspace{0.3cm}\hspace{1cm}
 \subfigure[]{
 \centering\hspace{-2.2cm}\label{subfig:graviton4pt}
\begin{fmffile}{graviton4pt}
\begin{fmfgraph*}(20,15)
\fmfleft{i2,i1}
\fmfright{o2,o1}
\fmfblob{.30w}{v}
\fmfv{decor.shape=circle,decor.filled=full,decor.size=0,l=$ \propto \langle TTTT\rangle$,l.a=0,l.d=8mm}{u}
\fmf{phantom}{v,u}
\fmf{dbl_wiggly}{i1,v}
\fmf{dbl_wiggly}{i2,v}
\fmf{dbl_wiggly}{v,o1}
\fmf{dbl_wiggly}{v,o2}
\end{fmfgraph*}
\end{fmffile}
 }\hspace{3.5cm}
 \subfigure[]{
 \centering\hspace{-2cm}\label{subfig:graviton3pt}
\begin{fmffile}{graviton3pt}
\begin{fmfgraph*}(25,15)
\fmfleft{i2,i1}
\fmfright{o2,o1}
\fmfv{decor.shape=circle,decor.filled=full,decor.size=3thick}{v1}
\fmfblob{.20w}{v2}
\fmfv{decor.shape=circle,decor.filled=full,decor.size=0,l=$ \propto \langle TTT\rangle$,l.a=0,l.d=6mm}{u}
\fmf{phantom}{v2,u}
\fmf{dbl_wiggly}{i1,v1}
\fmf{dbl_wiggly}{i2,v1}
\fmf{dbl_wiggly}{v1,v2}
\fmf{dbl_wiggly}{v2,o1}
\fmf{dbl_wiggly}{v2,o2}
\end{fmfgraph*}
\end{fmffile}
 }\hspace{3.1cm}
 \subfigure[]{
 \centering\hspace{-1.6cm}\label{subfig:graviton2pt}
\begin{fmffile}{graviton2pt}
\begin{fmfgraph*}(30,15)
\fmfleft{i2,i1}
\fmfright{o2,o1}
\fmfv{decor.shape=circle,decor.filled=full,decor.size=3thick}{v1}
\fmfv{decor.shape=circle,decor.filled=full,decor.size=3thick,l=$ \propto \langle TT\rangle$,l.a=0,l.d=5mm}{v3}
\fmfblob{.20w}{v2}
\fmf{dbl_wiggly}{i1,v1}
\fmf{dbl_wiggly}{i2,v1}
\fmf{dbl_wiggly}{v1,v2}
\fmf{dbl_wiggly}{v2,v3}
\fmf{dbl_wiggly}{v3,o1}
\fmf{dbl_wiggly}{v3,o2}
\end{fmfgraph*}
\end{fmffile}
 }\hspace{0.5cm}\vspace{0.3cm}
\begin{minipage}{5.5in}
\caption{\small
Contributions to the physical graviton-graviton scattering
amplitude at order $1/\MP^4$.
Curvy double lines denote gravitons, and the
blob denotes a correlation function of energy-momentum tensors
in the CFT.
The pseudo-amplitude corresponds to the contribution (a).
The contributions (b) and (c) are important for unitarity of the
physical graviton amplitude, but are not included in the
pseudo-amplitude.}
\end{minipage}
\label{fig:GravitonPropagator}
\end{figure}

We now give a summary of the ideas that enter into this sum rule.
We study the contribution of a 4D CFT to graviton-graviton
scattering.
This is related to correlation functions of the energy-momentum tensor
in the CFT,  defined by coupling the CFT to an arbitrary background
metric $g_{\mu\nu}$ and differentiating with respect to the metric.
For example, the connected 4-point function is given by
the time-ordered product
\[
\!\!\!\!\!\!\!\!\!
&i^4 \bra{0} \timeordering\bigl[ T^{\mu_1\nu_1}(x_1)
\cdots T^{\mu_4\nu_4}(x_4)\bigr]\ket{0}_\text{con}
\nn
&\quad
{}= \left.
\frac{2}{\sqrt{-g(x_1)}} \cdots \frac{2}{\sqrt{-g(x_4)}} \,
\frac{\de}{\de g_{\mu_1 \nu_1}(x_1)}
\cdots \frac{\de}{\de g_{\mu_4 \nu_4}(x_4)} \, iW_\text{CFT}[g_{\mu\nu}]
\right|_{g_{\mu\nu}  =  \eta_{\mu\nu}}
\eql{Tdefinition}
\!\!\!\!,
\]
where $W_\text{CFT}[g_{\mu\nu}]$ is the generating functional of
connected correlation functions, the quantum effective action.
For example, if the theory has a path integral formulation we have
\[
e^{i W_\text{CFT}[g_{\mu\nu}]} =
\myint d[\Phi] \,
e^{i S_\text{CFT}[\Phi,  g_{\mu\nu}]}.
\]
The probe limit corresponds to the $1/\MP^4$ contributions to the
graviton-graviton scattering amplitudes.
One of the contributions at this order is
proportional to the connected momentum-space 4-point function
\[
\eql{pseudoamp}
(2\pi)^4 \de^4(p_1 + \cdots + p_4)
\ggap i\scr{M}_{\la_1 \cdots \la_4}(p_1, \ldots, p_4)
= \bra{0} \text{T}[ \tilde{T}_{\la_1}(p_1)
\cdots \tilde{T}_{\la_4}(p_4)]\ket{0}_\text{con},
\]
where $\la_1, \ldots, \la_4$ are the graviton helicities
(each equal to $\pm 2$).
Here we have taken the Fourier transform
and contracted with graviton polarization
tensors
\[
\tilde{T}_\la(p) = \ep_\la^{\mu\nu}(p)
\int d^4x \ggap e^{i \gap p \cdot x} \ggap T_{\mu\nu}(x).
\eql{polarizations}
\]
There are additional contributions to the scattering amplitude
of order $1/\MP^4$ involving intermediate graviton propagators,
which are proportional to 2- and 3-point functions of $T^{\mu\nu}$;
see Fig.~\ref{fig:GravitonPropagator}.
The 2- and 3-point functions of the CFT are determined
by conformal invariance, so we will focus on \Eq{pseudoamp},
which we call a ``pseudo-amplitude.''
It is a contribution to a physical graviton scattering amplitude,
and it is Lorentz invariant (if the polarization tensors
are defined to be Lorentz covariant, as we will discuss below).
On the other hand, it does not obey the unitarity relations
obeyed by physical graviton scattering amplitudes, such as factorization
on graviton poles.
However, the CFT is a unitary theory by itself, and
\Eq{pseudoamp} is a well-defined Lorentz-invariant CFT correlation
function.
The fact that it is a contribution to a physical graviton amplitude
is important for this work only for motivation and physical intuition.
We are interested in \Eq{pseudoamp} from the point of view of
theory rather than phenomenology;
that is, we want to see what this quantity can teach us about the CFT itself.

In this paper we  will focus on the scale anomaly of the CFT.
A scale transformation is equivalent to a Weyl
transformation of the form
$\de g_{\mu\nu} = 2 \si g_{\mu\nu}$ where
$\si$ is independent of $x$.
In a CFT the anomaly in these transformations is by definition
local in the effective action:
\[
\eql{scaleanomaly}
\de_\sigma W_\text{CFT} = c \ggap \si \myint d^4 x \sqrt{-g} \,
W^{\mu\nu\rho\si} W_{\mu\nu\rho\si},
\]
where $W_{\mu\nu\rho\si}$ is the Weyl tensor.%
\footnote{The other Weyl anomaly in 4D CFT (the ``$a$ anomaly'')
is proportional to the Euler density (c.f. \Eq{WeylAnomalies}), but this is a surface term
for $\si = \text{constant}$, and therefore does
not contribute to the scale anomaly for correlation functions. In theories that have operators with special dimensions
(such as marginal operators) there may be additional contributions
to the scale anomaly (see {\it e.g.}~\Ref{Gomis:2015yaa, Nakayama:2017oye}).
We will not investigate this possibility in this work.}
\Eq{scaleanomaly} generates a dilatation of
the 4-point function of energy-momentum
tensors
\[
\left( \sum_{i \, = \, 1}^4 p_i \cdot \frac{\d}{\d p_i}
- 4 \right) &
\scr{M}_{\la_1 \cdots \la_4}(p_1, \ldots, p_4)
{}= c \ggap P_{\la_1 \cdots \la_4}(p_1, \ldots, p_4).
\eql{scaleanomalyT4}
\]
Here $P_{\la_1 \cdots \la_4}$ is a 4th order polynomial in the momenta,
reflecting the fact that the scale anomaly \Eq{scaleanomaly} is a
local term with 4 derivatives.
\Eq{scaleanomalyT4} assumes that the polarization vectors are chosen
to be scale invariant, namely
\[
p \cdot \frac{\d}{\d p} \ep_\la^{\mu\nu}(p) = 0.
\]
Na\"\i{}vely scale invariance (dimensional analysis) implies that
the \rhs\ of \Eq{scaleanomalyT4} should vanish,
but $c \ne 0$ in all unitary CFTs,
so there is always a nonvanishing scale anomaly.%
\footnote{Conformal Ward identities imply that the coefficient
of the 2-point function of the energy-momentum tensor is proportional
to $c$, so $c > 0$ in all unitary CFTs.}

The scale anomaly is associated with the presence of UV logarithms in the
momentum space correlation functions.
To get a simple form for these logarithms, we take the massless limit $p_i^2 \to 0$,
and the forward limit $t \to 0$, where $t$ is the Mandelstam invariant
$t = -(p_1 + p_3)^2$
(we use mostly plus metric).
With an appropriate Lorentz invariant choice of polarization
tensors (see \sec{polarization} below for details)
the forward amplitude is a function
only of a single variable, the Mandelstam invariant $s = -(p_1 + p_2)^2$.
In this limit we have for example
\[
\eql{Mfdppmm}
\scr{M}_{++--}(s) &= A \gap s^2 + B \gap s^2 \ln \frac{-s}{\La^2}
+ B' \gap s^2 \ln \frac{s}{\La^2},
\]
for some dimensionless real constants $A$, $B$, $B'$,
and UV cutoff $\La$.
(All momenta and helicities are ingoing,
so this is a contribution to graviton scattering with
helicity structure $++ \to ++$.)
The pseudo-amplitude is independent of $\La$ because
the explicit $\La$ dependence is canceled by the $\La$
dependence of $A$:
\[
\La \frac{\d}{\d\La} A = 2 \gap (B + B').
\]
By \Eq{scaleanomalyT4} we also have
\[
\left( s \frac{\d}{\d s} - 2 \right) \scr{M}_{++--}(s) = (B + B') s^2 \propto c s^2,
\]
where the constant of proportionality is independent of the CFT
(and is computed below).
We see that the quantities $B$ and $B'$ in \Eq{Mfdppmm}
together determine $c$.

We obtain a sum rule for $B$ using the fact that the
coefficient of $\ln(-s)$ in \Eq{Mfdppmm} determines the
imaginary part of the pseudo amplitude.
By unitarity of the CFT this is related to a positive sum over CFT
intermediate states:
\[
\eql{opticalthm}
\Im \scr{M}_{++--} = -B \gap \pi \gap s^2
= \tfrac 12 \sum_{\ket\psi \, \ne \, \ket{0}}
\left| \bra{\psi} \text{T}[ \tilde{T}_+(p_1) \tilde{T}_+(p_2) ] \ket{0}
\right|^2.
\]
This can be viewed as a version of the optical theorem.%
\footnote{%
We emphasize that this ``optical theorem'' follows from the unitarity of the
CFT, and does not depend on the existence of an $S$-matrix for the theory
of gravitons coupled to the CFT.
(This is discussed further in \S{}2 below.)
For example, we do not need to know if a UV completion of the
gravity theory exists.}
The vacuum state does not appear in the sum over states in \Eq{opticalthm}
because the pseudo-amplitude is a
connected correlation function.
To obtain a sum rule for $B'$, we note that
crossing symmetry relates pseudo-amplitudes with different helicity structures.
For example, $2 \leftrightarrow 4$ crossing gives
\[
\eql{Mfdpmpm}
\scr{M}_{+--+}(s) &= \scr{M}_{++--}(-s)
= A \gap s^2 + B \gap s^2 \ln\frac{s}{\La^2}
+ B' \gap s^2 \ln \frac{-s}{\La^2}.
\]
The imaginary part of this amplitude therefore determines $B'$,
allowing us to write $c$ as a sum over CFT states.

To compute sums over CFT intermediate states of the form \Eq{opticalthm},
we use the state-operator correspondence in momentum space,
which was developed in detail in \Ref{Gillioz:2016jnn}.
This leads to a sum rule of the form
\[
\eql{csumrule}
c = \sum_{\substack{\O \, \subset \, T \times T \\[3pt]
\O \, \ne \, \id,  T}}
\sum_{a, \, b}
\la_{TT\O}^{(a)} \ggap \la_{TT\O}^{(b)} \ggap f_{ab}({\O}),
\]
where the first sum is over all primary operators $\O$
appearing in the $TT$ OPE.
The identity operator does not appear in the sum because it corresponds
to the contribution of the vacuum state in \Eq{opticalthm}.
The energy-momentum tensor does not appear because we will show that
its contribution is itself proportional to $c$, and has been absorbed into the
\lhs.
The sum over $a$ and $b$ runs over the different
kinematic structures in the $TT\O$ 3-point function,
$\la_{TT\O}^{(a)}$ are the corresponding OPE coefficients,
and $f_{ab}(\O)$ is a positive-definite matrix
that depends only on the quantum numbers of $\O$.
The coefficients $f_{ab}(\O)$ are kinematic in the sense
that they are completely determined by conformal invariance.
We will give explicit  formulas for these coefficients in the case where
$\O$  has spin 0 or 2.
Because \Eq{csumrule} is a sum of positive terms, any finite
number of terms gives a lower bound on $c$, potentially a very
useful constraint in the conformal bootstrap program.

There are a number of technical points that must be understood
before claiming the validity of the sum rule \Eq{csumrule}.
Correlation functions of the energy-momentum tensors in momentum
space can have UV divergences and contact terms that are not
fixed by symmetries.
In addition, the polarization tensors must depend on reference momenta
in order to be Lorentz invariant.
We will discuss these points in detail and show that the sum
rule does not depend on them.
On the other hand, we do not have a complete
understanding of IR divergences, specifically possible divergences
in the massless and forward limits $p_i^2 \to 0$, $t \to 0$
described above.
We have checked that for free CFTs
(free scalars, fermions, and vectors) there are no IR divergences
in this limit, and the sum rule is valid for these theories.
The imaginary part of pseudo-amplitudes that enters into the optical
theorem is a total rate, which is expected to be an IR safe observable
for interacting theories.
However, IR divergences in the real part could also invalidate our sum
rule.
For example a crossing symmetric contribution of the form
\[
\scr{M}_{++--}(s, t, u) &\sim (s^2 + t^2 + u^2) \left[
\ln \frac{s}{\La^2} + \ln \frac{t}{\La^2} + \ln \frac{u}{\La^2}
\right]
\]
would give a finite contribution to the imaginary part
in the forward limit $t \to 0$.
This contribution does not have the form \Eq{Mfdppmm} in the forward
limit, and therefore invalidates the sum rule.
We note that the relationship between forward amplitudes and scale
anomalies was pioneered in the original proof of the $a$-theorem
\cite{Komargodski:2011vj}%
\footnote{There is a completely independent
proof for the $a$-theorem using properties
of entanglement entropy that does not require any such assumption
\cite{Casini:2017vbe}.}
and used in non-perturbative arguments that
scale implies conformal symmetry \cite{Luty:2012ww, Dymarsky:2013pqa}.
These works also rely on the assumption that a similar
forward amplitude involving the trace of the energy-momentum tensor
is free of IR divergences.
The validity of the sum rule for free theories
is very encouraging, but a better understanding of IR divergences
would be very reassuring.
We leave this for future work.

The sum rule \Eq{csumrule} is similar to a bound on the $TTT$
OPE coefficient $\nS$ that was recently derived in
\cite{Cordova:2017zej}.
This lower bound is given by a positive sum of $TT\O$ OPE
coefficients squared, where $\O$ is a scalar primary operator.
The bound is based on the average null energy condition,
which has been recently proven using several different methods
\cite{Faulkner:2016mzt, Hartman:2016lgu}.
The bound can also be obtained from requiring unitarity of
correlation functions in the Regge limit \cite{Meltzer:2017rtf}.
It would be interesting to understand the connection between our results
and this bound.

This paper is organized as follows.
In \S{\ref{sec:definingthesumrule}} we give precise definitions and
additional details about the quantities that enter into the sum rule:
we define the sum over CFT states,
derive the optical theorem,
discuss UV divergences and contact terms,
and define the polarization tensors.
In \S{\ref{sec:canomaly}} we carry out the computations needed to evaluate
the contributions of operators with spin 0 and 2 to the sum rule,
including the energy-momentum tensor.
The final form of our sum rule
is given there.
In \S{\ref{sec:freetheories}} we study free field CFTs.
We show that they are free from IR divergences,
use them to compute the contribution of states created by the
energy-momentum tensor, and check the convergence of the sum rule.
In \S{\ref{sec:conclusions}} we give our conclusions and outlook.

The reader interested in getting an overview of our results is encouraged
to skip to \sec{canomaly}, referring to \sec{definingthesumrule}
and the appendices as needed to fill in missing details.


\section{Defining the Sum Rule}
\scl{definingthesumrule}

\subsection{Completeness Relation for CFT States in Momentum Space}

The main technical tool that enables our results is the
Minkowski momentum space
completeness relation \cite{Gillioz:2016jnn}%
\footnote{In \Ref{Gillioz:2016jnn} the completeness relation is written with the
operators inserted at a finite imaginary time in Minkowski space.
The completeness relation is independent of this imaginary time,
and here we take the limit where the imaginary time goes to zero,
which leaves the usual $i\ep$ prescription defining the
Wightman ordering of the insertion of the operators.}
\[
\eql{completenessrelation}
\id = \sum_{\O}
    \int \frac{d^4k}{(2\pi)^4}  \th(-k^2)  \th(k^0)
    [\Pi^{-1}_\O(k)]_{\mu_1\ldots\mu_\ell,\nu_1\ldots\nu_\ell}
    \tilde{\O}^{\mu_1\ldots\mu_\ell}(k) \ket{0} \bra{0} \tilde{\O}^{\nu_1\ldots\nu_\ell}(-k).
\]
The $\th$ functions enforce the condition that physical states have
positive energy and timelike momentum ($k^2 < 0$).
The sum is over all primary operators $\O$, and
$\Pi^{-1}_\O(k)$ is the inverse of the tensor appearing in the
Wightman 2-point  function:
\begin{equation}
	\bra{0} \tilde{O}^{\mu_1 \cdots \mu_\ell}(-k')
    \tilde{O}^{\nu_1 \cdots \nu_\ell}(k) \ket{0} =
    (2\pi)^4  \delta^4(k' - k)  \th(-k^2)  \th(k^0)
    [\Pi_\O(k)]^{\mu_1\ldots\mu_\ell,\nu_1\ldots\nu_\ell}.
    \eql{Wightman2point}
\end{equation}
The tensor $\Pi^{-1}_\O(k)$ can be written as a sum over
suitably normalized polarization tensors of the form%
\footnote{This follows from the K\"all\'en-Lehmann representation
for the Wightman 2-point function
\begin{equation*}
\Pi^{\mu_1 \cdots \mu_\ell, \nu_1 \cdots \nu_\ell}(k)
= \th(k^0) \th(-k^2) \sum_\la \rho_\la(k^2) \bigl[
\hat{\ep}_\la^{\mu_1 \ldots \mu_\ell}(k) \bigr]^*
\hat{\ep}_\la^{\nu_1 \ldots \nu_\ell}(k)
\end{equation*}
where the hatted polarization tensors are orthonormal:
$(\hat{\ep}_{\la'})^* \cdot \hat{\ep}_\la = \de_{\la'\la}$.
The density of states $\rho_\la(k^2)$ is positive-definite,
so we can invert the 2-point function to obtain \Eq{Wightman2point}.}
\begin{equation}
	[\Pi^{-1}_\O(k)]_{\mu_1\ldots\mu_\ell,\nu_1\ldots\nu_\ell}
    = \sum_{\la} \left[ \epsilon^{\la}_{\mu_1\ldots\mu_\ell}(k) \right]^*
   	\epsilon^{\la}_{\nu_1\ldots\nu_\ell}(k),
\end{equation}
so that the completeness relation can be written in a more compact form
\begin{equation}
\eql{completenessrelationcompact}
	\id = \sum_{\O} \sum_{\la}
    \int \frac{d^4k}{(2\pi)^4} \gap \th(-k^2) \gap \th(k^0) \gap
    \tilde{\O}_{\la}(k) \ket{0} \bra{0} \tilde{\O}_{\la}(-k),
\end{equation}
where $\O_\la = \ep^\la_{\mu_1 \cdots \mu_\ell} \O^{\mu_1 \cdots \mu_\ell}$.

\Eq{completenessrelationcompact} is simple but quite nontrivial.
A detailed derivation is given in \Ref{Gillioz:2016jnn}.
(See also \Ref{Katz:2016hxp}.)
It follows from the operator-state correspondence, together with the
fact that the Fourier transform of a primary operator contains
the correct contribution of the descendants.
To understand the latter point, note that the Fourier transform
can be written
\[
\tilde{\O}(k) = \myint d^4 x \, e^{i k \cdot x} e^{x^\mu \d_\mu}
\O(0).
\]
This shows that $\tilde{\O}(k)$ is
the unique linear combination of descendants of $\O(0)$ with
momentum $k$.
Note that as an operator relation, \Eq{completenessrelationcompact} can only be
inserted between operators with a fixed (Wightman) ordering.
For such correlation functions, \Eq{completenessrelationcompact}
says that the conformal block is essentially given by the
product of the Fourier transform
of the 3-point functions with the primary operator being inserted.

In this paper we apply the completeness relation
to the sum over states on the \rhs\ of \Eq{opticalthm}, which will be derived
in the following subsection.
This gives
\begin{equation}
    \sum_{\ket\psi \, \ne \, \ket{0}}
    \left| \bra{\psi} \text{T}[ \tilde{T}_+(p_1) \tilde{T}_+(p_2) ] \ket{0} \right|^2
    = \sum_{\O} \sum_\la
	\bigl| \M_{T_1 T_2 \to \O_{\la}}(p_1, p_2)\bigr|^2,
    \eql{opticaltheorem}
\end{equation}
where we use the notation
\begin{equation}
	(2\pi)^4 \delta^4(p_1 + p_2 - k) \gap \M_{T_1 T_2 \to \O_{\la}}(p_1, p_2)
    = \bra{0} \tilde{\O}_{\la}(-k)
    \timeordering [ \tilde{T}_1(p_1) \tilde{T}_2(p_2) ] \ket{0}.
    \eql{threepointfunction}
\end{equation}
for the 3-point functions.
As the notation suggests, $\M_{T_i T_j \to \O_{\la}}$ is itself a
pseudo-amplitude describing the inverse decay of two massless gravitons
to a CFT state with mass $m^2 = s$.

\subsection{Optical Theorem}
We now turn to the ``optical theorem'' \Eq{opticalthm}.
This follows from the combinatoric identity
\[
\eql{algebraicoptical}
\sum_{k \, = \, 0}^n (-1)^k \!\!\!\!
\sum_{\si \, \in \, \Pi(k, n - k)} \!\!
\antitimeordering[ \O(x_{\si_1}) \cdots \O(x_{\si_k})]
\timeordering[ \O(x_{\si_{k+1}}) \cdots \O(x_{\si_n})] = 0,
\]
where sum runs over all partitions $\si$ of $1, \ldots, n$ into two
groups of size $k$ and $n-k$.
This identity is proved by writing out all the (anti-)time orderings and
checking that they cancel pairwise.
We apply this to the 4-point function \Eq{pseudoamp}
Fourier-transformed to momentum
space, with the kinematics
\begin{equation}
	p_i^2 = 0,
    \qquad
    s = - (p_1 + p_2)^2 > 0,
    \qquad
    t = - (p_1 + p_3)^2 \leq 0.
    \eql{kinematics}
\end{equation}
This corresponds to a physical $2 \to 2$ scattering process with incoming momenta $p_1$ and $p_2$ and outgoing momenta $-p_3$ and $-p_4$.
With this choice, most of the terms in \Eq{algebraicoptical} do not
contribute because the momentum flow between the time-ordered
and anti time-ordered products is unphysical.
The only partition that survives is the $s$-channel,
and we obtain
\begin{equation}
	(2\pi)^4 \gap \delta^4(p_1 + \ldots + p_4) \gap \Im \M(s,t)
    = \sfrac{1}{2} \gap \bra{0} \antitimeordering [ \tilde{T}_4(p_4) \tilde{T}_3(p_3) ]
    \timeordering [ \tilde{T}_2(p_2) \tilde{T}_1(p_1) ] \ket{0}.
    \eql{opticaltheorem:1}
\end{equation}
The completeness relation \eq{completenessrelation} can now be
inserted on the right-hand side of this equality to rewrite it
as a sum of products of 3-point functions.
We then take the forward limit $p_3 \to -p_1$, $p_4 \to -p_2$
($t \to 0$), and also choose the polarizations of
$\tilde{T}_3$ and $\tilde{T}_4$ to be respectively the same
as $\tilde{T}_1$ and $\tilde{T}_2$,
so that we have forward kinematics.
The \rhs\ of \Eq{opticaltheorem:1} becomes
a sum of squares, which gives \Eq{opticaltheorem} above.

\subsection{UV Divergences}
\scl{UVdivergences}

We now consider UV divergences in momentum space CFT
correlation functions.
The CFT correlation functions in position space
at finite separation are unambiguously determined by the CFT data.
However, the Fourier transform to momentum space involves the
integration of the position space correlation function over
coincident points, potentially introducing
ambiguities from UV divergences and contact terms.
We will discuss UV divergences in this subsection.
Contact terms are discussed in the following subsection.

UV divergences
are in one-to-one
correspondence with counterterms that are generally covariant,
local, and involve relevant or marginal operators.
In 4D, the possible UV divergent counterterms for the 4-graviton
pseudo-amplitude are%
\footnote{There are only two curvature-squared counterterms because
the Euler term $\sqrt{-g} \ggap E_4$ is a total derivative.}
\[
\eql{countertermsTTTT}
\De S \sim \myint d^4 x \sqrt{-g} \left[
\La^4 + \La^2 R + \ln \La \ggap R^2
+ \ln \La \ggap W^{\mu\nu\rho\si}W_{\mu\nu\rho\si} \right],
\]
where $\La$ is a UV cutoff scale.
However, because these counterterms are local they do not contribute
to the imaginary part of the pseudo-amplitude, and therefore do not
invalidate our sum rule.%
\footnote{In fact, the power divergences can be tuned away,
and the $R^2$ term breaks conformal invariance, and is therefore
absent by assumption.
The log-divergent $W^2$ term is associated with the
$c$ anomaly that is the subject of this paper.}
For example, in \Eq{Mfdppmm} these divergences would contribute
to the coefficient $A$, which is not important for us.

More dangerous are UV divergences that can contribute to the 3-point
functions on the \rhs\ of the optical theorem \Eq{opticaltheorem}.
These involve Fourier transforms of 3-point functions of the form
\[
\eql{TTOexample}
\bra 0 \scr{O}^{\al_1 \cdots \al_\ell}(z)
\text{T}[ T^{\mu\nu}(x) T^{\rho\si}(y)] \ket 0.
\]
In the Fourier transform of this quantity, there are
no UV divergences arising from the region
where $z \to x$ or $z \to y$.
Physically this is because the integral over $z$ is computing the sum
over final states with energy and momentum fixed by the total
momentum in the initial state created by the time-ordered product,
and such sums over states cannot have UV divergences because the
4-momentum of the final state is fixed.
More formally, this can be understood from the results of
\Ref{Gillioz:2016jnn},
which showed that we can compute the Fourier transform over $z$
in \Eq{TTOexample} with $z^0$ having a finite nonzero imaginary part.
The dependence on the imaginary part exactly cancels with the normalization
of the state, and so Fourier transform can be computed without integrating
over points where $z = x$ or $y$.

The only UV divergences in the Fourier transform of \Eq{TTOexample} therefore
arise from the coincident limit $x \to y$.
These UV divergences are in one-to-one correspondence with
relevant or marginal counterterms coupling the metric to $\O$.
In 4D, the only generally covariant relevant or marginal counterterms are
\[
\eql{countertermsTTO}
\De S \sim \myint d^4 x \sqrt{-g} \left[ \La^{4-\De_1} \gap \O_1
+ \La^{2 - \De_2} \gap R \ggap \O_2 \right],
\]
where $\O_{1,2}$ are primary scalar operators
with dimensions $\De_1 \le 4$, $\De_2 \le 2$.
(For $\De_1 = 4$ or $\De_2 = 2$ the powers of $\La$ become logarithms.)
Counterterms involving higher spin operators
or descendants are forbidden by
unitarity constraints on the dimensions,
together with the fact that we can neglect total derivative
terms.
In addition the operators $\O_{1,2}$ must be singlets
under all global symmetries of the CFT.
For example, we need not consider the operator $\bar{\psi}\psi$
in free fermion theory because it is not a singlet under chiral
symmetry.
We also need not consider the identity operator, since it does
not appear in the sum over states \Eq{csumrule}.

We begin with the counterterm $\O_1$ in \Eq{countertermsTTO}.
Note that this is a perturbation of the CFT dynamics even in flat
spacetime, and therefore represents a genuine breaking of conformal
invariance.
Even if $\O_1$ is an exactly marginal operator, the presence
of logarithmic divergences proportional to $\O_1$ breaks conformal symmetry.
Because $\O_1$ has non-local correlation functions,
such terms will give rise to non-local terms in the quantum effective
action that break conformal invariance.
Such terms are therefore absent by our assumption of conformal
invariance.
For example, if $\O_1$ is relevant ($\De_1 < 4$) then such UV divergent
terms must be tuned away to get conformal invariance.

We next consider the counterterm $R \ggap \O_2$ in \Eq{countertermsTTO}.
It does not affect the dynamics of the CFT in flat spacetime,
but it does change the definition of the energy-momentum tensor,
which follows from differentiating with respect to the metric.
The term $R \ggap \O_2$ is not Weyl invariant, and therefore leads to a breaking
of conformal invariance in correlation functions involving the energy-momentum
tensor.
(Effectively, it mixes $T^{\mu\nu}$ with the operator
$\d^\mu \d^\nu \scr{O}$, which violates conformal invariance because
the latter operator is a descendant.)
The situation is therefore quite similar to the counterterms
involving $\scr{O}_1$, namely such UV divergences must be absent by the
assumption of conformal invariance.
For example, in a free scalar conformal field theory
the term $\phi^2 R$ does not have a logarithmic divergence,
consistent with the fact that this theory is conformally
invariant.

\subsection{Contact Terms}
\scl{contactterms}

We now classify the possible contact terms that can occur in
the CFT correlation functions that appear in this work,
and explain why contact term ambiguities do not affect
the sum rule.
Although they cancel in the final results,
contact terms unavoidably appear in
intermediate steps, and it is essential to include them
correctly.

Contact terms are contributions to operator products that are
localized at coincident points, for example
\[
\eql{TTcontact}
\text{T} [ T^{\mu\nu}(x) T^{\rho\si}(y) ]
= \de^4(x - y) \O^{\mu\nu\rho\si}(x) + \cdots,
\]
where we absorb a possible coefficient into the
normalization of the operator $\O$.
We can ignore contact terms in position space simply by
staying away from coincident points, but the
Fourier transform to momentum space
includes integration over all points, and
contact terms cannot be ignored.

We first claim that the only contact terms that can appear
in correlation functions
are between operators within a time ordering.
This can be understood from the fact that both Wightman and
time-ordered products are defined by
analytic continuation from Euclidean correlation functions,
which may have contact terms at coincident points.
Wightman products are defined using an $i\ep$ prescription where
operators to the left have a larger negative imaginary time.
This prevents the times of Wightman-ordered
operators from coinciding, and therefore there are no contact
terms between such operators.%
\footnote{%
This is essentially the same argument used
above to show that UV divergences do not occur when
Wightman-ordered products approach each other.}
On the other hand, time-ordered products are defined by a common
Wick rotation, which allows the location of the analytically continued operators  to coincide.
This means that we only need to consider
contact terms between energy-momentum tensors in this work,
such as \Eq{TTcontact} above.
There are also possible contact terms where 3 or more
energy-momentum tensors coincide.
Such contributions do not appear in the imaginary part of the
pseudo-amplitude $\scr{M}$, and we do not need to consider them.

Conformal invariance requires that the operator $\O$ appearing
in the contact term \Eq{TTcontact} has dimension 4.
One operator that can always appear in such contact terms
is the energy-momentum tensor itself.
As we now explain, these contact terms are completely determined
by the definition \Eq{Tdefinition} of energy-momentum tensor
correlation functions.
\Eq{Tdefinition} is equivalent to writing the metric as
\[
\eql{hdefn}
g_{\mu\nu}(x) = \eta_{\mu\nu} + h_{\mu\nu}(x),
\]
and defining $T^{\mu\nu}$ by differentiating with respect to
the source $h_{\mu\nu}$.
Equivalently, the $T^{\mu\nu}$ correlation functions are the
coefficients of the expansion of the quantum effective
action in powers of $h_{\mu\nu}$:
\[
\begin{split}
\eql{linearizedgravity}
\!\!\!\!\!\!\!
i W_\text{CFT}[g_{\mu\nu}] &= \sum_{n \, = \, 1}^\infty
\frac{1}{n!} \myint d^4 x_1 \cdots d^4 x_n \,
h_{\mu_1 \nu_1}(x_1) \cdots h_{\mu_n \nu_n}(x_n)
\\[-5pt]
&\qquad\qquad\qquad\quad{} \times
\left( \frac{i}{2} \right)^n \!\!
\bra 0 \text{T}[ T^{\mu_1 \nu_1}(x_1)
\cdots T^{\mu_n \nu_n}(x_n) ] \ket{0}_\text{con}
\end{split}
\]
That is, $n$-point functions of $T^{\mu\nu}$ are the ``interaction
terms'' with $n$ powers of $h_{\mu\nu}$ in the quantum
effective action.
However, one could have instead made a different definition of the
energy momentum tensor, namely by taking
\[
\eql{hredefn}
h'_{\mu\nu} = h_{\mu\nu} + \al_1 h^\rho_\mu h_{\rho\nu}
+ \al_2 h h_{\mu\nu} + \al_3 h^2 \eta_{\mu\nu} + O(h^3),
\]
where $\al_1, \al_2, \ldots$ are arbitrary constants,
and we raise and lower indices with the flat metric in this
expression.
We can then define a new energy-momentum tensor $T'^{\mu\nu}$ as
the coefficient in the expansion in $h'_{\mu\nu}$,
similar to \Eq{linearizedgravity}.
By writing $h'_{\mu\nu}$ in terms of $h_{\mu\nu}$, we see
that correlation functions of the different energy-momentum
tensors agree except at coincident points.
This is the contact term ambiguity for operators $\O$ in
\Eq{TTcontact} that involve the energy-momentum tensor.

\Eq{hredefn} can be thought of as a redefinition of the graviton field,
and physical (on-shell) graviton amplitudes are invariant under
such field redefinitions.
On the other hand, our pseudo-amplitude is not invariant under
these redefinitions.
However, this ambiguity cancels on both sides of our sum rule
and therefore do not affect the final result, as we now explain.

In the pseudo-amplitude $\scr{M}(p_1, \ldots, p_4)$ defined in
\Eq{pseudoamp}
we need only worry about contact terms between 1 and 2, or between
3 and 4.
Contact terms between other pairs of momenta will have a vanishing
imaginary part, and therefore do not contribute to the optical theorem.
Contact terms between 1 and 2 do change the value of the amplitude,
but they also change the the \rhs\ of the optical theorem, and the
difference cancels.
This is because the contact terms such as \Eq{TTcontact} can be viewed
as an operator statement that appears on both sides
of the sum rule.
Since the sum rule is the insertion of a complete set
of states, the addition of the same operator on both sides of the sum
rule does not affect the result.
In fact, on the \rhs\ of the sum rule the contact terms only affect
the contribution from  the states $\tilde{T}^{\mu\nu}(p) \ket{0}$,
since states created by  the energy-momentum tensor are orthogonal
to states created by other primary operators.
Once the contribution of the energy-momentum tensor is
included, the sum rule has no contact term ambiguities term by term.

Although the final result is independent of contact terms,
we must be careful to compute both sides of the sum rule
with the same contact terms.
For free field CFTs, this is a matter of using the Feynman rules
for the fields $h_{\mu\nu}$ defined in \Eq{hdefn}.
For general CFTs, we will see that the contact
terms that enter on the \rhs\ of the sum rule
are completely fixed by using the Ward identities for
conservation and conformal invariance, which encode the
definition \Eq{Tdefinition} as well as the ordering
(time-ordered or Wightman) of the correlation functions.
In a number of cases we have used both methods and checked that
they agree.
This will be discussed in more detail below.

We next consider contact terms  that
involve operators $\O$ on the \rhs\ of \Eq{TTcontact}
that are not equal to the energy-momentum tensor.
By conformal invariance, these can only occur when there
are operators with special dimensions, and therefore do not
appear in generic CFTs.
Such contact terms are associated with terms in the action
that couple the operator $\O$ to the source for $T^{\mu\nu}$,
namely the metric.
The only such term compatible with conformal invariance is
\[
\eql{contacttermpert}
\De S = \myint d^4 x \sqrt{-g} \ggap \rho \ggap \O,
\]
where $\O$ is an exactly marginal operator, and $\rho$
is an arbitrary coupling.%
\footnote{In the previous subsection, we argued that logarithmic UV
divergences in such terms violate conformal invariance.}
This gives rise to contact terms of the form
\[
\text{T} [ T^{\mu\nu}(x) T^{\rho\si}(y) ]
= \rho \de^4(x - y) (\eta^{\mu\rho}
\eta^{\nu\si} + \eta^{\mu\si} \eta^{\nu\rho})
\O(x) + \ldots
\]
If the theory contains exactly marginal operators,
there may be additional contact terms, as well as additional scale anomalies,
and the analysis of our sum rule is more complicated.
To avoid these complications, we will not consider exactly marginal
operators in this work.

\subsection{Polarization Tensors}
\scl{polarization}
We now give a precise definition of the polarization tensors appearing in the
pseudo-amplitudes.
From the point of view of graviton scattering amplitudes, a
pseudo-amplitude such as \Eq{pseudoamp} is not a natural observable.
For example, it is not invariant under spacetime diffeormorphisms,
the gauge group of gravity.
This is closely related to the fact that physical polarization
tensors for  gravitons are not  well-defined Lorentz invariant
functions of the graviton momentum \cite{Weinberg:1995mt}.
However, we can define Lorentz invariant polarization tensors by
allowing them to depend on an additional ``reference'' momentum.
This makes the pseudo-amplitude a well-defined Lorentz invariant
observable in the CFT.
There is some arbitrariness in the choice of the polarization
tensors, but this arbitrariness cancels on both sides of our sum rule.

To define the polarization tensors, we first define spin-1 polarization
vectors.
We assume that all momenta are massless ($p_i^2 = 0$, $i = 1, 2, 3, 4$),
so these can be written in terms of Weyl spinors as
\[
\ep^\mu_+(p|q) = \frac{1}{\sqrt{2}} \frac{\tilde\la \bar{\si}^\mu r}
{r\la},
\qquad
\ep^\mu_-(p|q) = \frac{1}{\sqrt{2}} \frac{\tilde{r} \bar{\si}^\mu \la}
{\tilde\la \tilde{r}},
\eql{polarizationvectors}
\]
where $p^\mu = \frac12\tilde\la \bar{\si}^\mu \la$ is the 4-momentum
of the
``massless particle'' coupling to the external operator,
and $q^\mu = \frac12 \tilde{r} \bar{\si}^\mu r$ is an arbitrary massless
reference momentum.
The normalization factors in the denominators
ensure that these are properly normalized and
scale invariant for any choice of the reference momentum.
We then define spin-2 polarization tensors by
\[
\ep^{\mu\nu}_\pm(p|q) = \ep^\mu_\pm(p|q) \ep^\nu_\pm(p|q).
\]
Practical calculations are simplified by noting that
\[
p_\mu \ep_\pm^{\mu}(p|q) = q_\mu \ep_\pm^{\mu}(p|q) = 0,
\]
which uniquely determines the polarization vectors up to a phase,
which cancels in our sum rule.

We see that a complete definition of the pseudo-amplitudes appearing
in the sum rule requires a choice of reference momentum for the
polarization tensors.
We define
\[
\eql{Mppmmdefn}
\begin{split}
\scr{M}_{++--}(p_1, p_2, p_3, p_4)
&= \ep_+^{\mu_1\nu_1}(p_1|p_2) \ep_+^{\mu_2\nu_2}(p_2|p_1)
\ep_-^{\mu_3\nu_3}(p_3|p_4) \ep_-^{\mu_4\nu_4}(p_4|p_3)
\\
&\qquad{} \times
\scr{M}_{\mu_1\nu_1 \cdots \mu_4 \nu_4}(p_1, p_2, p_3, p_4),
\end{split}
\]
where
\[
\begin{split}
\!\!\!\!\!
(2\pi)^4 & \de^4(p_1 + \cdots + p_4)
i \scr{M}_{\mu_1\nu_1 \cdots \mu_4 \nu_4}(p_1, p_2, p_3, p_4)
\\
&\quad{}
= \myint d^4 x_1 \cdots d^4 x_4
e^{i(p_1 \cdot x_1 + \cdots + p_4 \cdot x_4)} \gap
\bra{0} \text{T} [ T_{\mu_1\nu_1}(x_1) \cdots T_{\mu_4\nu_4}(x_4) ]
\ket{0}.
\end{split}
\]
We can define the forward limit in terms of the helicity spinors by
\[
\eql{fwdlimspinor}
\begin{array}{l@{\qquad\qquad}l@{\qquad\qquad}l}
\tilde{\la}_1 \to \la_1^*,
& \la_3 \to \la_1,
& \la_4 \to \la_2,
\\
\tilde{\la}_2 \to \la_2^*,
& \tilde{\la}_3 \to -\la_1^*,
& \tilde{\la}_4 \to -\la_2^*.
\end{array}
\]
In this limit, the momenta are real and
everything is a function of the two spinors $\la_{1,2}$,
and we have
\[
\begin{split}
\scr{M}_{++--}(p_1, p_2, -p_1, -p_2)
&= \ep_+^{\mu_1\nu_1}(p_1|p_2) \ep_+^{\mu_2\nu_2}(p_2|p_1)
\bigl[ \ep_+^{\mu_3\nu_3}(p_1|p_2) \ep_+^{\mu_4\nu_4}(p_2|p_1) \bigr]^*
\\
&\qquad{} \times
\scr{M}_{\mu_1\nu_1 \cdots \mu_4 \nu_4}(p_1, p_2, -p_1, -p_2).
\end{split}
\]
With the definition \Eq{Mppmmdefn},
we see that the forward pseudo-amplitude has a positive imaginary
part given by the optical theorem, as desired.

We then define the crossed pseudo-amplitude by $2 \leftrightarrow 4$
crossing:
\[
\scr{M}_{+--+}(p_1, p_2, p_3, p_4)
= \scr{M}_{++--}(p_1, p_4, p_3, p_2).
\]
In the forward limit \eq{fwdlimspinor}, this is given by
\[
\!\!
\scr{M}_{+--+}(p_1, p_2, -p_1, -p_2)
&= \ep_+^{\mu_1\nu_1}(p_1|{p_2}) \ep_-^{\mu_2\nu_2}(p_2|{p_1})
\bigl[ \ep_+^{\mu_3\nu_3}(p_1|{p_2}) \ep_-^{\mu_4\nu_4}(p_2|{p_1}) \bigr]^*
\nn
&\qquad{} \times
\scr{M}_{\mu_1\nu_1 \cdots \mu_4 \nu_4}(p_1, p_2, -p_1, -p_2),
\]
which also has a positive imaginary part.

Changing the reference momentum changes the polarization vectors
by a ``gauge transformation''
$\ep^{\mu\nu}(p) \to \ep^{\mu\nu}(p) + p^{(\mu} \xi^{\nu)}$
for some $\xi^\mu$ that can depend on $p^\mu$ as well as the reference
momentum.
This gives a contribution to the sum rule
proportional to $p_\mu \tilde{T}^{\mu\nu}(p)$, the Fourier
transform of  $\d_\mu T^{\mu\nu}(x)$.
This changes the pseudo-amplitude by a contact term,
and does not affect the sum rule, as explained
in \sec{contactterms} above.
In fact, a basis for the transverse traceless polarization tensors
is given by
\[
\epsilon_+^\mu(p) \epsilon_+^\nu(p), \ \
\epsilon_-^\mu(p) \epsilon_-^\nu(p), \ \
p_{\vphantom{+}}^{(\mu} \epsilon_+^{\nu)}(p), \ \
p_{\vphantom{-}}^{(\mu} \epsilon_-^{\nu)}(p), \ \
p^\mu p^\nu,
\]
where $\ep_\pm^\mu(p)$ are the helicity eigenstates used previously.
Except for the first two, all of these are ``pure gauge,'' and give pure
contact term contributions on the \rhs\ of the sum rule. In other words, the sum rule is saturated by contributions from the state $\ket{T}$.
We have verified explicitly that these additional ``trivial'' sum rules are
satisfied, providing an additional check on our normalizations and treatment of contact
terms.


\section{Evaluation of the Sum Rule}
\scl{canomaly}

In this section, we carry out the evaluation of the sum rule for $c$.
We first review the properties of the pseudo-amplitudes that enter
into the sum rule.
We then outline how conformal Ward identities can be used to
determine the  contribution of arbitrary operators in terms of their
OPE coefficients.
The details are relegated to Appendix~\ref{sec:conformalblocks}, including
the calculation of the contribution of operators of spin 0 and 2,
and of the energy-momentum tensor.
We conclude this section by giving the final form of the
sum rule and making some comments about the result.

\subsection{Ingredients for the Sum Rule}
We first review all the properties of the pseudo-amplitudes $\scr{M}_{++--}$
and $\scr{M}_{+--+}$ that are needed to derive the sum rule.

\begin{enumerate}
\item
The pseudo-amplitudes $\scr{M}_{++--}(p_1, \ldots, p_4)$
and $\scr{M}_{+--+}(p_1, \ldots, p_4)$ have a finite real and
imaginary part in the massless limit $p_i^2 \to 0$.%
\footnote{Note that these are the only independent helicity
amplitudes for our purposes:
all others are related to these by complex conjugation or the crossing
$1 \leftrightarrow 2$, which act trivially in the forward limit.}
\item
The resulting massless pseudo-amplitudes are Lorentz-invariant functions
of the momenta $p_1, \ldots, p_4$.%
\footnote{Helicity is not Lorentz invariant for massive momenta,
so the helicity amplitudes are Lorentz invariant only in the massless
limit.}
\item
The massless pseudo-amplitudes are real-analytic functions of complex
momenta. This follows from the standard analyticity properties of correlation
functions in quantum field theory.
\item
The pseudo-amplitudes have a finite real and imaginary part in the
forward limit $p_3 \to -p_1$, $p_4 \to -p_2$.
\item
The crossing symmetry $2 \leftrightarrow 4$ relates the two helicity
pseudo-amplitudes.
In the forward limit, this implies
\[
\scr{M}_{++--}(s) = \scr{M}_{+--+}(-s).
\]
\item
The imaginary part of the helicity pseudo-amplitudes is a positive sum
over CFT states:
\[
\Im \scr{M}_{+\pm -\mp }(s)
&= \sfrac 12 \sum_{\ket\psi \, \ne \, \ket{0}}
\left| \bra{\psi} \text{T}[ \tilde{T}_+(p_1) \tilde{T}_\pm(p_2) ] \ket{0}
\right|^2.
\]
\end{enumerate}
Except for the absence of IR divergences (points 1 and 4),
all of these properties were established
in the previous section.
The absence of IR divergences will be demonstrated for free field
theories in \sec{freetheories}.
Together, the above statements imply that
\[
\Im \scr{M}_{+\pm -\mp }(s) = A \gap s^2 + B \gap s^2 \ln \frac{\mp s}{\La^2}
+ B' \gap s^2 \ln \frac{\pm s}{\La^2},
\]
where $A$, $B$ and $B'$ are real.
The imaginary part of the pseudo-amplitudes is given by the coefficient of
$\ln(-s)$, so we can write both $B$ and $B'$ as a positive
sum over states using the optical theorem.
This is our sum rule.

\subsection{Conformal Blocks from Ward Identities}

We now outline how the Ward identities can be used to compute the
contributions of individual operators in the sum over states.
(Details are in Appendix \ref{sec:conformalblocks}.)
Because the CFT states are labeled by primary operators, this contribution can
be thought of as a conformal block in momentum space.
These conformal blocks are essentially the squares of 3-point functions
in momentum space, and these are completely fixed up to OPE coefficients
by conformal Ward identities.

It is convenient to write the 2- and 3-point functions that
we must compute in terms of matrix elements of the states
\[
\ket{\O^{\al_1 \cdots \al_\ell}} =
\O^{\al_1 \cdots \al_\ell}(i\epsilon, \pvec{0}) \ket{0},
\eql{CFTstate}
\]
where $\O$ is an arbitrary spin-$\ell$ operator (with $\ell$ even).
The operator is inserted at an infinitesimal positive imaginary time
($\ep > 0$) to give the proper Wightman ordering of correlation functions.
We have
\[
\bra{0} \tilde{\O}^{\al_1 \cdots \al_\ell}(k')
\tilde{\O}^{\be_1 \cdots \be_\ell}(k) \ket{0}
&= (2\pi)^4 \de^4(k' + k)
\bra{\O^{\al_1 \cdots \al_\ell}} \tilde{\O}^{\be_1 \cdots \be_\ell}(k) \ket{0}
\eql{OO}
\\
\begin{split}
\bra{0} \tilde{\O}^{\al_1 \cdots \al_\ell}(k)
\text{T}[ \tilde{T}^{\mu\nu}(p_1) \tilde{T}^{\rho\si}(p_2)] \ket{0}
&= (2\pi)^4 \de^4(k + p_1 + p_2)
\\
&\qquad\quad{} \times
\bra{\O^{\al_1 \cdots \al_\ell}}
\text{T}[ \tilde{T}^{\mu\nu}(p_1) \tilde{T}^{\rho\si}(p_2)] \ket{0},
\eql{TTO}
\end{split}
\]
and therefore, for physical momenta $k$ we have  (see~\Eq{Wightman2point})
\[
\Pi_\O^{\al_1 \cdots \al_\ell,\be_1 \cdots \be_\ell}(k)
&= \bra{\scr{O}^{\al_1 \cdots \al_\ell}}
\scr{O}^{\be_1 \cdots \be_\ell}(k)
\ket{0}.
\eql{Piredefn}
\]
The 2- and 3-point functions appearing in \Eqs{TTO}
and \eq{Piredefn}
are completely determined by conformal
invariance, up to operator normalization and OPE coefficients.
We can therefore use Ward identities to completely determine the
3-point functions, again up to OPE coefficients.
Specifically, we will need the Ward identities for
the conservation and tracelessness of the energy-momentum tensor,
as well as the conformal Ward identities.
The operator ordering is essential in these matrix elements,
and is reflected in the contact term structure, as explained in
\sec{contactterms} above.
The contact term structure is also reflected in the Ward
identities.
For example, the conservation Ward identities for the 3-point
functions we need are given in position space by
\[
& \frac{\d}{\d x_1^\mu}
\bra{\O^{\al_1 \cdots \al_\ell}}
\text{T}[ T^{\mu\nu}(x_1) T^{\rho\si}(x_2)] \ket 0
= 0,
\qquad
(\O \ne T)
\eql{WIOTT}
\\
\begin{split}
& \frac{\d}{\d x_1^\mu}
\bra{T^{\al\be}} \text{T}[ T^{\mu\nu}(x_1) T^{\rho\si}(x_2)] \ket 0
\\
&\qquad\quad{}
= \de^4(x_1 - x_2) \left[
\frac{\d}{\d x_{2\nu}} \bra{T^{\al\be}} T^{\rho\si}(x_2) \ket 0 \right.
- \eta^{\nu\rho} \frac{\d}{\d x_2^\tau}
\bra{T^{\al\be}} T^{\si\tau}(x_2) \ket 0
\\
&\qquad\qquad\qquad\qquad\qquad\quad{}
\left. - \eta^{\nu\si} \frac{\d}{\d x_2^\tau}
\bra{T^{\al\be}} T^{\rho\tau}(x_2) \ket 0
\right]\!,
\eql{WITTT}
\end{split}
\]
and
\[
\frac{\d}{\d z^\al} \bra 0
T^{\al\be}(z)
\text{T}[ T^{\mu\nu}(x_1) T^{\rho\si}(x_2)] \ket 0
&= 0.
\eql{WITzTT}
\]
The presence or absence of contact terms on the \rhs{}s of these
equations can be
understood as follows.
There are no contact terms between operators with Wightman
ordering (see \sec{contactterms} above),
so the only possible contact terms in these identities are
between the energy-momentum tensors in the time-ordered product.
The only operators that can appear in the contact term between
two energy-momentum tensors are the identity operator and
the energy-momentum tensor itself.%
\footnote{We assume that the theory does not have exactly marginal
operators, as discussed in \sec{contactterms}.}
Neither of these contributes a contact term in \Eqs{WIOTT}
or \eq{WITzTT},
while the energy-momentum tensor gives rise to the contact
terms on the \rhs\ of \Eq{WITTT}.
When Fourier transformed to momentum space, the contact terms
become polynomials in the momenta.
This yields the conservation Ward identities given in
Eqs.~(\ref{eq:WardIdentity:transverse:TTT}--\ref{eq:WardIdentity:transverse:TTO})
in  Appendix~\ref{sec:conformalblocks}.
Analogous results hold for the tracelessness and conformal Ward identities.

The procedure to determine the 3-point functions \Eq{TTO} is
then the following.
We write the Ward identities for conservation, tracelessness,
and conformal invariance in momentum space, taking care to
use the correct contact term structure as explained above.
We solve the Ward identities explicitly for the case where $\O$ is a scalar
operator or a spin-2 operator (2-index symmetric traceless tensor).
The latter case includes the energy-momentum tensor, which
must be treated with special care.

This in principle gives the result, but we still want to
relate the unknown coefficients in momentum space to a conventional
definition of the $TT\O$ OPE coefficients in position space.
For the case where $\O$ is the energy-momentum tensor, the
Ward identities in momentum space
imply that the contribution
of the energy-momentum tensor is proportional to $c$ itself
and determining the coefficient.
For other operators with generic dimension,
there are no contact term ambiguities, and it is straightforward
to perform the Fourier transform from position space to momentum space.
In practice, since we are only normalizing one or two OPE coefficients,
we only need to compute the Fourier transform
for several independent scalar contractions of the correlation function,
greatly simplifying the algebra.
We expect that the results obtained by analytic continuation
in the dimensions are correct even for special integer dimensions,
with the possible exception of exactly marginal scalar operators,
which have additional subtleties.

In this way, we obtain the momentum-space 3-point functions
\Eq{TTO} in terms of OPE coefficients defined in position space.
The calculations are carried out explicitly only for operators
with spin 0 or 2, but the same methods can be used for arbitrary symmetric tensor
operators, at the price of additional calculational
complexity.
The 2-point functions \Eq{Piredefn} are also determined by the
momentum space Ward identities in a similar manner.

The final sum rule is obtained by writing sum rules for both
$\Im \scr{M}_{++--}$ and $\Im \scr{M}_{+--+}$ using the
optical theorem \Eq{opticaltheorem}, where the pseudo-amplitudes are defined
in terms of the polarization vectors defined in \sec{polarization}.
We use the result
\[
\eql{ImMc}
	\Im \M_{++--} +\Im \M_{+--+} &= 5 \gap \pi \gap c \gap s^2,
\]
obtained for example from any of the free field CFTs.
We therefore define
\begin{equation}
	\Im \M_{+\pm-\mp}
    = 5 \gap \pi \gap s^2 \gap
    \sum_{\O \, \ne \, \id} \sum_{a,\, b}
    \la_{TT\O}^{(a)} \gap \la_{TT\O}^{(b)} \gap
	f^{(\pm)}_{ab}(\O),
    \eql{ImM}
\end{equation}
so that
\begin{equation}
	c =
	\sum_{\O \, \ne \, \id} \sum_{a,\, b}
    \la_{TT\O}^{(a)} \gap \la_{TT\O}^{(b)} \gap
	\bigl[ f^{(+)}_{ab}(\O) + f^{(-)}_{ab}(\O) \bigr].
    \eql{sumrulealmost}
\end{equation}
The contribution of the energy-momentum tensor
(obtained either from Ward identities, or any of the free-field
CFTs) is found to be
\[
	\sum_{a,\, b}
    \la_{TTT}^{(a)} \gap \la_{TTT}^{(b)} \gap
	\bigl[ f^{(+)}_{ab}(T) + f^{(-)}_{ab}(T) \bigr]
    = \sfrac 35 c,
    \eql{Tcontribution}
\]
and therefore the final form of our
sum rule is
\begin{equation}
	c = \sum_{\O \, \neq \, \id, \, T}
	\sum_{a,\, b} \la_{TT\O}^{(a)} \gap \la_{TT\O}^{(b)} \gap
	f_{ab}(\O),
    \eql{themainresult}
\end{equation}
where
\[
f_{ab}(\O) = \sfrac 52
\bigl[ f^{(+)}_{ab}(\O) + f^{(-)}_{ab}(\O) \bigr].
\]
For the case where $\O$ is a scalar operator, we have
\begin{equation}
    f(\O) = \frac{ 9  \pi^3  2^{2 \Delta + 2}
	\sin^2\left( \frac{\pi}{2}  \Delta \right)}
	{(\Delta - 6)^2  (\Delta - 4)^2
	\Delta^4  (\Delta + 2)^4} \ggap
	\frac{\Gamma\left( \frac{\Delta - 1}{2} \right)
    \Gamma\left( \frac{\Delta + 1}{2} \right)}
	{\Gamma\left( \frac{\Delta + 4}{2} \right)^2},
	\eql{fscalar}
\end{equation}
where the normalization of the $\la_{TT\O}$ OPE coefficients is defined
to be that of \Ref{Cordova:2017zej}.
For spin-2 operators there are 2 independent $TT\O$ OPE coefficients,
and the result is more complicated.
It is given in \Eq{conformalblock:spin2} in
Appendix \ref{sec:conformalblocks}.

\begin{figure}
	\centering
	\includegraphics[width=11cm]{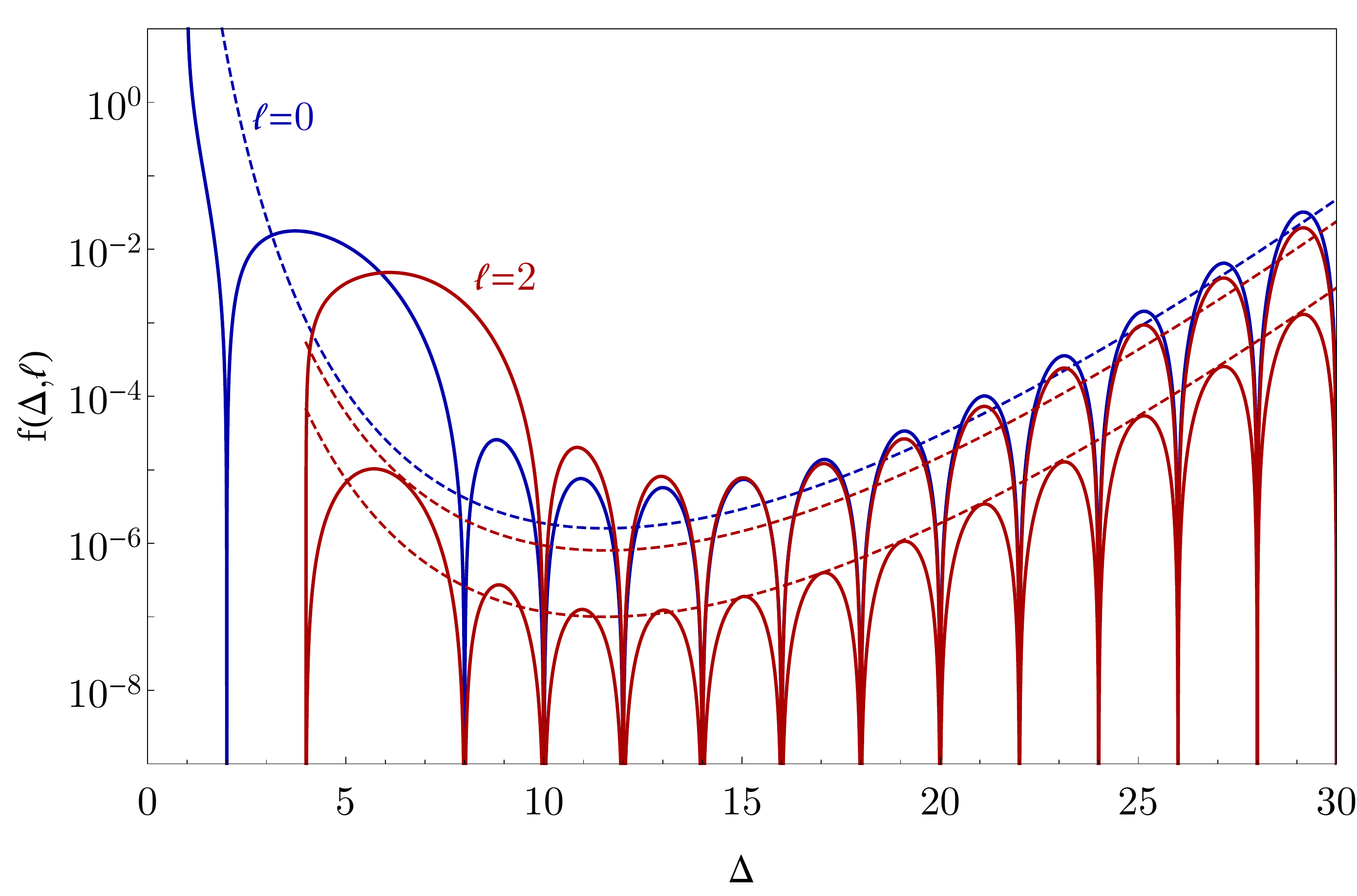}
    \begin{minipage}{5.5in}
	\caption{\small
    Numerical value of the conformal blocks \eqref{eq:fscalar}
    for an operator of spin zero and
    \eqref{eq:conformalblock:spin2} for spin two,
    showing the eigenvalues of the $2 \times 2$ matrix in the latter case.
    The dashed lines correspond to the amplitude of the asymptotic expressions
    \eqref{eq:conformalblock:scalar:asymptotics}
    and~\eqref{eq:conformalblock:spin2:asymptotics} with the $\sin^2$ factored out.}.
	\label{fig:blocks}
    \end{minipage}
\end{figure}

The numerical values of the coefficients are plotted in
Fig.~\ref{fig:blocks} as a function of the dimension $\De$ of the
operator $\O$, along with asymptotic approximations that
are valid for large $\De$.


\subsection{Discussion of Results}

We now discuss some features of the functions $f(\O)$ appearing in our sum
rule \Eq{themainresult}.
Some of these features are shared by the sum rule of \Ref{Gillioz:2016jnn}
and the bound of \Refs{Cordova:2017zej, Meltzer:2017rtf}.
\begin{enumerate}
\item
For scalar $\O$, $f(\O)$ diverges as $\De \to 1$ (the unitarity limit):
\[
f(\O) \sim \frac{1}{\De - 1}.
\]
This means that for any scalar operator with dimension $\De = 1 + \ep$,
the $TT\O$ OPE coefficient must vanish at least as fast as $\sqrt{\ep}$
as $\ep \to 0$.
\item
For scalar $\O$ we have a double zero as $\De \to 2$:
\[
f(\O) \sim (\De - 2)^2.
\]
This means that operators with dimension 2 do not contribute to the
sum rule, and the contribution of operators with dimension near 2
are highly suppressed.
This zero ensures that the operator $\phi^2$
in free scalar theory does not contribute to the sum rule, despite
the fact that the OPE coefficient $TT\phi^2$ is nonzero,
as explained in~\sec{freetheories} below.
In fact, the coefficient of the double zero can be used
to check the normalization of $f(\O)$,
as explained in~\S\ref{appsec:Delta2}.
\item
For scalar $\O$, we have double zeros as $\De \to 8, 10, 12, \ldots\,$:
\[
f(\O) \sim \bigl( \De - (8 + 2n) \bigr)^2,
\qquad n = 0, 1, 2, \ldots
\]
The presence of these zeros can be understood from the fact that
the sum rule must be valid in generalized free field theories.
A generalized free field theory is obtained from a large-$N$ theory by
taking the $N \rightarrow \infty$ limit ``from the beginning'', as follows.
If $c \sim N$, we define $\hat{T} = T/\sqrt{N}$, and our sum rule is
\[
\hat{c} = {\rm anomaly\ coefficient\ in}\
\langle \hat{T}\hat{T}\hat{T}\hat{T}\rangle_{\rm con}
= \sum_{\hat{\cal O}} \lambda_{\hat{T}\hat{T}\hat{\cal O}}^2 f(\hat{\cal O}),
\]
where we normalize the operators so that $\avg{\hat{\scr{O}}\hat{\scr{O}}} \sim N^0$.
The connected correlation function vanishes in the $N \to \infty$ limit,
as does its anomaly, so $\hat{c} = 0$.
The only operators $\hat{\cal O}$ appearing in the $\hat{T}\hat{T}$ OPE
are double-trace operators $\sim T \d^{2n} T$.
For such operators the OPE coefficients $\hat{T}\hat{T}\hat{\cal O}$ are
nonzero, so the coefficients $f(\hat{\cal O})$ must vanish,
as we have found from explicit computation.
In this sense the sum rule is trivially satisfied for generalized free field
theories.

\item
It is also interesting to take the large-$N$ limit of our sum rule
in a way that keeps the information in the connected correlation function.
To do this, we compute our sum rule for large $N$, and take the
$N \to \infty$ limit at the end.
At large $N$ a generic single trace operator
$\scr{O}$ has $\lambda_{TT\cal O} \sim N$ and
$f({\cal O}) \sim 1/{\langle {\cal OO} \rangle} \sim 1/N$,
so the contribution of every such term in the sum is $\de c \sim N$,
consistent with $c \sim N$.
Things are more interesting in a large-$N$ holographic theory.
In such a theory all the single trace operators in the $TT$ OPE
have dimensions larger than $\Delta_{\rm gap} \gg 1$
($\Delta_{\rm gap}$ may grow with $N$).
For a double trace operator ${\cal O} \sim T \partial^{2n} T$,
we have $\lambda_{TT{\cal O}} \sim N^2$ and $\langle{\cal OO}\rangle \sim N^2$,
and $f({\cal O}) \sim  1/N^4$.
This gives a contribution $\delta c \sim N^0$ to our sum rule,
which is negligible in the large-$N$ limit (in agreement with our
discussion of the generalized free field limit above).
However, our sum rule can still hold in such a large-$N$ theory.
For example, it may be saturated at large $N$ by contributions from the
single-trace operators with dimensions above $\Delta_{\rm gap}$.
In theories where $\Delta_{\rm gap}$ grows with $N$, the number of double
trace operators below the gap grows with $N$, and it is conceivable that
the sum of their contributions is of order $N$.
The various scenarios are beyond the scope of our present work, but it is
clear that large-$N$ counting alone does not invalidate our sum rule
in holographic theories.
It would be interesting to see what can be learned from the AdS description
of such a theory, and we make some brief comments on this in the conclusions.
\item
For $\O$ with spin 2, both eigenvalues of
$f_{ab}(\O)$ have double zeros at
$\De = 10, 12, 14, \ldots\,$,
while at $\De = 8$  one of the eigenvalue has a double zero.
Some or all of these zeros can be understood from the
fact that the double trace operators $T \d^{2 + 2n} T$ appear in
the  spin-2 terms in the OPE of the disconnected part of
the  $\avg{TTTT}$ correlation function.
A definite statement would require a classification of the
spin-2 double trace operators appearing in the $TT$ OPE, which
we do not attempt here.
\item
For $\O$ with spin 2,
both eigenvalues of $f_{ab}(\O)$ vanish as $\De \to 4$.
One of the eigenvalues is a double zero,
the other is a simple zero. These zeros can be understood from free CFTs, as explained in
\sec{Tgeneral} below.
\item
It is interesting to note that
even though for a spin-2 operator $f(\O)$ has zeros at the unitarity bound,
higher-spin conserved currents with $\ell \geq 4$
do contribute to the sum rule.
This can be seen from the free field calculations of the next
section.
\item
In order for the sum over $\O$ to converge, the $TT\O$ OPE coefficients must
be bounded for large $\De$.
We have
\[
f(\O) \sim \frac{4^\Delta}{\Delta^{16}} \sin^2\left(\pi\Delta/2\right) .
\]
A necessary condition for convergence is therefore
\[
\left|\la_{TT\O}\right|^2
\lsim \frac{\Delta^{15}}{4^\Delta} \frac{1}{\sin^2\left(\pi\Delta/2\right)} .
\]
This is consistent with the implication of the OPE convergence found in \Refs{Pappadopulo:2012jk, Rychkov:2015lca}.
Note that this bound becomes very weak near
integer values of $\De$.

\item
Our function $f(\scr{O})$ is similar to a related quantity
appearing in the bound derived in \Ref{Cordova:2017zej}:
\[
	\nS \ge \sum_{\O \, = \, \text{scalar}} \la_{TT\O}^2 f_{\text{CMT}}(\O),
\]
while our result gives the bound
\[
\nS + 6 \nF + 12 \nV \ge \sum_{\O \, = \, \text{scalar}}
\la_{TT\O}^2 \frac{f(\O)}{c_\text{scalar}}
\]
The relation between the coefficients in these bounds is
\begin{equation}
	\frac{1}{c_\text{scalar}}  f(\O) = \frac{5}{96}  (\Delta - 2)^2  f_\text{CMT}(\O).
\end{equation}
\end{enumerate}

The fact that the zeros of the function $f(\O)$ are generically double zeros follows
from the fact that this function is given by a square of a momentum-space 3-point
function, which has zeros at special values of the dimension.
The exception is the single zero that occurs for spin-2 operators at $\De = 4$.
This is due to the fact that the tensor $\Pi_{\scr{S}}$
that normalizes the states has a zero eigenvalue
at $\De = 4$.

\begin{figure}
	\centering
	\includegraphics[width=11cm]{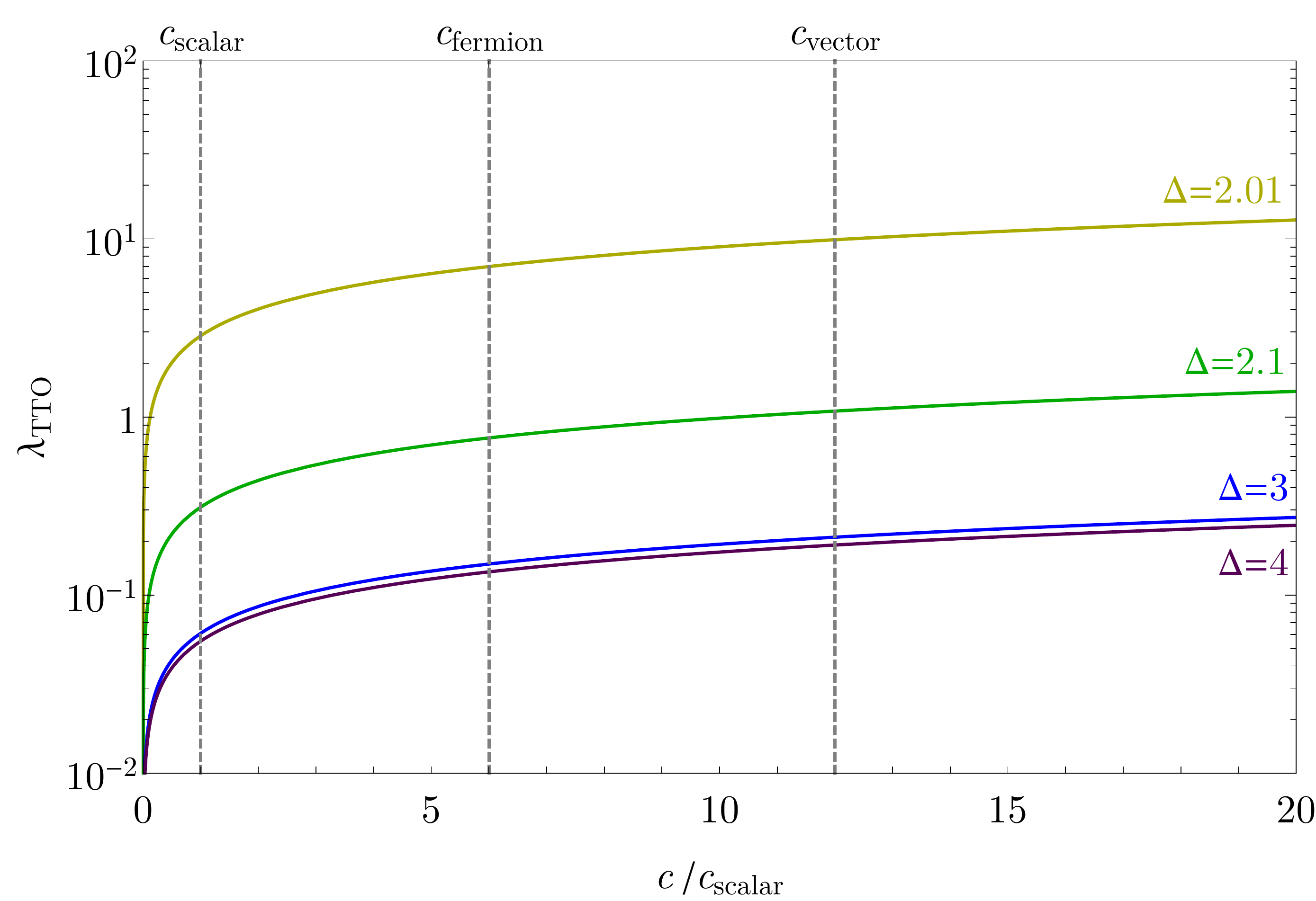}
    \begin{minipage}{5.5in}
	\caption{
    Upper bound on the size of the OPE coefficient
    for a scalar operator of dimension $\Delta$
    entering the $T \times T$ OPE,
    for a given value of the $c$ anomaly.
    The bound is weaker for operators of dimension close to 2.
    This can be alternatively be understood as a lower bound on $c$
    for a fixed value of the OPE coefficient.}
	\label{fig:bound}
    \end{minipage}
\end{figure}

Because our result gives $c$ as a sum of positive terms,
any lower bound on the $TT\O$ OPE coefficients immediately gives
a lower bound on $c$.
This is the kind of result that is usually obtained
using the numerical bootstrap,
but here it is obtained analytically.
To give some idea of the normalization, we present such a ``bootstrap
plot'' in Fig.~\ref{fig:bound}.
The anomaly coefficient $c$ is a natural input parameter for the
conformal bootstrap.
Given a value for $c$, our sum rule bounds the value of all $TT\O$
OPE coefficients,
however with ``blind spots'' at special
double-trace values of the scaling dimension as discussed above.
It is our hope that this sum rule will be a useful input into the
program of bootstrapping the energy-momentum tensor.


\section{Free Theories}
\scl{freetheories}

In this section we investigate our sum rule in the case of free CFTs,
namely massless free scalars, fermions, and vectors.
In these theories we can check that the pseudo-amplitudes appearing in our
sum rule have finite massless and forward limits.
We also use free field theories to fix the constant of proportionality
between the imaginary part of the pseudo-amplitude and $c$,
give an independent way of computing
the contribution of the energy-momentum tensor to the sum rule,
and check of the normalization of the contribution of scalar
operators.

\subsection{IR Finiteness}

As explained in \sec{canomaly}, a crucial assumption in the
derivation of our sum rule is that the pseudo-amplitude $\mathcal{M}$
defined in~\Eq{pseudoamp} is finite under both the massless limits
$m^2\equiv -p_i^2\to 0$ and the forward scattering limit
$t\equiv-(p_1+p_3)^2\to 0$.
It is not sufficient to require that the imaginary
part is finite (see \sec{intro}).

We have verified that this IR finiteness
holds for all the three free  CFTs.
To perform this check, we used the method of expansion by
regions~\cite{Smirnov:2002pj}.
We find that all the regions in all the relevant diagrams are free from
IR divergences provided that the polarization tensors are chosen to be
transverse and traceless.
Details of this computation are given in Appendix~\ref{appsec:IRfree}.
With this check, we know that the sum rule is valid for free theories.

\subsection{Scattering Rates}

We begin by computing the imaginary part of the pseudo-amplitudes
$\Im\mathcal{M}_{+\pm-\mp}(s)$.
This is proportional to the total scattering rate
$\sigma \left(h_+ h_\pm\to \text{CFT} \right)$, where $h$ denotes the
graviton field.
In free theories, this computation can be carried out using the usual Feynman
rules, given in the Appendix~\ref{appsec:IRfree}.
The results for a real scalar are
\begin{subequations}
\begin{align}
\eql{ScalarRates}
2\Im\mathcal{M}_{++--}^\text{scalar} &= \frac{1}{{320\pi }}{s^2} , \\
2\Im\mathcal{M}_{+--+}^\text{scalar} &= \frac{1}{{480\pi }}{s^2} .
\end{align}
\end{subequations}
For a free Dirac fermion we have
\begin{subequations}
\begin{align}
2\Im\mathcal{M}_{++--}^\text{fermion} &= \frac{3}{{160\pi }}{s^2} , \\
2\Im\mathcal{M}_{+--+}^\text{fermion} &= \frac{1}{{80\pi }}{s^2} ,
\end{align}
\end{subequations}
and for a free vector theory we have
\begin{subequations}
\begin{align}
2\Im\mathcal{M}_{++--}^\text{vector} &= \frac{3}{{80\pi }}{s^2} , \\
2\Im\mathcal{M}_{+--+}^\text{vector} &= \frac{1}{{40\pi }}{s^2} .
\eql{Vectorrates}
\end{align}
\end{subequations}
These result are all consistent with the statement that the sum of
the two scattering rates is proportional to the $c$-anomaly,
by \Eq{cnsnfnv}.
In fact, we can use any of the three free theories described above to fix the constant of proportionality in \Eq{ImMc}.

\subsection{Contributing Operators}\scl{contributing}

We now discuss which operators $\O$ can appear on the \rhs\ of the sum rule
in the case of free field theories.
To get a non-zero contribution, both the OPE coefficient $\la_{TT\O}$
and the coefficient $f(\O)$ in~\Eq{csumrule} need to be nonzero.
In free theories, the energy-momentum operator $T$
is parity-even and contains 2 powers of the free fields ($\phi$, $\psi$, or $A_\mu$).
Therefore, in order to appear in the $T \times T$ OPE, $\O$ must be an $\ell = \text{even}$
operator made of either 2 or 4 powers of the free field.
If $\O$ has four powers of the field, the 3-point function $\langle TT\O\rangle$ can
be factorized into a product of 2-point functions.
In this case, $\la_{TT\O}$ can be nonzero only if $\O\sim \partial^{2n}TT$
$(n \in \mathbb{N})$.
Such operators have scaling dimensions $\Delta = 8 + 2 \gap n$.
We find in \sec{conformalblocks} that the coefficients $f(\O)$ have zeros precisely at
these dimensions, which are the ``blind spots'' of the sum rules.
Therefore, no operator with four powers of the field contributes to the sum.
We conclude that $\O$ must have 2 powers of the free field.

Another observation is that the scalar operators $\phi^2$, $\bar\psi\psi$, and $F_{\mu\nu}F^{\mu\nu}$ do not contribute to the sum rule.
This is easy to understand for the fermion and vector theory cases, as chiral symmetry
forbids their presence in the $T \times T$ OPE~\cite{Cordova:2017zej}.
On the other hand, in the free scalar theory, $\la_{TT\phi^2}$ is non-zero.
However, we find by direct computation in free field theory, or using the results
of \sec{conformalblocks}, that $f(\phi^2)$ is zero in this case.

To summarize, the contributing operators $\O$ are built from two powers of the
field and spin-even, but not $\phi^2$, $\bar\psi\psi$, or $F_{\mu\nu}F^{\mu\nu}$.

\subsection{A Complete Example: The Free Scalar}
\scl{scalarexample}

We now verify the sum rule explicitly in the free scalar theory.
We showed above that the primary operators contributing to the sum rule
have the form
$\O^{\al_1\cdots\al_\ell}\sim \phi \partial^{\al_1}\cdots\partial^{\al_\ell} \phi$
with $\ell$ even.
The primary operators are easily seen to be symmetric traceless spin-$\ell$ tensors.
Because they have dimension $\De = \ell + 2$, they saturate the unitarity bounds,
and are therefore conserved currents.

\subsubsection*{Partial wave expansion}
A consequence of $\O^{\al_1\cdots\al_\ell}$ being conserved currents is that the sum rule expansion coincides with the usual partial wave expansion. This is because the three-point functions must satisfy the Ward identity
\begin{equation}
	(p_1+p_2)_{\al_i} \gap \bra{\O^{\al_1\cdots\al_\ell}}
    \timeordering [ \tilde{T}_1(p_1) \tilde{T}_2(p_2) ] \ket{0} = 0.
\end{equation}
In the center-of-mass frame where $(p_1 + p_2)_\al = \sqrt{s} \gap (1, \pvec{0})$,
this Ward identity implies that only the spatial components of the three-point function
are non-zero, \textit{i.e.}~those with tensor indices $\al_i = 1,2,3$.
Therefore, the three-point function forms a spin-$\ell$ representation of the spatial
rotation group $SO(3)$.
There is no $s$-wave contribution, corresponding to the fact that the operator $\phi^2$
does not contribute to the pseudo-amplitude.

The correspondence between the OPE and the partial wave expansion can be used to compute all
of the contributions to the sum rule in the free scalar case.
The imaginary part of the pseudo-amplitude is computed as the scattering rate of the process
$hh \to \phi\phi$, whose amplitude $\mathcal{M}_{hh\to\phi\phi}(\theta,\varphi)$ can be
expanded in terms of the spherical harmonics $Y_{\ell m}(\theta,\varphi)$,
where $\theta$ and $\varphi$ are the scattering angles in the center-of-mass frame.
Then each spherical harmonic corresponds exactly to one operator of the free scalar theory.
Concretely, we consider the cases where the initial gravitons have helicities $++$ and $+-$.
We obtain
(see \S\ref{appsec:ImaginaryPart} for details):
\begin{subequations}
\begin{align}
\M_{h_+ h_+ \to \phi \phi }\left( {\theta ,\varphi } \right)
&= \frac{s}{{16}}\left( {3  {{\cos }^2}\theta  - 1} \right)
\nn
&= \frac{s}{4} \sqrt{\frac{\pi}{5}} \, Y_{20}(\theta ,\varphi),
\eql{PWA1}
\\
\M_{h_+ h_- \to \phi \phi }\left( {\theta ,\varphi } \right)
&= \frac{s}{{16}} \, { e^{4i  \varphi }}\left( {1 - {{\cos }^2}\theta } \right)
\nn
&= 6 \sqrt\pi s \sum\limits_{\text{even } \ell \, = \, 4}^\infty
{\sqrt {\left( {2\ell  + 1} \right)\frac{{\left( {\ell  - 4} \right)!}}{{\left( {\ell  + 4} \right)!}}} \,
\ggap Y_{\ell 4}(\theta ,\varphi)} . \eql{PWA2}
\end{align}
\end{subequations}
From \Eq{PWA1}, we see that for initial helicity $++$,
only spin-2 operators contribute to the scattering rate.
In the free scalar theory, this corresponds to the contribution of energy-momentum tensor,
so $T$ saturates this scattering rate.
This result can also be obtained by direct computation of the $T$ contribution, as we show below.
Using \Eq{ScalarRates}, this shows that $T$ contributes as $\frac{3}{5}c$ in our sum rule.
As we will explain in \sec{Tgeneral}, this result can be extended to general CFTs.

For initial helicity $+-$ \Eq{PWA2} implies that an infinite number of operators
with even spin $\ell\ge 4$ contribute to the rate.
In fact, we can identify the contribution of each operator to be
\begin{align}
	\left| \M_{T_+ T_- \to \phi^2 \partial^\ell} \right|^2 &= \frac{1}{\pi} \int{\frac{d\Omega}{4\pi} \left| \mathcal{M}_{h_+h_-\to\phi\phi}(\theta, \varphi) \right|^2} = \frac{9}{\pi}  s^2
    \left( {2\ell  + 1} \right)\frac{{\left( {\ell  - 4} \right)!}}{{\left( {\ell  + 4} \right)!}} \nonumber \\
    &= 6  \left(2\ell+1\right)
    \frac{6! \left(\ell-4\right)!}{\left(\ell+4\right)!}
    \times \Big( 2\Im\mathcal{M}_{+-}^\text{scalar} \Big),
    \eql{partialwaveexpansion}
\end{align}
for $\ell=4, 6, 8, \ldots$.

\subsubsection*{Operator product expansion}
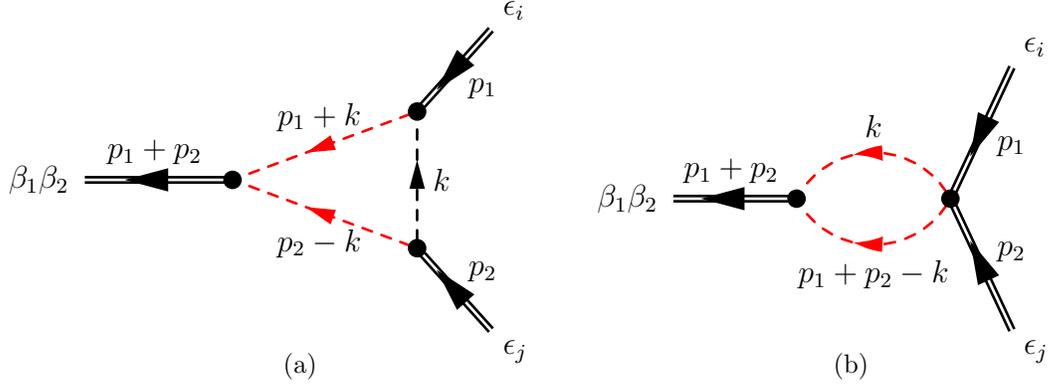
\begin{figure}[t]
\centering
\vspace{0.3cm}
 \subfigure[]{
 \centering\label{subfig:Scalar3ptTriangle}
\begin{fmffile}{Scalar3ptTriangle}
\begin{fmfgraph*}(60,40)
\fmfleft{i}
\fmfright{o2,o1}
\fmflabel{$\beta_1\beta_2$}{i}
\fmflabel{$\epsilon_i$}{o1}
\fmflabel{$\epsilon_j$}{o2}
\fmfv{decor.shape=circle,decor.filled=full,decor.size=3thick}{v1}
\fmfv{decor.shape=circle,decor.filled=full,decor.size=3thick}{v2}
\fmfv{decor.shape=circle,decor.filled=full,decor.size=3thick}{v}
\fmf{dbl_plain_arrow,label=$p_1+p_2$,label.side=right}{v,i}
\fmf{dbl_plain_arrow,label=$p_1$,label.side=left}{o1,v1}
\fmf{dbl_plain_arrow,label=$p_2$,label.side=right}{o2,v2}
\fmf{scalar,tension=0.4,label=$k$,label.side=right}{v2,v1}
\fmf{scalar,tension=0.4,f=(1,,0,,0),label=$p_1+k$,label.side=right}{v1,v}
\fmf{scalar,tension=0.4,f=(1,,0,,0),label=$p_2-k$,label.side=left}{v2,v}
\end{fmfgraph*}
\end{fmffile}
 }\hspace{1cm}
 \subfigure[]{
\centering\label{subfig:Scalar3ptBubble}
\begin{fmffile}{Scalar3ptBubble}
\begin{fmfgraph*}(50,35)
\fmfleft{i}
\fmfright{o2,o1}
\fmflabel{$\beta_1\beta_2$}{i}
\fmflabel{$\epsilon_i$}{o1}
\fmflabel{$\epsilon_j$}{o2}
\fmfv{decor.shape=circle,decor.filled=full,decor.size=3thick}{vi}
\fmfv{decor.shape=circle,decor.filled=full,decor.size=3thick}{vo}
\fmf{dbl_plain_arrow,label=$p_1+p_2$,label.side=right}{vi,i}
\fmf{dbl_plain_arrow,label=$p_1$,label.side=left}{o1,vo}
\fmf{dbl_plain_arrow,label=$p_2$,label.side=right}{o2,vo}
\fmf{scalar,right=0.6,tension=0.4,f=(1,,0,,0),label=$k$,label.side=right}{vo,vi}
\fmf{scalar,left=0.6,tension=0.4,f=(1,,0,,0),label=$p_1+p_2-k$,label.side=left}{vo,vi}
\end{fmfgraph*}
\end{fmffile}
 }\vspace{0.3cm}
 \begin{minipage}{5.5in}
 \caption{\small
 Feynman diagrams for the three-point function $\bra{T^{\beta_1\beta_2}} \timeordering[ \tilde{T}_i(p_1) \tilde{T}_j(p_2)] \ket{0}$ in the free scalar theory. Here we use straight double lines for $T$ operators and dashed singled lines for scalars. Red lines indicate Wightman propagators, black lines ordinary Feynman propagators.}
 \label{fig:Scalar3ptDiagrams}
 \end{minipage}
\end{figure}

\begin{figure}[t]
\vspace{0.2cm}
 \centering\label{subfig:Scalar2ptBubble}
\begin{fmffile}{Scalar2ptBubble}
\begin{fmfgraph*}(60,35)
\fmfleft{i}
\fmfright{o}
\fmflabel{$\al_1\al_2$}{i}
\fmflabel{$\beta_1\beta_2$}{o}
\fmfv{decor.shape=circle,decor.filled=full,decor.size=3thick}{vi}
\fmfv{decor.shape=circle,decor.filled=full,decor.size=3thick}{vo}
\fmf{dbl_plain_arrow,label=$p_1+p_2$,label.side=right}{vi,i}
\fmf{dbl_plain_arrow,label=$p_1+p_2$,label.side=right}{o,vo}
\fmf{scalar,right=0.6,tension=0.4,f=(1,,0,,0),label=$k$,label.side=right}{vo,vi}
\fmf{scalar,left=0.6,tension=0.4,f=(1,,0,,0),label=$p_1+p_2-k$,label.side=left}{vo,vi}
\end{fmfgraph*}
\end{fmffile}
 \vspace{0.2cm}
 \begin{minipage}{5.5in}
 \caption{\small Feynman diagram for two-point function $\bra{0} \tilde{T}^{\al_1\al_2}(-p_1-p_2) \tilde{T}^{\beta_1\beta_2}(p_1+p_2) \ket{0}$ in the free scalar theory. Straight double lines are for the $T$ operator and dashed singled lines for scalars. Red lines indicate Wightman propagators.} \label{fig:Scalar2ptDiagrams}
 \end{minipage}
\end{figure}
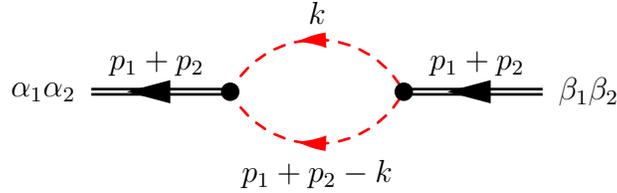

Now we use the OPE~\Eq{opticalthm} and compare it with the results of
the partial wave expansion.
The lowest-dimensional operator that appears in the sum is the energy-momentum tensor.
According to \Eq{completenessrelation}, its contribution is
\begin{align}
\begin{split}
	\!\!\!\!\!
    \left| \M_{T_1 T_2 \to T} (p_1, p_2) \right|^2 &=
	[\Pi_T^{-1}(p_1 + p_2)]_{\al_1\al_2, \beta_1\beta_2}
    \\
    &\qquad {}\times
	\bra{0} \antitimeordering[ \tilde{T}_1(-p_1) \tilde{T}_2(-p_2) ]
	\ket{T^{\al_1\al_2}}
	\bra{T^{\beta_1\beta_2}}
	\timeordering[ \tilde{T}_1(p_1) \tilde{T}_2(p_2)] \ket{0} .
    \end{split}
\end{align}
We compute the three-point function
$\bra{T^{\beta_1\beta_2}} \timeordering[ \tilde{T}_1(p_1) \tilde{T}_2(p_2)] \ket{0}$
and the two-point function
$[\Pi_T(p_1+p_2)]^{\al_1\al_2,\beta_1\beta_2} \sim \bra{0} \tilde{T}^{\al_1\al_2}(-p_1-p_2) \tilde{T}^{\beta_1\beta_2}(p_1+p_2) \ket{0}$
by evaluating the contributing Feynman diagrams, which are shown respectively in
Figs.~\ref{fig:Scalar3ptDiagrams} and~\ref{fig:Scalar2ptDiagrams}.
Note that these loop diagrams are all UV finite due to the presence of Wightman propagators,
marked in red in the figures.
A straightforward (but tedious) calculation yields
\begin{align}
\left| \M_{T_1 T_2 \to T} \right|^2
= \frac{{{s^2}}}{{135 \times 64\pi }}
&{\epsilon_1^{{\mu _1}{\nu _1}}}{\epsilon_2^{{\mu _2}{\nu _2}}}(\epsilon_1^{{\mu _3}{\nu _3}})^*(\epsilon_2^{{\mu _4}{\nu _4}})^*{\eta_{{\mu _1}{\mu_2}}}{\eta _{{\mu _3}{\mu _4}}}
\nonumber \\
& \times
\bigl[ {64({{\eta _{{\nu _1}{\nu_3}}}{\eta _{{\nu _2}{\nu _4}}} + {\eta _{{\nu _1}{\nu _4}}}{\eta _{{\nu _2}{\nu _3}}}}) - 37{\eta_{{\nu_1}{\nu_2}}}{\eta _{{\nu _3}{\nu _4}}}} \bigr] ,
\end{align}
which separates into the two helicity structures as
\begin{subequations}
\begin{align}
\left| \M_{T_+ T_+ \to T} \right|^2
&= \frac{{{s^2}}}{{320\pi }} , \\
\left| \M_{T_+ T_- \to T} \right|^2
&= 0 .
\end{align}
\end{subequations}
As anticipated from the partial wave expansion,
we see indeed that the $T$ contribution saturates the scattering rate for initial helicities $++$,
and gives no contribution in the case $+-$.

Contributions from the other operators can be evaluated in the same way.
For example, it can be verified that the scalar operator $\phi^2$ does not contribute
to either helicity structure, and that the spin-4 operator $\partial^4\phi^2$ gives
\begin{subequations}
\begin{align}
\left| \M_{T_+ T_+ \to \partial^4\phi^2 } \right|^2 &= 0 , \\
\left| \M_{T_+ T_- \to \partial^4\phi^2 } \right|^2
&= \frac{9}{{4480\pi }}{s^2} = \frac{27}{28} \times \Big( 2\Im\mathcal{M}_{+-}^\text{scalar} \Big).
\eql{spin4contribution}
\end{align}
\end{subequations}
Note that the fractional contribution of the spin-four operator precisely matches
\Eq{partialwaveexpansion} with $\ell = 4$.
If we use the partial expansion further, we can obtain the contribution of higher-spin
operators:
\begin{equation}
	\frac{\left| \M_{T_+ T_- \to \phi^2 \partial^\ell} \right|^2}
	{2\Im\mathcal{M}_{+-}^\text{scalar}}
	= \frac{27}{28},\, \frac{13}{420},\, \frac{17}{4620},\cdots
	\qquad \text{for } \ell=4,6,8,\ldots
\end{equation}
We can check that this converges to the expected value.
We can also check the rate of convergence:
\begin{equation}
	\sum_{\text{even}~\ell \, = \, 4}^{\ell_\text{max}}
	\frac{\left| \M_{T_+ T_- \to \phi^2 \partial^\ell} \right|^2}
	{2\Im\mathcal{M}_{+-}^\text{scalar}}
	- 1 = - \frac{6!  (\ell_\text{max} - 2)!}{(\ell_\text{max} + 4)!}
	\sim \ell_\text{max}^{-6}.
\end{equation}
The convergence of this sum is very fast, but not exponentially fast, unlike the position-space OPE at separated points~\cite{Pappadopulo:2012jk, Rychkov:2015lca}.

\subsection[T contribution in General CFTs]{$T$ Contribution in General CFTs}
\scl{Tgeneral}

The contribution of the energy-momentum tensor to the scattering rates can be computed
similarly in the theories of free fermions and free vectors,
and we also find in both cases that it saturates the $++$ scattering rate,
and therefore gives a contribution of $3c/5$ to the sum rule \eq{ImM}.

In fact, we can argue that this result extends to arbitrary CFTs, as follows.
The tensor structures of the 3 free CFTs form a basis for the 3 independent
tensor structures in the $TTT$ 3-point function.
That is, we can write the 3-point function in a general CFT as
\[
\avg{TTT} = \nS \avg{TTT}_\text{s} + \nF \avg{TTT}_\text{f}
+ \nV \avg{TTT}_\text{v},
\]
where the subscripts s, f, v refer to real scalars,
Dirac fermions, and free vectors.
In this way $\nS$, $\nF$, and $\nV$ become the OPE coefficients in
general CFTs.

In a free theory of real scalars, Dirac fermions, and vectors,
these OPE coefficients simply count the number of each type of
free field.%
\footnote{A free Weyl fermion corresponds to $\nF = \frac 12$  in our conventions.}
In such theories, there is a separate conserved 2-index symmetric
tensor for each field, while the energy-momentum tensor is given
by the sum of these:
\[
T = T_{\text{s}} + T_\text{f} + T_\text{v},
\]
where the sum over the different free fields is implicit.
The states created by these conserved tensors are mutually orthogonal,
so we have (see \Eq{sumrulealmost})
\[
\sum_{\O \, = \, T_\text{s}, T_\text{f}, T_\text{v}}
\sum_{a,\, b} \la_{TT\O}^{(a)} \la_{TT\O}^{(b)} \bigl[
& f^{(+)}_{ab}(\O) + f^{(-)}_{ab}(\O) \bigr]
\nn
&= \sum_{a,\, b} \la_{TTT}^{(a)} \la_{TTT}^{(b)} \bigl[
f^{(+)}_{ab}(T) + f^{(-)}_{ab}(T) \bigr]
\nn
&= \sfrac 35 c_\text{free},
\eql{TTTfree}
\]
where the function $c_\text{free}$
is given in \Eq{cnsnfnv}.
On the other hand, for  general CFT, we have
\[
\sum_{a,\, b} \la_{TTT}^{(a)} \la_{TTT}^{(b)} \bigl[
f^{(+)}_{ab}(T) + f^{(-)}_{ab}(T) \bigr]
= \frac{\text{quadratic function of $\nS$, $\nF$, $\nV$}}{C_T}.
\eql{TTTgeneral}
\]
This function of the three OPE coefficients $\nS, \nF, \nV$ must reduce to the
\rhs\ of \Eq{TTTfree} at non-negative integer values.
The only possibility is for the numerator on the \rhs\ of~\Eq{TTTgeneral} to be
proportional to $c^2$, so that
\Eq{TTTfree} to holds for general CFTs.
Furthermore, spin-2 operators with $\De = d = 4$ that are not the
energy-momentum tensor must give a vanishing contribution to the sum
rule.
Both of these features are also found in the computation of the
previous section, which was based on the Ward identities in a general
CFT.

\section{Conclusions and Outlook}
\scl{conclusions}

We have presented a sum rule for $c$ in 4D CFTs based on inserting a
complete set of states in a 4-point function of energy-momentum
tensors in Minkowski momentum space.
This 4-point function can be viewed as a contribution to a
graviton-graviton scattering amplitude, and the completeness
relation is a version of the optical theorem.
The sum rule can also be thought of as an OPE, since the intermediate states
are labeled by primary operators.
This work realizes the idea of \Ref{Gillioz:2016jnn} relating scale
(dilatation) anomaly coefficients to a positive sum over states.
We have given explicit expressions for the coefficients for operators of spin
0 and 2, and showed that our results satisfy a number of
consistency checks.

Our derivation of the sum rule depends on the unproven assumption that the
pseudo-amplitude we study is free of IR divergences in the massless and
forward limits.
This has been checked only in the case of free field CFTs, and a
better understanding of this question would be very helpful.
Conceptually, the scale anomaly can be thought of as the dependence
on a UV regulator, which should have nothing to do with IR divergences.
It may be worthwhile to look for a different derivation of the sum
rule in which IR divergences are not an issue.

The sum rule is a relationship among CFT data of the same form
as the bootstrap equations, specifically the crossing relations
for the $\avg{TTTT}$ correlation function.
The strategy of the numerical bootstrap is to numerically identify
linear combinations of bootstrap constraints such that the remainder
from omitted operators has a definite sign, so that neglecting
the remainder gives rigorous inequalities.
Our sum rule is a relation of this kind where all terms are positive
as a direct consequence of unitarity.
Although the methods used here to derive the sum rule are very
special to the 4D scale anomaly we have studied, it suggests that other
positive sum rules may be obtained analytically.
More prosaically, we hope that our sum rule will prove to be a useful
constraint in the program of bootstrapping the energy-momentum tensor.
It remains to be seen whether this constraint is redundant with
the constraints that are accessible via the numerical bootstrap.

There are a number of directions for future work.
As already mentioned above, we would like to have a better understanding
of the possible IR divergences in the 4-point functions of energy-momentum
tensors.
It would be very interesting to test our sum rule
in some specific CFTs, but this requires a 4-dimensional CFT where the
$TT\scr{O}$ OPE coefficients are all known, as well as extending our
calculation of the coefficients $f(\scr{O})$ to operators of arbitrary spin.
Checking the IR convergence in some case requires knowing the
4-point function of energy-momentum tensors in momentum space.
We are not aware of any theories in the literature where these results are
available, but perhaps this work can serve as a motivation to compute
these quantities, for example in $\scr{N} = 4$ super Yang-Mills theory
at large $N$.
It may also be interesting to consider our sum rule in holographic theories.
As discussed in the body of the paper, in such theories the contribution
of individual single double-trace operators (corresponding to supergravity
states in AdS) is negligible in the large-$N$ limit,
but it is possible that these states add up to give a sizeable contribution.
If not, it would imply that our sum rule requires contributions from the
single trace operators above the dimension gap (corresponding to string
states in AdS), even for a large gap.
In theories with maximal gap (``pure gravity'' in AdS) this would imply that
black hole states make important contributions to $c$.
The ideas in this paper can also be applied to other kinds of scale anomalies
in CFTs, for example in theories with global symmetries, supersymmetry,
or exactly marginal operators in various spacetime dimensions.
We can also try to extend this work to find a sum rule for the
$a$-anomaly in 4D CFT.
More generally, our work illustrates the usefulness of momentum-space techniques
in the study of conformal field theory, and we hope that these
will find further applications in the study of quantum field theory.


\subsection*{Acknowledgments}
We have benefited from discussions and encouragement from many people, including
L.~Dixon,
A.~Dymarsky,
J.~Kaplan,
M.~Mojaza,
J.~Penedones,
R.~Rattazzi,
A.~Vichi,
and M.~Walters.
M.G.~is supported by the Swiss National Science Foundation through the NCCR SwissMAP and formerly under grant number P300P2154559.
X.L.~and M.A.L.~are supported by
the Department of Energy under grant DE-FG02-91ER406746.


\appendix
\titleformat{\section}{\centering\normalsize\bfseries}{Appendix~\thesection:~}{0pt}{}

\section{Conventions}

We use mostly plus spacetime metric and the GR conventions
of Wald \cite{Wald:1984rg}.
The energy-momentum tensor is defined by differentiation with
respect to the metric
\[
\frac{\de W[g_{\mu\nu}]}{\de g_{\mu\nu}(x)} = \sfrac 12 \sqrt{-g} \ggap
\avg{ T^{\mu\nu}(x)}_g,
\]
where $W[g_{\mu\nu}]$ is the quantum effective action.
The anomaly under a Weyl transformation $\de g_{\mu\nu} = 2\si g_{\mu\nu}$
is given by
\[
\de_\si W[g_{\mu\nu}] = \myint d^4 x \si \bigl[
c W^{\mu\nu\rho\si} W_{\mu\nu\rho\si}
+ a E_4 \bigr], \eql{WeylAnomalies}
\]
where the square of the Weyl tensor is given by
\[
W^{\mu\nu\rho\si} W_{\mu\nu\rho\si}
= R^{\mu\nu\rho\si} R_{\mu\nu\rho\si}
- 2 R^{\mu\nu} R_{\mu\nu} + \tfrac{1}{3} \gap R^2, \eql{Wsqure}
\]
and the Euler density by
\[
E_4 = R^{\mu\nu\rho\si} R_{\mu\nu\rho\si}
- 4 R^{\mu\nu} R_{\mu\nu} + R^2. \eql{E4}
\]
With this normalization, in free field theories we have~\cite{Duff:1977ay}
\[
\eql{cnsnfnv}
c_\text{free} &= \frac{1}{(4\pi)^2} \frac{\nS + 6 \gap \nF + 12 \gap \nV}{120},
\\
\eql{ansnfnv}
a_\text{free} &= \frac{1}{(4\pi)^2} \frac{\nS + 11 \gap \nF + 62 \gap \nV}{360},
\]
where $\nS$ is the number of real scalars,
$\nF$ is the number of Dirac fermions,
and $\nV$ is the number of vectors.
The normalization of the 2-point function of energy-momentum tensors $C_T$
is defined in \Eq{Tnormalization}, and is related to $c$ by
\begin{equation}
\eql{CTc}
	C_T = \frac{640}{\pi^2}  c.
\end{equation}


\section{Computation of the Conformal Blocks in Momentum Space}
\scl{conformalblocks}

In this appendix we compute the ``conformal blocks'' $f(\O)$ that appear
as coefficients in our sum rule~\Eq{csumrule}.
We will obtain explicit results for scalar operators ($\ell = 0$) and for traceless symmetric spin-2 tensors ($\ell = 2$), including
the energy-momentum tensor itself.
Specifically, we compute the Wightman 2- and 3-point functions
(see \Eqs{CFTstate}, \eq{OO}, and \eq{TTO})
\begin{equation}
	\bra{\O^{\al_1\ldots\al_\ell}}
    \tilde{\O}^{\beta_1\ldots\beta_\ell}(k) \ket{0},
    \qquad
    	\bra{\O^{\al_1\ldots\al_\ell}}
    \timeordering[ \tilde{T}^{\mu\nu}(p_1) \tilde{T}^{\rho\sigma}(p_2) ] \ket{0}.
    \eql{23pt}
\end{equation}
The method used here is general and can in principle be used to compute the functions $f(\O)$ for other types of operators, but the complexity of the
calculation increases significantly for higher representations.

This appendix is organized as follows.
We first state our conventions for the normalization of operators
and OPE coefficients.
We then compute the 2- and 3-point functions where the energy-momentum
tensor is replaced by scalar operators;
this amounts to reviewing the results of Ref.~\cite{Gillioz:2016jnn}.
We then use Ward identities to fix completely the correlators of operators with spin, up to a few coefficients.
Finally we compute these coefficients in terms of OPE coefficients
defined in position space using the results for scalar correlators.

\subsection{Position Space 2-point Functions and Normalization of Operators}

For scalar operators, we use the standard normalization of the
Wightman two-point function
\begin{equation}
	\bra{\O} \O(x) \ket{0} = \frac{1}{(x^2_\text{W})^{\De}},
    \eql{operatornormalization}
\end{equation}
where the Wightman ordering is
imposed by an $i\ep$ prescription
(see \Eq{CFTstate})
\[
\ket{\O} = \O(i\ep, \pvec{0}) \ket{0},
\]
which implies the standard Wightman $i\ep$ prescription
\[
x^2_\text{W} \equiv -(x^0 + i\ep)^2 + \pvec{x}^2. \eql{x2Wightman}
\]
Note that $x^2_\text{W} \ne (-x)^2_\text{W}$ due to the $i\ep$
prescription.
We will drop the subscript W hereafter,
unless it is not clear from the context.

We also consider traceless symmetric spin-two operators, which we denote by $\scr{S}^{\mu\nu}$.
The conventional normalization for these is
\begin{equation}
	\bra{ \scr{S}^{\mu\nu}} \scr{S}^{\rho\sigma}(x) \ket{0}
    = \frac{\I^{\mu\nu\rho\sigma}(x)}
    {(x^2)^\De }
\end{equation}
where
\begin{equation}
	\I^{\mu\nu\rho\sigma}(x) = \frac{1}{2}
    \left[ \I^{\mu\rho}(x) \I^{\nu\sigma}(x)
    + \I^{\mu\sigma}(x) \I^{\nu\rho}(x) \right]
    - \frac{1}{d} \ggap \eta^{\mu\nu} \eta^{\rho\sigma}
\end{equation}
and
\begin{equation}
	\I^{\mu\nu}(x) = \eta^{\mu\nu} - \frac{2 x^\mu x^\nu}{x^2}.
\end{equation}
Note that we have kept the space-time dimension $d$ general.
We will continue this as far as possible in this appendix,
although we are interested in $d = 4$ at the end.

The normalization of the energy-momentum tensor is special, since it is
fixed by the fact that it defines the conserved energy and momentum.
We therefore have
\begin{equation}
	\bra{ T^{\mu\nu}}  T^{\rho\sigma}(x) \ket{0}
    = C_T \frac{\I^{\mu\nu\rho\sigma}(x)}{(x^2)^d}.
    \eql{Tnormalization}
\end{equation}
The constant $C_T$ is related to $c$ by conformal Ward identities,
see \Eq{CTc}.

\subsection{Position Space 3-point Functions and OPE Coefficients}
\scl{OPEcoefficients}

Now that we have fixed the normalization of operators,
we can define the OPE coefficients in terms of the tensor
structures that enter the 3-point function \Eq{TTO}.
We mostly follow the conventions of \Ref{Cordova:2017zej}, which builds
on the seminal works of Refs.~\cite{Osborn:1993cr, Costa:2011mg}
(see also \Ref{Dymarsky:2013wla}).
The 3-point functions can be written in terms of the quantities
\begin{align}
\begin{split}
	V_1^\mu &= \frac{x_1^2 x_{12}^\mu - x_{21}^2 x_1^\mu}
	{x_2^2},
	\\
	V_2^\mu &= \frac{x_2^2 x_{12}^\mu + x_{21}^2 x_2^\mu}
	{x_1^2},
	\\
	V_3^\mu &= \frac{x_2^2 x_1^\mu - x_1^2 x_2^\mu}
	{x_{21}^2},
\end{split}
\end{align}
and the 2-index tensors
\begin{align}
\begin{split}
	H_{12}^{\mu\nu} &= x_{21}^2 \eta^{\mu\nu}
	- 2 x_{12}^\mu x_{12}^\nu,
	\\
	H_{13}^{\mu\nu} &= x_1^2 \eta^{\mu\nu} - 2 x_1^\mu x_1^\nu,
	\\
	H_{23}^{\mu\nu} &= x_2^2 \eta^{\mu\nu} - 2 x_2^\mu x_2^\nu,
\end{split}
\end{align}
where $x_{ij} \equiv x_i - x_j$.
For a traceless symmetric tensor operator $\O^{\al_1 \ldots \al_\ell}$
with scaling dimension $\Delta$ and spin $\ell$,
conformal invariance requires the
position-space correlation function (compare \Eq{23pt}) to be of the form
\begin{equation}
	\bra{\O^{\al_1 \ldots \al_\ell}}
	T^{\mu\nu}(x_1) T^{\rho\sigma}(x_2) \ket{0}
	= \frac{\left( \text{tensors built out of the~} V_i
	\text{~and~} H_{ij} \right)}
	{(x_1^2)^{(\De +\ell)/2} (x_2^2)^{(\De+\ell)/2}
	(x_{21}^2)^{d+2-(\De+\ell)/2}},
\end{equation}
where the numerator is such that the $V_i$, and $H_{ij}$ have indices corresponding respectively to the operators inserted at the points $x_i$
and $x_j$ (identifying $x_3 = 0$).
The examples $\ell = 0, 2$ will now be discussed in detail.

For a scalar operator $\O$, the most general form of the three-point function
consistent with conformal symmetry is
\begin{equation}
	\bra{\O} T^{\mu\nu}(x_1) T^{\rho\sigma}(x_2) \ket{0}
	= \frac{t^{\mu\nu\rho\sigma}}
	{(x_1^2)^{\De/2} (x_2^2)^{\Delta/2} (x_{21}^2)^{d+2-\Delta/2}}
    \eql{OTTposn}
\end{equation}
where the tensor $t^{\mu\nu\rho\sigma}$ is a linear combination of three terms,
\begin{align}
	t^{\mu\nu\rho\sigma} = \frac{1}{4} \Big[ &
	\al_1 V_1^\mu  V_1^\nu V_2^\rho V_2^\sigma
	+ \al_2 H_{12}^{\mu\rho} V_1^\nu V_2^\sigma
	\nonumber \\
	& + \al_3 H_{12}^{\mu\rho} H_{12}^{\nu\sigma}
	+ \text{permutations} - \text{traces} \Big].
\end{align}
Permutations and traces are understood to be among indices of the same
operators, \textit{e.g.}~$(\mu \leftrightarrow \nu)$ and $(\rho \leftrightarrow \sigma)$.
The three coefficients $\al_1$, $\al_2$ and $\al_3$ are not
independent: requiring conservation of the energy-momentum tensor adds 2
constraints, so that the correlator eventually depends on a single OPE
coefficient.
A solution to these constraints is for instance the choice of \Ref{Cordova:2017zej},
\begin{align}
	\al_1 &= \la_{TT\O},
    \eql{lambdaTTO}
    \\
	\al_2 &= \frac{2  (d-1) \left( \Delta - d \right) - 4}
	{(d-2) \left( \Delta + 2 \right)}
	\la_{TT\O},
    \\
	\al_3 &= \frac{(d-1) \left( \Delta - d \right)^2 - 2  d}
	{(d-2)  \Delta \left( \Delta + 2 \right)}
	\la_{TT\O}.
\end{align}
Note that all three coefficients remain of order unity even when
$\Delta$ is large.

In the case of a traceless symmetric spin-2 operator, the most general  conformal invariant correlator has the form
\begin{equation}
	\bra{\scr{S}^{\al\beta}} T^{\mu\nu}(x_1) T^{\rho\sigma}(x_2) \ket{0}
    = \frac{t^{\mu\nu\rho\sigma\al\beta}}
    {(x_1^2)^{\Delta/2+1} (x_2^2)^{\Delta/2+1}
    (x_{21}^2)^{d+1-\Delta/2}},
\end{equation}
where
\[
\begin{split}
\!\!\!\!\!\!\!\!\!
	t^{\mu\nu\rho\sigma\al\beta} = \frac{1}{8} \Big[ &
	\al_1 V_1^\mu V_1^\nu V_2^\rho V_2^\sigma
	V_3^\al V_3^\beta
	+ \al_2 \big( H_{13}^{\mu\al} V_2^\rho
	+ H_{23}^{\rho\al} V_1^\mu \big)
	V_1^\nu V_2^\sigma V_3^\beta
	\\
	& + \al_3 H_{12}^{\mu\rho}
	V_1^\nu V_2^\sigma V_3^\al V_3^\beta
	+ \al_4 \big( H_{13}^{\mu\al} V_2^\rho
	+ H_{23}^{\rho\al} V_1^\mu \big)
	H_{12}^{\nu\sigma} V_3^\beta
	\\
	& + \al_5 H_{13}^{\mu\al} H_{23}^{\rho\beta}
	V_1^\nu V_2^\sigma
	+ \al_6 H_{12}^{\mu\rho} H_{12}^{\nu\sigma}
	V_3^\al V_3^\beta
	\\
	& + \al_7 \big( H_{13}^{\mu\al} H_{13}^{\nu\beta}
	V_2^\rho V_2^\sigma + H_{23}^{\rho\al} H_{23}^{\sigma\beta}
	V_1^\mu V_1^\nu \big)
	+ \al_8 H_{12}^{\mu\rho} H_{13}^{\nu\al}
	H_{23}^{\sigma\beta}
	\\
	& + \text{permutations} - \text{traces} \Big].
    \eql{OPEcoeffs:TTS}
\end{split}
\]
There are 6 constraints from the conservation of the energy-momentum tensor,
and therefore only two independent OPE coefficients.
We choose them to be
\[
	\lambda_{TT\scr{S}}^{(1)}
	= \alpha_1 + 2 \alpha_2 + 4 \alpha_7,
	\qquad
	\lambda_{TT\scr{S}}^{(2)}
	= -5 \alpha_1 - 12 \alpha_2 + \alpha_3
	- 9 \alpha_5 + 4 \alpha_7 - 6 \alpha_8.
    \eql{lambdaTTS:def}
\]
With this choice, the coefficients $\alpha_i$ are
(specializing to $d = 4$)
\[
\begin{split}
	\alpha_1 &=
    -\frac{4 (\Delta^2 - 29 \Delta - 24) \lambda_{TT\scr{S}}^{(1)}
    + 3 \Delta (\Delta - 8) \lambda_{TT\scr{S}}^{(2)}}
    {2 (\Delta + 2) (\Delta + 4)},
    \\
	\alpha_2 &=
    \frac{2 (5 \Delta^2 - 78 \Delta - 24) \lambda_{TT\scr{S}}^{(1)}
    + (\Delta - 8) (5 \Delta - 2) \lambda_{TT\scr{S}}^{(2)}}
    {4 (\Delta + 2) (\Delta + 4)},
    \\
	\alpha_3 &=
    -\frac{2(\Delta^2 - 114 \Delta + 184) \lambda_{TT\scr{S}}^{(1)}
    + (\Delta - 8) (7 \Delta - 10) \lambda_{TT\scr{S}}^{(2)}}
    {4 (\Delta + 2) (\Delta + 4)},
    \\
	\alpha_4 &=
    \frac{2 (3 \Delta^2 -54 \Delta + 88) \lambda_{TT\scr{S}}^{(1)}
    + 3(\Delta - 2) (\Delta - 8) \lambda_{TT\scr{S}}^{(2)}}
    {4 (\Delta + 2) (\Delta + 4)},
    \\
	\alpha_5 &=
    - \frac{2 \Delta (3 \Delta - 32) \lambda_{TT\scr{S}}^{(1)}
    + (3 \Delta^2 - 20 \Delta + 8) \lambda_{TT\scr{S}}^{(2)}}
    {4 (\Delta + 2) (\Delta + 4)},
    \\
	\alpha_6 &=
    \frac{2 (\Delta^2 + 17 \Delta - 56) \lambda_{TT\scr{S}}^{(1)}
    - (\Delta - 8) (2 \Delta - 5) \lambda_{TT\scr{S}}^{(2)}}
    {4 (\Delta + 2) (\Delta + 4)},
    \\
	\alpha_7 &=
    - \frac{2 (\Delta^2 - 13 \Delta + 8) \lambda_{TT\scr{S}}^{(1)}
    + (\Delta - 1) (\Delta - 8) \lambda_{TT\scr{S}}^{(2)}}
    {4 (\Delta + 2) (\Delta + 4)},
    \\
	\alpha_8 &=
    -\frac{2 (3 \Delta^2 - 39 \Delta + 68) \lambda_{TT\scr{S}}^{(1)}
    + (3 \Delta^2 - 27 \Delta + 44) \lambda_{TT\scr{S}}^{(2)}}
    {4 (\Delta + 2) (\Delta + 4)}.
\end{split}
\eql{alpharelns}
\]
The definition \Eq{lambdaTTS:def} was chosen so that all the
coefficients $\al_i$ are finite for all $\De$ allowed by
unitarity (as well as $\De \to \infty$), as long as
$\la_{TT\scr{S}}^{(1, 2)}$ are finite.

\Eqs{alpharelns} are not valid for the case where $\scr{S}$ is
the energy-momentum tensor, corresponding to $\De = d = 4$.
In this case there are algebraic degeneracies in the constraints above,
and we only have 5 independent constraints.
This is in agreement with the well-known fact that there are 3 independent
$TTT$ OPE coefficients in $d = 4$.
Using this approach is cumbersome for this case because of the existence
of contact terms, as discussed in \sec{contactterms}.
We will instead use the Ward
identities in momentum space to fix
the contribution of the energy-momentum tensor.

\subsection{Fourier Transform of Scalar Correlators}
\scl{ScalarFourierIntegrals}

For scalar correlators, the Fourier transform into momentum space can be
performed straightforwardly. We have
\begin{equation}
	\bra{\O} \tilde{\O}(k) \ket{0}
    = \myint d^dx  \frac{e^{i k \cdot x}}
    {(x_\text{W}^2)^{\Delta}}
    \equiv \Pi_\O(k) \theta(k^0)\theta(-k^2),
\end{equation}
where
\begin{equation}
	\Pi_\O(k)
    = \frac{2^{d - 2 \Delta + 1} \pi^{(d+2)/2}}
    {\Gamma\left( \Delta \right)
    \Gamma\left( \Delta - \frac{d-2}{2} \right)} \ggap
    (-k^2)^{\Delta - d/2} .
    \eql{Pi:scalar}
\end{equation}
For the three-point function, we consider three scalar operators $\O_i$ with scaling dimensions $\Delta_i$, for which we obtain
\begin{align}
	\bra{\O_3} \timeordering[ \tilde{\O}_1(p_1) \tilde{\O}_2(p_2) ] \ket{0}
    &= \myint d^dx_1  d^dx_2
    \frac{e^{i (p_1 \cdot x_1 + p_2 \cdot x_2)} \la_{\O_1\O_2\O_3}}
    {\left(x_{1,\text{W}}^2\right)^{\Delta_{31,2}/2} \left(x_{2,\text{W}}^2\right)^{\Delta_{32,1}/2}
    \left(x_{12,\text{F}}^2\right)^{\Delta_{12,3}/2}}
    \nonumber \\
    &\equiv \la_{\O_1\O_2\O_3}\mathcal{F}\left( \Delta_{31,2}, \Delta_{32,1}; \Delta_{12,3} \right) ,
    \eql{3ptFourierIntegral}
\end{align}
where $\Delta_{ij,k} = \left( \Delta_i + \Delta_j - \Delta_k \right)/2$.
We have restored the $i\ep$ here, using \Eq{x2Wightman} and defining
\begin{equation}
	x^2_\text{F} \equiv x^2 + i \epsilon .
\end{equation}
The function $\mathcal{F}$ is a function symmetric in its first two arguments,
given by
\begin{equation}
	\mathcal{F}(\Delta_1, \Delta_2; \Delta_3)
    = \frac{-2i \pi^{d+1}
    \Gamma\left( \Delta_1 + \Delta_3 - \frac{d}{2} \right)
    \Gamma\left( \Delta_2 + \Delta_3 - \frac{d}{2} \right)
    \Gamma\left( \frac{d}{2} - \Delta_3 \right)}
    {\Gamma\left( \Delta_1 \right)
    \Gamma\left( \Delta_2 \right)
    \Gamma\left( \Delta_3 \right)
    \Gamma\left( \Delta_\text{tot} - \frac{d}{2} \right)
    \Gamma\left( \Delta_\text{tot} - d + 1 \right)}
    \left( \frac{s}{4} \right)^{\Delta_\text{tot} - d},
    \eql{F}
	\end{equation}
where $\Delta_\text{tot} = \Delta_1 + \Delta_2 + \Delta_3$.
Note that the result can be divergent, for example if $\Delta_3 = d/2 + n$
(with $n \in \mathbb{N}$).

\subsection{Conformal Ward Identities}
\scl{WardIdentities:conformal}

We use conformal Ward identities to determine the 2- and 3-point functions of general tensor operators in momentum space, \Eq{23pt}.
The conformal generators act in momentum space as
\begin{subequations}
\[
	\big[ P_\mu, \tilde{\O}(p) \big] &= p_\mu \ggap \tilde{\O}(p),
	\\
	\big[ D, \tilde{\O}(p) \big] &= -i
	\left( -p^\nu \frac{\partial}{\partial p^\nu}
	+ \Delta - d \right) \tilde{\O}(p),
	\\
	\big[ M_{\mu\nu}, \tilde{\O}(p) \big] &=
	-i \left( p_\mu \frac{\partial}{\partial p^\nu}
	- p_\nu \frac{\partial}{\partial p^\mu}
	+ \Sigma_{\mu\nu} \right) \tilde{\O}(p),
	\\
	\big[ K_\mu, \tilde{\O}(p) \big] &=
	\left( - 2 p^\nu \frac{\partial^2}{\partial p^\mu \partial p^\nu}
	+ p_\mu \frac{\partial^2}{\partial p_\nu \partial p^\nu} \right.
    \nn
    &\qquad\quad{} \left.
	+ 2 (\Delta - d) \frac{\partial}{\partial p^\mu}
	+ 2 \Sigma_{\mu\nu} \frac{\partial}{\partial p_\nu} \right) \tilde{\O}(p),
\]
\end{subequations}
where $\Sigma_{\mu\nu}$ is the spin operator, acting on a spin-$\ell$ tensor as
\begin{equation}
	\Sigma_{\mu\nu}  \tilde{\O}^{\al_1 \ldots \al_\ell}
	= \sum_{i \, = \, 1}^\ell \left( \delta_\mu^{\al_i}
	\tilde{\O}^{\al_1 \ldots~\ldots \al_\ell}_{~~~~\nu}
	- \delta_\nu^{\al_i}
	\tilde{\O}^{\al_1 \ldots~\ldots \al_\ell}_{~~~~\mu} \right).
\end{equation}
The states $\ket{\O^{\al_1 \cdots \al_\ell}}$ that appear in the matrix
elements \Eq{23pt} transform as%
\footnote{These states correspond to inserting a position-space
operator at $x = 0$,
and the resulting momentum-space correlation functions have the
momentum-conserving delta function factored out.
An alternative approach is to work with the full momentum space correlation
function, which includes the momentum-conserving delta function.
In this case, the differential operators must act on the
delta function.
This approach is discussed in \Ref{Maldacena:2011nz}.}
\begin{align}
	D  \ket{\O^{\al_1 \ldots \al_\ell}}
	 & = - i \Delta  \ket{\O^{\al_1 \ldots \al_\ell}},
	\\
	M_{\mu\nu} \ket{\O^{\al_1 \ldots \al_\ell}}
	 & = - i \Sigma_{\mu\nu} \ket{\O^{\al_1 \ldots \al_\ell}},
	\\
	K_\mu \ket{\O^{\al_1 \ldots \al_\ell}}
    & = 0.
\end{align}
Conformal Ward identities follow then from invariance of the vacuum state, which in turns implies, \textit{e.g.} for special conformal transformations,
\begin{equation}
	\bra{\O^{\al_1\ldots\al_\ell}}
    \big[ K_\tau, \tilde{T}^{\mu\nu}(p_1) \big]
    \tilde{T}^{\rho\sigma}(p_2) \ket{0}
	+ \bra{\O^{\al_1\ldots\al_\ell}}
    \tilde{T}^{\mu\nu}(p_1) \big[ K_\tau,
    \tilde{T}^{\rho\sigma}(p_2) \big] \ket{0}
	= 0.
    \eql{Wardidentities:def}
\end{equation}
This Ward identity is satisfied in the above form where there is no time ordering
of the momentum-space operators.
In this case, there are no UV divergences or contact terms
as discussed in \sec{UVdivergences} and \sec{contactterms}.
(The absence of UV divergences also implies the absence of anomalies.)
When considering time-ordered correlators, we have to worry about
these subtleties.
However, the only case where there are contact terms
is when $\O$ is the energy-momentum tensor itself.
There are no conformal anomalies in these correlation functions,
because conformal anomalies are purely local and therefore do not
contribute to correlation functions where some operators are Wightman
ordered.
Our correlator~\eq{23pt} therefore
obeys non-anomalous Ward identities with no additional
contact terms like \Eq{Wardidentities:def} above, namely
\begin{equation}
\begin{split}
	\sum_{i \, = \,1}^2 & \left[ - 2 p_i^\omega \frac{\partial^2}
	{\partial p_i^\tau \partial p_i^\omega}
	+ p_{i\tau} \frac{\partial^2}{\partial p_{i\omega} \partial p_i^\omega}
	+ 2  \Sigma_{\tau\omega}^{(i)}
	\frac{\partial}{\partial p_{i\omega}} \right]
    \\
    &\qquad\qquad\qquad\qquad{} \times
	\bra{\O^{\al_1 \ldots \al_\ell}}
    \timeordering[ T^{\mu\nu}(p_1) T^{\rho\sigma}(p_2) ]
	\ket{0} = 0,
	\eql{WardIdentity:K}
\end{split}
\end{equation}
where the $\Sigma^{(i)}_{\tau\omega}$ only acts on the indices of the operator carrying momentum $p_i$.
This equation is quite non-trivial to solve, and its discussion is postponed to \sec{Wardidentites:3pt}.

In the case of Wightman two-point functions, conformal invariance is simple enough to allow for a direct solution. For our traceless symmetric spin-two operator $\scr{S}$, we obtain
\[
\begin{split}
	[ \Pi_\scr{S}(k) ]^{\mu\nu\rho\sigma}
    &= \tilde{C}_\scr{S} \bigg[
    \frac{1}{2} \left( \eta^{\mu\rho} \eta^{\nu\sigma}
    + \eta^{\mu\sigma} \eta^{\nu\rho} \right)
    - \frac{\Delta (\Delta + 1) - d}
    {\Delta (\Delta - 1) d}
    \eta^{\mu\nu} \eta^{\rho\sigma}
    \\
    &\qquad\quad - \frac{2 \Delta - d}{2\Delta} \left(
    \eta^{\mu\rho}   \frac{k^\nu k^\sigma}{k^2}
    + \eta^{\mu\sigma}   \frac{k^\nu k^\rho}{k^2}
    + \eta^{\nu\rho}  \frac{k^\mu k^\sigma}{k^2}
    + \eta^{\nu\sigma}  \frac{k^\mu k^\rho}{k^2} \right)
    \\
    &\qquad\quad + \frac{2 \Delta - d}{\Delta (\Delta - 1)}
    \left( \eta^{\mu\nu} \frac{k^\rho k^\sigma}{k^2}
    + \eta^{\rho\sigma} \frac{k^\mu k^\nu}{k^2} \right)
    \\
    &\qquad\quad + \frac{(2 \Delta - d) (2 \Delta - d - 2)}
    {\Delta (\Delta - 1)}
    \frac{k^\mu k^\nu k^\rho k^\sigma}{(k^2)^2} \bigg] s^{\Delta - d/2},
    \eql{Pi:S}
\end{split}
\]
where $k = p_1 + p_2$, and $\tilde{C}_\scr{S}$ is a coefficient that is
not fixed by conformal invariance.
The inverse of this tensor is the object that enters in the completeness
relation~\Eq{completenessrelation} and thus in the computation of the conformal blocks.
It is given by
\[
\begin{split}
\!\!\!
	[ \Pi_\scr{S}^{-1}(k) ]^{\mu\nu\rho\sigma}
    &= \frac{s^{d/2 - \Delta}}{\tilde{C}_\scr{S}} \bigg[
    \frac{1}{2} \left( \eta^{\mu\rho}  \eta^{\nu\sigma} + \eta^{\mu\sigma} \eta^{\nu\rho} \right)
    \\
    &\qquad\quad{} - \frac{2 \Delta - d}{2 (\Delta - d)} \left(
    \eta^{\mu\rho}  \frac{k^\nu k^\sigma}{k^2}
    + \eta^{\mu\sigma}  \frac{k^\nu k^\rho}{k^2}
    + \eta^{\nu\rho} \frac{k^\mu k^\sigma}{k^2}
    + \eta^{\nu\sigma} \frac{k^\mu k^\rho}{k^2} \right)
    \\
    &\qquad\quad{} + \frac{(2 \Delta - d) (2 \Delta - d + 2)}
    {(\Delta - d) (\Delta - d + 1)}
    \frac{k^\mu k^\nu k^\rho k^\sigma}{(k^2)^2} \bigg].
    \eql{Pi:S:inverse}
\end{split}
\]
Note that the expression for $\Pi_\scr{S}$ also holds in the case of the
energy-momentum tensor by setting $\Delta = d$, or in a simpler form
\[
\begin{split}
	[ \Pi_T(k) ]^{\mu\nu\rho\sigma}
    &= \tilde{C}_T  s^{d/2}
    \left[ \frac{1}{2} \left( \tilde{\I}^{\mu\rho}(k)
    \tilde{\I}^{\nu\sigma}(k)
	+ \tilde{\I}^{\mu\sigma}(k) \tilde{\I}^{\nu\rho}(k) \right) \right.
    \\
    &\qquad\qquad\qquad\qquad{} \left.
	- \frac{1}{d-1} \tilde{\I}^{\mu\nu}(k) \tilde{\I}^{\rho\sigma}(k) \right]
	\eql{TT}
\end{split}
\]
where $\tilde{C}_T = 4 \pi \gap c$ and
\begin{equation}
	\tilde{\I}^{\mu\nu}(k) = \eta^{\mu\nu} - \frac{k^\mu  k^\nu}{k^2}.
\end{equation}
The inverse of \Eq{Pi:S:inverse} appears to be ill-defined in the
limit $\Delta \to d$.
This is the case of the conserved energy-momentum tensor, and its inverse
is not unique because it is transverse, $k_\mu \tilde{T}^{\mu\nu}(k) = 0$.
We can invert it on the transverse space, since transverse
contributions vanish by \Eq{WardIdentity:transverse:TTT:2}.
We therefore obtain
\begin{equation}
	[ \Pi_T^{-1}(k) ]^{\mu\nu\rho\sigma} = \frac{1}{\tilde{C}_T s^{d/2}}
    \frac{1}{2} \left( \eta^{\mu\rho} \eta^{\nu\sigma} + \eta^{\mu\sigma} \eta^{\nu\rho} \right).
    \eql{Pi:T:inverse}
\end{equation}

\subsection{Conservation Ward Identities}
\scl{WardIdentities:conservation}

In addition to the conformal Ward identities, there are identities
encoding the fact that the energy-momentum tensor is a traceless,
conserved current, {\it i.e.}~that it belongs into a short representation of
the conformal algebra.
The precise form of these Ward identities depend on our definition of
the energy-momentum  tensor.
With our choice \eq{Tdefinition} the Ward identities are summarized by~\cite{Coriano:2017mux}
\[
\nabla_\mu \bra{0} T^{\mu\nu}(x) \ket{0}_g &= 0,
\eql{conservationWard}
\\
g_{\mu\nu} \bra{0} T^{\mu\nu}(x) \ket{0}_g
    &= c \ggap W^{\mu\nu\rho\si} W_{\mu\nu\rho\si} + a E_4,
\eql{traceWard}
\]
with $\nabla_\mu$ the covariant derivative in the metric $g_{\mu\nu}$.
The Ward identities for higher correlation functions follow by repeated differentiation with
respect to the metric $g_{\mu\nu}$ or other source fields.

For example, the trace Ward identity \Eq{traceWard} gives in position space
\begin{subequations}
\[
	\eta_{\mu\nu} \gap \bra{0} T^{\al\beta}(z)
    \timeordering[ T^{\mu\nu}(x_1) T^{\rho\sigma}(x_2) ] \ket{0}
    & = 2 \gap  \delta^4(x_1 - x_2) \gap \bra{0} T^{\al\beta}(z) T^{\rho\sigma}(x_2) \ket{0},
    \\
    \eta_{\al\beta} \gap \bra{0} T^{\al\beta}(z)
    \timeordering[ T^{\mu\nu}(x_1) T^{\rho\sigma}(x_2) ] \ket{0} & = 0,
    \\
	\eta_{\mu\nu} \gap \bra{0} \O^{\al_1 \ldots \al_\ell}(z)
    \timeordering[ T^{\mu\nu}(x_1) T^{\rho\sigma}(x_2) ] \ket{0} & = 0.
    \qquad (\O \neq T)
\]
\end{subequations}
Note that the anomaly terms on the \rhs\ of \Eq{traceWard} do not contribute, because
they are $O(h_{\mu\nu}^2)$, and therefore only appear if there are 3 or more
energy-momentum tensors in the same time ordered product.

In momentum space, these become
\begin{subequations}
\[
    \eta_{\mu\nu} \gap \bra{T^{\al\beta}}
    \timeordering[ \tilde{T}^{\mu\nu}(p_1) \tilde{T}^{\rho\sigma}(p_2) ] \ket{0} &
    = 2 \bra{T^{\al\beta}} \tilde{T}^{\rho\sigma}(p_1 + p_2) \ket{0},
	\eql{WardIdentity:trace:TTT}
    \\
    \eta_{\al\beta} \gap \bra{T^{\al\beta}}
    \timeordering[ \tilde{T}^{\mu\nu}(p_1) \tilde{T}^{\rho\sigma}(p_2) ] \ket{0} & = 0.
	\eql{WardIdentity:trace:TTT:2}
	\\
	\eta_{\mu\nu} \gap \bra{\O^{\al_1\ldots\al_\ell}}
    \timeordering[ \tilde{T}^{\mu\nu}(p_1) \tilde{T}^{\rho\sigma}(p_2) ] \ket{0} & = 0.
    \qquad (\O \neq T)
	\eql{WardIdentity:trace}
\]
\end{subequations}
Similar Ward identities can be derived
for the divergence of the energy-momentum tensor using \Eq{conservationWard}:
\begin{subequations}
\begin{align}
    (p_1)_{\mu} \bra{T^{\al\beta}}
    \timeordering[ \tilde{T}^{\mu\nu}(p_1) \tilde{T}^{\rho\sigma}(p_2) ] \ket{0} & =
    (p_2)^{\nu} \bra{T^{\al\beta}} \tilde{T}^{\rho\sigma}(p_1 + p_2) \ket{0}
	\nonumber \\
	& \qquad{}
    - (p_2)_\la \eta^{\nu\rho}
	\bra{T^{\al\beta}} \tilde{T}^{\sigma\la}(p_1 + p_2) \ket{0}
	\nonumber \\
	& \qquad{}
    - (p_2)_\la   \eta^{\nu\sigma}
	\bra{T^{\al\beta}} \tilde{T}^{\rho\la}(p_1 + p_2) \ket{0}.
	\eql{WardIdentity:transverse:TTT}
\\
    (p_1 + p_2)_{\al} \bra{T^{\al\beta}} \timeordering [\tilde{T}^{\mu\nu}(p_1)
	\tilde{T}^{\rho\sigma}(p_2)] \ket{0} & = 0,
	\eql{WardIdentity:transverse:TTT:2}
\\
	(p_1)_{\mu} \bra{\O^{\al_1\ldots\al_\ell}}
    \timeordering[ \tilde{T}^{\mu\nu}(p_1) \tilde{T}^{\rho\sigma}(p_2) ] \ket{0} & = 0.
    \qquad (\O \neq T)
	\eql{WardIdentity:transverse:TTO}
\end{align}
\end{subequations}
Some of these identities involve the Wightman 2-point function given by Eq.~\eqref{eq:TT} on the right-hand side.
Note that Ward identities for momentum-space 3-point functions have been obtained in Refs.~\cite{Bzowski:2013sza, Bzowski:2017poo}; they differ from ours, due to our specific ordering of operators.

\subsection{Using Ward Identities to Compute 3-point Functions}
\scl{Wardidentites:3pt}

We now show how to use the Ward identities listed above
to determine the 3-point functions~\Eq{TTO} up to OPE coefficients.
This is greatly simplified by the fact that we need the result only
for massless external momenta.

The Ward identities for translation, Lorentz, and dilatation symmetry
can be solved simply by writing the 3-point function in terms of appropriate
invariants.
Translation invariance implies that the 3-point function is a function of only
two momenta $p_{1,2}$.
Lorentz invariance requires that it is a Lorentz tensor made from
$p_{1,2}^\mu$ and $\eta^{\mu\nu}$.
Dilatation invariance implies that this tensor has the appropriate dimension.

The only remaining Ward identity is the one for special conformal transformations,
which must be solved explicity.
Although we are interested in the 3-point function for $p_{1,2}^2 = 0$, we
must consider $p_{1,2}^2 \ne 0$ because the special conformal Ward identity
relates configurations with different values of $p_{1,2}^2$.
We are interested in kinematics with $s = -(p_1 + p_2)^2 > 0$, so
we can write the correlator in terms of the dimensionless quantities
\begin{equation}
	\hat{p}_1^\mu = \frac{p_1^\mu}{\sqrt{s}},
	\qquad
	\hat{p}_2^\mu = \frac{p_2^\mu}{\sqrt{s}},
	\qquad
	\hat{m}_1^2 = \frac{p_1^2}{s},
	\qquad
	\hat{m}_2^2 = \frac{p_2^2}{s}.
    \eql{dimensionlessquantities}
\end{equation}
The integral \eqref{eq:TTO} can then be parametrized as
\[
\begin{split}
	\bra{\O^{\al_1 \ldots \al_\ell}}
    &\timeordering[ \tilde{T}^{\mu\nu}(p_1) \tilde{T}^{\rho\sigma}(p_2) ] \ket{0}
    \\
    &\qquad\qquad{}
	= s^{\Delta/2} \sum_{a \, = \, 1}^n
	h_a(\hat{m}_1^2, \hat{m}_2^2)
	t_a^{(\mu\nu)(\rho\sigma)\al_1\ldots\al_\ell}
	(\hat{p}_1^\mu , \hat{p}_2^\mu, \eta^{\mu\nu}),
	\eql{tensor:mostgeneral}
    \end{split}
\]
where the $h_a$ are arbitrary functions of the two variables $\hat{m}^2_1$
and $\hat{m}^2_2$, and the $t_a$ are all possible tensors constructed from
the dimensionless quantities in \Eq{dimensionlessquantities}.
Because we eventually want to take $p_1^2 = p_2^2 \to 0$, we only need
to know the $h_a\left( 0, 0 \right)$.
However, the special conformal Ward identity \Eq{WardIdentity:K}
relates the $h_a(0,0)$ with derivatives $\partial_{\hat{m}^2_i} h_a(0,0)$. Explicitly, the Ward identity becomes
\begin{align}
	& \sum_{a \, = \,1}^n \sum_{i \, = \, 1}^2 \bigg[ 2  s^{\Delta/2-1}
	\partial_{\hat{m}^2_i} h_a(0,0)
	\left( (d-2)  p_{i\tau}
	+ 2  p_i^\la  \Sigma_{\tau\la}^{(i)} \right)
	t_a^{(\mu\nu)(\rho\sigma)\al_1\ldots\al_\ell}
	\nonumber \\
	& + h_a(0,0) \left( - 2 p_i^\la \frac{\partial^2}
	{\partial p_i^\tau \partial p_i^\la}
	+ p_{i\tau} \frac{\partial^2}{\partial p_{i\la} \partial p_i^\la}
	+ 2  \Sigma_{\tau\la}^{(i)}
	\frac{\partial}{\partial p_{i\la}} \right)
	\left( s^{\Delta/2}
	t_a^{(\mu\nu)(\rho\sigma)\al_1\ldots\al_\ell} \right) \bigg] = 0.
	\eql{WardIdentity:K:2}
\end{align}
There are no second derivative terms of the form $\partial_{\hat{m}^2_i} \partial_{\hat{m}^2_j} h_a(0,0)$.\footnote{This is because the differential operator in \Eq{WardIdentity:K}
is related to the Todorov operator that preserves the condition $p_i^2 = 0$~\cite{Dobrev:1975ru, Dymarsky:2017yzx}.}
\Eq{WardIdentity:K:2} provides a set of linear constraints involving both $\partial_{\hat{m}^2_i} h_a(0,0)$ and $h_a(0,0)$. But we can eliminate $\partial_{\hat{m}^2_i} h_a(0,0)$ and obtain a set of constraints among $h_a(0,0)$ themselves.
To do this, we note that contracting \Eq{WardIdentity:K:2} with $p_2^\tau$,
the terms involving $\partial_{\hat{m}^2_2} h_a(0,0)$ vanish, giving us a
relation between $\partial_{\hat{m}^2_1} h_a(0,0)$ and $h_a(0,0)$.
The tensor multiplying $\partial_{\hat{m}^2_1} h_a(0,0)$ in this expression
is invertible and the resulting linear equations have a
unique solution.
Similarly, contracting with $p_1^\tau$ allows us to fix
$\partial_{\hat{m}^2_2} h_a(0,0)$.
The equations are rather complicated, but are straightforward to solve using
computer algebra.
We will give several explicit examples below.

\subsubsection*{Scalar states}

In the case where $\O$ is a scalar operator, the most general tensor
$t^{(\mu\nu)(\rho\sigma)}$ built out of $\hat{p}_{1,2}^\mu$ and $\eta^{\mu\nu}$
that is symmetric in both pairs of indices has 21 independent tensor structures.
After imposing the special conformal Ward identity
and considering on-shell momenta $p_{1,2}^2 = 0$,
only 6 linearly independent tensor structures remain,
of which 5 are symmetric under the exchange of the two energy-momentum tensors.
Imposing the transversality and trace conditions \Eqs{WardIdentity:trace} and \eq{WardIdentity:transverse:TTO}, only one unique linear combination of the 5 previous structures remains.
It can be written as
\begin{align}
	\bra{\O} \timeordering [\tilde{T}^{\mu\nu}(p_1) \tilde{T}^{\rho\sigma}(p_2) ] \ket{0} &
	\nonumber \\
	= \tilde{\la}_{TT\O} s^{\Delta/2} \bigg\{&
	\left( \eta^{\mu\nu} + 2 \hat{p}_1^{\mu} \hat{p}_2^{\nu}
	+ 2 \hat{p}_2^{\mu} \hat{p}_1^{\nu} \right)
	\left( \eta^{\rho\sigma} + 2 \hat{p}_1^{\rho} \hat{p}_2^{\sigma}
	+ 2 \hat{p}_2^{\rho} \hat{p}_1^{\sigma} \right)
	\nonumber \\
	& {} - \frac{d-2}{2} \bigg[ \left( \eta^{\mu\rho}
	+ 2 \hat{p}_2^{\mu} \hat{p}_1^{\rho}
	- \xi_1  \hat{p}_1^{\mu} \hat{p}_2^{\rho} \right)
	\left( \eta^{\nu\sigma}
	+ 2 \hat{p}_2^{\nu} \hat{p}_1^{\sigma}
	- \xi_1 \hat{p}_1^{\nu} \hat{p}_2^{\sigma} \right)
	\nonumber \\
	& \qquad\qquad {}+ \left( \rho \leftrightarrow \sigma \right) \bigg]
	+ \xi_2
	\hat{p}_1^{\mu} \hat{p}_1^{\nu}
	\hat{p}_2^{\rho} \hat{p}_2^{\sigma} \bigg\}.
	\eql{TTscalar}
\end{align}
This is a polynomial in the momenta because we are evaluating the
correlation function for $p_{1,2}^2 = 0$;
for general momenta it would be a much more complicated function.
The numerical factors $\xi_1$ and $\xi_2$ are given by
\begin{align}
	& \xi_1 = \frac{\Delta (\Delta - d) + 2  (d-6)}{(d-2)^2},
	\\
	& \xi_2 = \frac{1}{(d-1)(d-2)^3 (d-4)^2}  \Big[
	(3 d^2 - 12 d + 8)  \Delta^2 (\Delta - d)^2
	\nonumber \\
	& \hspace{5.4cm}
	+ 2  (d^4 - 2d^3 - 44d^2 + 160 d -112)
	\Delta (\Delta - d)
	\nonumber \\
	& \hspace{5.4cm}
	+ 4  (d-1)(d-4)(d-6)(3d^2 - 6d - 8)) \Big].
\end{align}
Note that $\xi_2$ diverges as $d \to 4$.
However, this multiplies a tensor structure that vanishes when we contract
with transverse polarization tensors, and therefore does not affect our results.
Also note that \Eq{TTscalar} is invariant under the symmetry
$\Delta \to d - \Delta$ up to the overall scale factor.
This is not an accident; it reflects the fact that for each operator $\O$,
there exist an non-local ``shadow'' operator with scaling dimension
$d-\Delta$ that satisfies the same Ward identities as $\O$~\cite{Ferrara:1972uq, SimmonsDuffin:2012uy}.

For the physical graviton polarizations in $d = 4$, we obtain
\begin{align}
	\bra{\O} \timeordering [\tilde{T}_+(p_1) \tilde{T}_+(p_2)] \ket{0}
	&= 2 \tilde{\la}_{TT\O} s^{\Delta/2},
    \\
    \bra{\O} \timeordering [\tilde{T}_+(p_1) \tilde{T}_-(p_2)] \ket{0} &= 0.
\end{align}
The vanishing of the 3-point function for $+-$ initial
helicity can be understood from angular momentum conservation.
For the $++$ initial helicity, the
conformal block can then be directly computed from the square of
the three-point function, weighted with the inverse of
$\Pi_\O(p_1 + p_2)$ given in Eq.~\eqref{eq:Pi:scalar}, to get
\begin{equation}
	\la_{TT\O}^2 f(\O)
	= \la_{TT\O}^2 \tfrac{5}{2} f^{(+)}(\O)
	= \big| \tilde{\la}_{TT\O} \big|^2 \frac{2^{2 \Delta - 5} \Gamma(\Delta)
	\Gamma(\Delta - 1)}{\pi^4}.
    \eql{fscalar:lambdatilde}
\end{equation}
The coefficient $\tilde{\la}_{TT\O}$ is arbitrary.
It will be related to the position-space OPE coefficient in the following section.

\subsubsection*{Spin-2 states}

We now consider states created by the 2-index traceless symmetric tensor $\scr{S}$.
Imposing symmetry in each pair of indices as well as under the exchange of the two time-ordered operators leads to 77 structures. The Ward identity for special conformal transformations brings this number down to 17 tensor structures. Imposing the transversality and trace conditions
Eqs.~(\ref{eq:WardIdentity:trace:TTT}--\ref{eq:WardIdentity:transverse:TTO})
reduces this number further to only 2 structures.

If we only consider the transverse and traceless part of this three-point function,
\textit{i.e.}~if we ignore all terms proportional to $p_1^\mu$, $p_1^\nu$, $p_2^\rho$, $p_2^\sigma$, $(p_1+p_2)^\al$, $(p_1+p_2)^\beta$, as well as $\eta^{\mu\nu}$, $\eta^{\rho\sigma}$ or $\eta^{\al\beta}$, then the result can be written
\[
\begin{split}
	\bra{\scr{S}^{\al\beta}} \timeordering [\tilde{T}^{\mu\nu}(p_1) \tilde{T}^{\rho\sigma}(p_2)] \ket{0}
	= s^{\Delta/2} \Big[ & \tilde{\la}_{TT\scr{S}}^{(1)}  t_1^{(\mu\nu)(\rho\sigma)(\al\beta)}
	+ \tilde{\la}_{TT\scr{S}}^{(2)}   t_2^{(\mu\nu)(\rho\sigma)(\al\beta)}
    \\
	& + (\text{longitudinal/trace parts}) \Big],
	\eql{TTS}
\end{split}
\]
where
\begin{subequations}
\begin{align}
	t_1^{\mu\nu\rho\sigma\al\beta} =&
    \left[ \eta^{\mu\rho} + 2 \hat{p}_2^\mu \hat{p}_1^\rho \right]
    \left[ \eta^{\nu\sigma} + 2 \hat{p}_2^\nu \hat{p}_1^\sigma \right]
    \left( \hat{p}_1 - \hat{p}_2 \right)^\al
    \left( \hat{p}_1 - \hat{p}_2 \right)^\beta,
    \eql{t1}
    \\
	t_2^{\mu\nu\rho\sigma\al\beta} =&
    \left[ \eta^{\mu\rho} + 2 \hat{p}_2^\mu \hat{p}_1^\rho \right]
    \left[ \eta^{\nu\al} + \hat{p}_2^\nu (\hat{p}_1 - \hat{p}_2)^\al \right]
    \left[ \eta^{\sigma\beta} - \hat{p}_1^\sigma (\hat{p}_1 - \hat{p}_2)^\beta \right],
    \eql{t2}
\end{align}
\end{subequations}
and symmetrization in each pair of indices is understood in Eq.~\eqref{eq:TTS}.
Again, the number of tensor structures (2) is in agreement with the position-space result of \sec{OPEcoefficients}.

This result can be used to compute the conformal blocks in terms of the momentum-space OPE coefficients
$\tilde{\la}_{TT\scr{S}}^{(1,2)}$, and we get (in $d = 4$)
\begin{align}
	\sum_{a,\, b = 1}^2 \la_{TT\scr{S}}^{(a)}
	\la_{TT\scr{S}}^{(b)} f_{ab}(\scr{S})
	&= \sum_{a,\, b = 1}^2 \tilde{\la}_{TT\scr{S}}^{(a)}
	\tilde{\la}_{TT\scr{S}}^{(b)}
	\frac{ 2^{2 \Delta - 9} (\Delta + 1) \Gamma(\Delta - 1)^2}
	{\pi^4 (\Delta - 1) (\Delta - 3)}
    \nonumber \\
    & \qquad \times
    \left( \begin{array}{cc}
    	-3 \Delta (\Delta - 4) - 8 &
    	\Delta ( \Delta - 4 ) + 4 \\
        \Delta ( \Delta - 4 ) + 4 &
        -\Delta ( \Delta - 4 ) - 2
    \end{array} \right)_{ab}.
	\eql{fspin2:lambdatilde}
\end{align}

\subsubsection*{Energy-momentum states}

In the special case in which the spin-2 operator is the energy momentum tensor,
the Ward identities are modified due to the presence of contact terms,
\Eqs{WardIdentity:trace:TTT} and \eq{WardIdentity:transverse:TTT}.
In addition to the two structures of \Eq{TTS}, the 3-point function
admits therefore a third tensor structure proportional to $\tilde{C}_T$
(and hence $c$) through its appearance in the two-point function of $T$, namely
\[
\begin{split}
	\bra{T^{\al\beta}} \timeordering [\tilde{T}^{\mu\nu}(p_1) \tilde{T}^{\rho\sigma}(p_2)] \ket{0}
	= s^2 \Big[ & \tilde{\la}_{TTT}^{(1)}  t_1^{(\mu\nu)(\rho\sigma)(\al\beta)}
    + \tilde{\la}_{TTT}^{(2)}  t_2^{(\mu\nu)(\rho\sigma)(\al\beta)}
    \\
    &{}+ \tilde{C}_T  t_c^{(\mu\nu)(\rho\sigma)(\al\beta)}
     \\
	& {}+ (\text{longitudinal/trace parts}) \Big],
	\eql{TTT}
\end{split}
\]
where $t_1$ and $t_2$ are given in \Eqs{t1} and \eq{t2}, while $t_c$ is
given by
\[
\begin{split}
	t_c^{\mu\nu\rho\sigma\al\beta} = & \ggap  2 \eta^{\mu\rho} \eta^{\nu\al} \eta^{\sigma\beta}
    + 2 \eta^{\mu\rho} \left( \eta^{\nu\al} \hat{p}_1^\sigma
    - \eta^{\sigma\al} \hat{p}_2^\nu \right) \left( \hat{p}_1 - \hat{p}_2 \right)^\beta
     \\
    & - 2 \left( \eta^{\mu\al} \eta^{\nu\beta} \hat{p}_1^\rho \hat{p}_1^\sigma
    + \eta^{\rho\al} \eta^{\sigma\beta} \hat{p}_2^\mu \hat{p}_2^\nu \right)
     \\
   	& + 4  \eta^{\mu\al} \eta^{\rho\beta} \hat{p}_2^\nu \hat{p}_1^\sigma
    - \frac{1}{2} \eta^{\mu\rho} \eta^{\nu\sigma}
    \left( \hat{p}_1 - \hat{p}_2 \right)^\al \left( \hat{p}_1 - \hat{p}_2 \right)^\beta.
\end{split}
\]
Projecting these tensors onto physical graviton polarizations, we obtain%
\footnote{We have eliminated all dependence on polarization vectors
on the \rhs\ by use of the identity
\begin{equation*}
	\epsilon_+^{(\mu}(p_1 | p_2) \epsilon_+^{\nu)}(p_2 | p_1)
    = e^{i\th}
    \left( \frac{1}{2} \eta^{\mu\nu} - \frac{p_1^{(\mu}
    p_2^{\nu)}}{p_1 \cdot p_2} \right),
\end{equation*}
The phase $\th$ can be thought of as an arbitrary phase in the definition
of the polarization tensors, which cancels in our sum rule because
it involves the square of the 3-point function.}
\[
\!\!\!\!\!\!\!\!\!\!
\begin{split}
	\bra{T^{\al\beta}} \timeordering [\tilde{T}_+(p_1) \tilde{T}_+(p_2)] \ket{0} &=
	\frac{1}{2} \left( \tilde{C}_T
	- \tfrac{2}{3} \tilde{\la}_{TTT}^{(1)}
	+ \tfrac{1}{3} \tilde{\la}_{TTT}^{(2)} \right) s^2
	\\
	& \quad \times
	\left[ \eta^{\al\beta}
	+ 4 \left( \hat{p}_1^\al \hat{p}_2^\beta
	+ \hat{p}_2^\al  \hat{p}_1^\beta \right)
	- 2 \left( \hat{p}_1^\al \hat{p}_1^\beta
	+ \hat{p}_2^\al \hat{p}_2^\beta \right) \right],
\end{split}
    \\
    \bra{T^{\al\beta}} \timeordering [\tilde{T}_+(p_1) \tilde{T}_-(p_2)] \ket{0} &= 0.
\]
For general $d$, \Eq{TTT} depends on 3 unknown constants, in agreement with
the fact that there are 3 independent OPE coefficients in general $d$
\cite{Osborn:1993cr}.
However, in $d = 4$ we find that only one linear combination of the OPE
coefficients appears in these results.
This is analogous to the vanishing of the momentum space OPE coefficients
at special values of $\De$.
Specifically, we find that in all 4D CFTs we have
$\tilde{\la}_{TTT}^{(2)} = 2  \tilde{\la}_{TTT}^{(1)}$.
There are two ways to prove this.
First, the contribution of a generic spin-2 operator
vanishes in the limit $\De \to 4$.
Since this corresponds to the \Eq{TTT} with $\tilde{C}_T = 0$ (compare \Eq{TTS}), we conclude that the contribution from
the energy-momentum tensor is proportional to
$\tilde{C}_T = 4 \pi c$.
Alternatively, the two OPE coefficients $\tilde{\la}_{TTT}^{(1,2)}$
can be computed in each of the 3 free theories, and
subsequently expressed in terms of $\nS$, $\nF$ and $\nV$
(see \sec{Tgeneral}).
In fact, we obtain
\[
\tilde{\la}_{TTT}^{(2)} = 2  \tilde{\la}_{TTT}^{(1)}
= 4\pi (c - a),
\]
where $a$ is the anomaly coefficient given in
\Eq{ansnfnv}.

The conformal block can therefore be directly computed as a function of $c$ only, using the inverse of the two-point function in Eq.~\eqref{eq:Pi:T:inverse}, and we obtain
\begin{equation}
	\left| \M_{T_+ T_+ \to T}(p_1, p_2) \right|^2 = 6 \pi  c  s^2,
    \qquad
    \left| \M_{T_+ T_- \to T}(p_1, p_2) \right|^2 = 0,
\end{equation}
or equivalently (see \Eq{Tcontribution})
\begin{equation}
	\sum_{a, \, b = 1}^3 \la_{TTT}^{(a)} \la_{TTT}^{(b)}
    \bigl[ f^{(+)}_{ab}(T) + f^{(-)}_{ab}(T) \bigr] = \sfrac{3}{5}  c.
\end{equation}
This agrees with the result of the free field theory calculation in
\sec{freetheories}.

\subsection{Relation to Position Space OPE Coefficients}

In the previous section, we have presented a method that allows to evaluate the 3-point function~\eq{TTO} up to unknown momentum-space OPE coefficient, and it only remains to relate them with the definitions of \sec{OPEcoefficients}, in order to obtain the conformal blocks for operators that are not the energy-momentum tensor itself.

\subsubsection*{Scalar operator}
We consider the following Lorentz scalar quantity
\begin{equation}
	Q(s) = \left[ p_2^\mu  p_1^\rho - (p_1 \cdot p_2) \eta^{\mu\rho} \right] \eta^{\nu\sigma}
	\bra{\O} \timeordering[ \tilde{T}_{\mu\nu}(p_1) \tilde{T}_{\rho\sigma}(p_2) \ket{0}.
\end{equation}
The tensor in square brackets is transverse, so this is a well-defined object
in $d \to 4$.
On the one hand, from the result of the Ward identity analysis and in particular \Eq{TTscalar}, it must be equal to
\begin{equation}
	Q(s) = -\tilde{\la}_{TT\O}  \frac{\Delta (\Delta - 4) + 20}{8}
     s^{(\Delta + 2)/2} .
    \eql{TTO:scalarcontraction}
\end{equation}
On the other hand, we can compute this same quantity by direct Fourier transform
of \Eq{OTTposn} from position to momentum space.
Note that this Fourier transform in general is ambiguous due to contact terms,
but as we have argued repeatedly above, there are no contact terms in this case.
For generic values of the the dimension of $\O$, the Fourier transform is unambiguous,
and for other values we can define the Fourier transform by analytic continuation.
We expect subtleties only in the case of exactly marginal operators, as discussed in
\sec{contactterms}.
The integrals can be performed using integration by parts to relate all terms
to the function $\scr{F}$ defined in \Eq{3ptFourierIntegral}.
The result is
\begin{equation}
	Q(s) = \la_{TT\O}
    \frac{3  i  \pi^4  \left[ \Delta (\Delta - 4) + 20 \right]
    \sin\left( \frac{\pi \Delta}{2} \right)}
    {2^{\Delta} (\Delta - 6) (\Delta - 4) \Delta (\Delta + 2)
    \Gamma\left( \frac{\Delta + 4}{2} \right)}
    s^{(\Delta + 2)/2}.
\end{equation}
This can be equated with \Eq{TTO:scalarcontraction} to fix $\tilde{\la}_{TT\O}$
in terms of $\la_{TT\O}$.
Note that $\tilde{\la}_{TT\O}$ vanishes when $\Delta = 2$,
as well as when $\Delta = 2d + 2n$ with $n \in \mathbb{N}$.

The 3-point function with initial graviton helicities $++$ for the energy-momentum tensors
is then given by
\begin{equation}
	\bra{\O} \timeordering[ \tilde{T}^+(p_1) \tilde{T}^+(p_2)] \ket{0}
	= \la_{TT\O}
	\frac{3  i  \pi^4
	\sin\left( \frac{\pi}{2}  \Delta \right)}
	{2^{\Delta-4}  (\Delta - 4)  (\Delta - 6) \Delta (\Delta + 2)
	\Gamma\left( \frac{\Delta + 4}{2} \right)^2} s^{\Delta/2} .
\end{equation}
The 3-point function with helicities $+-$ vanishes by angular momentum conservation.

The conformal block is then directly obtained from \Eq{fscalar:lambdatilde}:
\begin{equation}
	f(\O)
    = \sfrac{5}{2}  f^{(+)}(\O)
    = \frac{ 9  \pi^3  2^{2 \Delta + 2}
	\sin^2\left( \frac{\pi}{2}  \Delta \right)}
	{(\Delta - 6)^2  (\Delta - 4)^2
	\Delta^4  (\Delta + 2)^4}
	\frac{\Gamma\left( \frac{\Delta - 1}{2} \right)
    \Gamma\left( \frac{\Delta + 1}{2} \right)}
	{\Gamma\left( \frac{\Delta + 4}{2} \right)^2}.
	\eql{conformalblock:scalar}
\end{equation}
This is the result reported in \Eq{fscalar}.
Note that this number grows exponentially fast at asymptotically large scaling dimensions (see Fig.~\ref{fig:blocks}), with the asymptotic form given by
\begin{equation}
	f(\O) \sim 3^2  2^6  \pi^3
    \sin^2\left( \frac{\pi}{2}  \Delta \right)
	\frac{4^{\Delta}}{\Delta^{16}}
	\qquad ( \Delta \gg 1 ).
    \eql{conformalblock:scalar:asymptotics}
\end{equation}

\subsubsection*{Traceless Symmetric Spin-2 Operator}

In the case of a traceless, symmetric, spin-two operator $\scr{S}^{\mu\nu}$,
the relation between the OPE coefficients defined in \Eq{lambdaTTS:def}
and the quantities $\tilde{\lambda}_{TT\scr{S}}^{(1,2)}$ appearing in
\Eq{fspin2:lambdatilde} can be worked out in a similar manner from
scalar contractions. One finds
\[
\begin{split}
    \left(\begin{array}{c}
    	\tilde{\lambda}_{TT\scr{S}}^{(1)} \\[1em]
        \tilde{\lambda}_{TT\scr{S}}^{(2)}
    \end{array}\right)
    &= \frac{i \pi^4 \sin\left( \frac{\pi}{2}  \Delta \right)}
    {2^{\Delta - 2} (\Delta - 6) (\Delta - 4) \Delta  (\Delta + 2)
    \Gamma\left( \frac{\Delta + 6}{2} \right)^2}
    \\
    & \qquad \times
    \left(\begin{array}{cc}
    	\frac{\Delta^3 + 20 \Delta^2 - 8 \Delta - 48}{\Delta - 8} &
        - \frac{2 \Delta^2 + \Delta - 4}{2} \\[1em]
        - \frac{3 \Delta^3 - 30 \Delta^2 - 104 \Delta + 96}{\Delta - 8} &
        - \frac{3 \Delta^2 + 6 \Delta - 8}{2}
    \end{array}\right)
	\left(\begin{array}{c}
    	\lambda_{TT\scr{S}}^{(1)} \\[1em]
        \lambda_{TT\scr{S}}^{(2)}
    \end{array}\right),
\end{split}
\]
and therefore
\[
\begin{split}
	f_{ab}(\scr{S})
	&= \frac{ 2^{2 \Delta -1} \pi^3
	\sin^2\left( \frac{\pi}{2}  \Delta \right)}
	{(\Delta - 6)^2 (\Delta - 4) (\Delta - 1) \Delta^3 (\Delta + 2)^4 (\Delta + 4)^2}
	\frac{\Gamma\left( \frac{\Delta - 3}{2} \right)
    \Gamma\left( \frac{\Delta + 3}{2} \right)}
	{\Gamma\left( \frac{\Delta + 6}{2} \right)^2}
    \\
    & \qquad \times
    \left( \begin{array}{cc}
    	\frac{9 \Delta^5 + 9 \Delta^4 + 20 \Delta^3 + 1116 \Delta^2 + 4752 \Delta - 3456}
        {(\Delta - 8)^2} &
        -\frac{18 \Delta^3 + 65 \Delta^2 + 96 \Delta - 144}{\Delta - 8}
        \\[1em]
        -	\frac{18 \Delta^3 + 65 \Delta^2 + 96 \Delta - 144}{\Delta - 8} &
        \frac{9 \Delta^3 + 27 \Delta^2 + 16 \Delta - 48}{8}
    \end{array} \right)_{ab},
\end{split}
\eql{conformalblock:spin2}
\]
$f_{ab}(\scr{S})$ is a positive-definite matrix provided that the unitarity bound $\Delta \geq 4$
is satisfied.
Both its eigenvalues vanish when $\De = 10 + 2 n$ with $n \in \mathbb{N}$,
as well as $\De = 4$, where the unitarity bound is saturated.
A single eigenvalue vanishes when $\Delta = 8$.
The asymptotic form of the two eigenvalues at large $\De$ is
\begin{equation}
	f_{ab}(\scr{S}) \sim 3^2  2^6  \pi^3
    \sin^2\left( \frac{\pi}{2}  \Delta \right)
    \frac{4^{\Delta}}{\Delta^{16}}
	\left( \begin{array}{cc}
		\frac{1}{16} & 0
        \\
        0 & \frac{1}{2}
    \end{array} \right)_{ab}
	\qquad ( \Delta \gg 1 ).
    \eql{conformalblock:spin2:asymptotics}
\end{equation}
Fig.~\ref{fig:blocks} illustrates the behavior of the eigenvalues of $f(\scr{S})$
as a function of the scaling dimension $\Delta$.

The analysis can be extended to higher spin operators, at the cost of additional complexity.
For instance, with a traceless, symmetric tensor of spin four,
there are three tensor structures and as many OPE coefficients.
One notable difference with respect to the previous cases is that both
polarizations $\M_{++--}$ and $\M_{+--+}$ can contribute to the conformal blocks.
This has been verified in the free scalar theory, in which the spin-four conserved
current gives a non-zero contribution to the sum, see \Eq{spin4contribution}.


\section{IR Finiteness for Free Theories}
\label{appsec:IRfree}

In this appendix, we give some details about the verification of the IR
finiteness  of the pseudo-amplitude in the free CFTs.
As discussed in \S{1}, it is not sufficient to check the
finiteness of the imaginary part of the amplitude.
This check is much easier to perform, since it only involves the
computation of the total rate $hh \to \phi\phi, \bar\psi\psi, AA$,
and we have verified that these are all finite;
the results are given in
Eqs.~(\ref{eq:ScalarRates}--\ref{eq:Vectorrates}).
To check the finiteness of the real part as well is more
delicate, and requires detailed examination of the loop diagrams.
Instead of giving the details for all 3 free CFTs, we will take the free scalar
theory as demonstrating example.
Similar derivations have been made for the case of free fermions and vectors,
and we will make some brief remarks about these at the end.

This appendix is organized as follows.
We first give the Feynman rules of the free scalar theory in \S\ref{appsec:FeynmanRules}.
We then show the IR finiteness of the pseudo-amplitude in \S\ref{appsec:IRfiniteness}.
We give an explicit computation of the imaginary part of the pseudo-amplitude in
\S\ref{appsec:ImaginaryPart},
and present a check of the normalization of our sum rule for operators near $\De = 2$
in \S\ref{appsec:Delta2}

\subsection{Feynman rules}\label{appsec:FeynmanRules}

\begin{figure}[t]
\centering\hspace{-1.8cm}
 \subfigure[]{
 \centering\label{subfig:ScalarV1}
\begin{fmffile}{ScalarV1}
\begin{fmfgraph*}(35,45)
\fmfleft{i}
\fmfright{d6,o2,d5,d4,d3,d2,o1,d1}
\fmflabel{$h^{\mu\nu}$}{i}
\fmflabel{$\phi$}{o1}
\fmflabel{$\phi$}{o2}
\fmfv{decor.shape=circle,decor.filled=full,decor.size=4thick,l=$ =V_{h\phi\phi}^{\mu\nu}(p_1,,p_2)$,l.a=0,l.d=8mm}{v}
\fmf{dbl_plain_arrow,label=$p_1+p_2$,label.side=left}{i,v}
\fmf{scalar,label=$p_1$,label.side=left}{v,o1}
\fmf{scalar,label=$p_2$,label.side=right}{v,o2}
\end{fmfgraph*}
\end{fmffile}
 }\hspace{3cm}
 \subfigure[]{
 \centering\label{subfig:ScalarV2}
\begin{fmffile}{ScalarV2}
\begin{fmfgraph*}(35,45)
\fmfleft{e1,i2,e2,e3,i1,e4}
\fmfright{d6,o2,d5,d4,d3,d2,o1,d1}
\fmflabel{$h^{\mu\nu}$}{i1}
\fmflabel{$h^{\rho\sigma}$}{i2}
\fmflabel{$\phi$}{o1}
\fmflabel{$\phi$}{o2}
\fmfv{decor.shape=circle,decor.filled=full,decor.size=4thick,l=$ =V_{hh\phi\phi}^{\mu\nu,,\rho\sigma}(p,,q,,p_1,,p_2)$,l.a=0,l.d=8mm}{v}
\fmf{dbl_plain_arrow,label=$p$,label.side=left}{i1,v}
\fmf{dbl_plain_arrow,label=$q$,label.side=right}{i2,v}
\fmf{scalar,label=$p_1$,label.side=left}{v,o1}
\fmf{scalar,label=$p_2$,label.side=right}{v,o2}
\end{fmfgraph*}
\end{fmffile}
 }\hspace{2cm}
 \begin{minipage}{5.5in}
 \caption{\small Feynman rules for the vertices of the free scalar theory.}
 \label{fig:ScalarVertices}
 \end{minipage}
\end{figure}
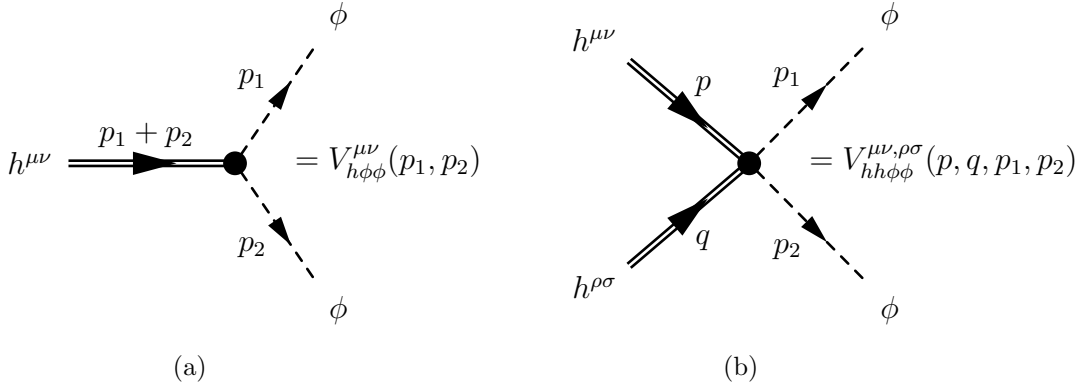

We obtain the Feynman rules by expanding the action
in powers of the perturbed graviton field $h_{\mu\nu} \equiv g_{\mu\nu} - \eta_{\mu\nu}$.
The interaction vertices we will need are those involving 1 or 2 gravitons.
There are diagrams involving a 3-graviton vertex, but we will show that this contribution
vanishes identically without requiring the detailed form of the 3-graviton vertex.
The 1- and 2-graviton vertices follow from expanding all quantities to second order in $h_{\mu\nu}$:
\begin{subequations}
\[
{g^{\mu \nu }} &= {\eta ^{\mu \nu }} - {h^{\mu \nu }} + {h^{\mu \rho }}h_\rho ^\nu , \\
\sqrt {-g} &= 1 + \frac{1}{2}h + \frac{1}{8}\left( {{h^2} - 2{h_{\alpha \beta }}{h^{\alpha \beta }}} \right) ,
\\
\begin{split}
R &=
( {{\partial ^\mu }{\partial ^\nu } - {\eta ^{\mu \nu }}{\partial ^2}})
{h_{\mu \nu }} + {h_{\mu \nu }}
({{\partial ^\mu }{\partial ^\nu }h + {\partial ^2}{h^{\mu \nu }} - 2{\partial ^\mu }{\partial _\rho }{h^{\nu \rho }}})
- \sfrac{1}{4} {{\partial ^\mu }h} {{\partial _\mu }h}
\\
&\quad{} + \sfrac{3}{4} ( {{\partial ^\rho }{h^{\mu \nu }}} )
 ( {{\partial _\rho }{h_{\mu \nu }}} ) +
 ({{\partial _\mu }h} {{\partial _\nu }{h^{\mu \nu }}}
- \sfrac{1}{2}  {{\partial ^\rho }{h^{\mu \nu }}}
{{\partial _\mu }{h_{\nu \rho }}}
- \left( {{\partial ^\mu }{h_{\mu \rho }}} \right)\left( {{\partial _\nu }{h^{\nu \rho }}} \right),
\end{split}
\]
\end{subequations}
where indices are now raised and lowered using the flat metric.
After collecting terms, we find that the 1-graviton
vertex is given by (see Fig.~\ref{fig:ScalarVertices})
\[
\begin{split}
iV_{h\phi \phi }^{\mu \nu }\left( {{p_1},{p_2}} \right)
&=  \left( {p_1^\mu p_2^\nu  - \frac{1}{2}{\eta ^{\mu \nu }}{p_1}\cdot{p_2}} \right)
\\
&\qquad\quad{}
- \frac{{d - 2}}{{4\left( {d - 1} \right)}}
\left[ {{{\left( {{p_1} + {p_2}} \right)}^\mu }{{\left( {{p_1} + {p_2}} \right)}^\nu } - {\eta ^{\mu \nu }}{{\left( {{p_1} + {p_2}} \right)}^2}} \right]  ,
\end{split}
\]
while the 2-graviton vertex is given by
\[
& V_{hh\phi \phi }^{\mu \nu ,\rho \sigma }\left( {p,q,{p_1},{p_2}} \right)
\nonumber \\
& \quad
= i\left\{ \begin{array}{l}
 - \frac{1}{2}{\eta ^{\rho \sigma }}p_1^\mu p_2^\nu  - \frac{1}{2}{\eta ^{\mu \nu }}p_1^\rho p_2^\sigma  + {\eta ^{\nu \sigma }}\left( {p_1^\mu p_2^\rho  + p_1^\rho p_2^\mu } \right) + \frac{1}{4}\left( {{\eta ^{\mu \nu }}{\eta ^{\rho \sigma }} - 2{\eta ^{\mu \rho }}{\eta ^{\nu \sigma }}} \right)\left(p_1\cdot p_2 \right) \\
 + \frac{{d - 2}}{{4\left( {d - 1} \right)}}\left[ \begin{array}{l}
{\eta ^{\rho \sigma }}\left( {{q^\mu }{q^\nu } + \frac{1}{2}{p^\mu }{p^\nu } + {p^\mu }{q^\nu }} \right) + {\eta ^{\mu \nu }}\left( {{p^\rho }{p^\sigma } + \frac{1}{2}{q^\rho }{q^\sigma } + {q^\rho }{p^\sigma }} \right)\\
 - 2{\eta ^{\nu \sigma }}\left( {{p^\mu }{p^\rho } + {q^\mu }{q^\rho } + {p^\mu }{q^\rho } + \frac{1}{2}{q^\mu }{p^\rho }} \right)\\
 - \frac{1}{2}\left( {{\eta ^{\mu \nu }}{\eta ^{\rho \sigma }} - 3{\eta ^{\mu \rho }}{\eta ^{\nu \sigma }}} \right)\left( p\cdot q \right) - \frac{1}{2}\left( {{\eta ^{\mu \nu }}{\eta ^{\rho \sigma }} - 2{\eta ^{\mu \rho }}{\eta ^{\nu \sigma }}} \right)\left( {{p^2} + {q^2}} \right)
\end{array} \right]
\end{array} \right\} .
\]

\subsection{IR Finiteness of the Pseudo-amplitude}\label{appsec:IRfiniteness}

\begin{figure}[t]
\centering
 \subfigure[Box]{
 \centering\label{subfig:Scalar4ptBox}
\begin{fmffile}{Scalar4ptBox}
\begin{fmfgraph*}(45,35)
\fmfleft{i2,i1}
\fmfright{o2,o1}
\fmflabel{$h^{\mu_1\nu_1}$}{i1}
\fmflabel{$h^{\mu_2\nu_2}$}{i2}
\fmflabel{$h^{\mu_3\nu_3}$}{o1}
\fmflabel{$h^{\mu_4\nu_4}$}{o2}
\fmfv{decor.shape=circle,decor.filled=full,decor.size=3thick}{v1}
\fmfv{decor.shape=circle,decor.filled=full,decor.size=3thick}{v2}
\fmfv{decor.shape=circle,decor.filled=full,decor.size=3thick}{v3}
\fmfv{decor.shape=circle,decor.filled=full,decor.size=3thick}{v4}
\fmf{dbl_plain_arrow,label=$p_1$,label.side=right}{i1,v1}
\fmf{dbl_plain_arrow,label=$p_2$,label.side=left}{i2,v2}
\fmf{dbl_plain_arrow,label=$p_3$,label.side=left}{o1,v3}
\fmf{dbl_plain_arrow,label=$p_4$,label.side=right}{o2,v4}
\fmf{scalar,tension=0.6,label=$k$,label.side=left}{v2,v1}
\fmf{scalar,tension=0.6,label=$k+p_1$,label.side=left}{v1,v3}
\fmf{scalar,tension=0.6,label=$k+p_1+p_3$,label.side=left}{v3,v4}
\fmf{scalar,tension=0.6,label=$k-p_2$,label.side=left}{v4,v2}
\end{fmfgraph*}
\end{fmffile}
 }\hspace{1.5cm}
 \subfigure[Triangle]{
 \centering\label{subfig:Scalar4ptTriangle}
\begin{fmffile}{Scalar4ptTriangle}
\begin{fmfgraph*}(45,35)
\fmfleft{i2,i1}
\fmfright{o2,o1}
\fmflabel{$h^{\mu_1\nu_1}$}{i1}
\fmflabel{$h^{\mu_2\nu_2}$}{i2}
\fmflabel{$h^{\mu_3\nu_3}$}{o1}
\fmflabel{$h^{\mu_4\nu_4}$}{o2}
\fmfv{decor.shape=circle,decor.filled=full,decor.size=3thick}{v1}
\fmfv{decor.shape=circle,decor.filled=full,decor.size=3thick}{v2}
\fmfv{decor.shape=circle,decor.filled=full,decor.size=3thick}{v3}
\fmf{dbl_plain_arrow,label=$p_1$,label.side=right}{i1,v1}
\fmf{dbl_plain_arrow,label=$p_2$,label.side=left}{i2,v2}
\fmf{dbl_plain_arrow,label=$p_3$,label.side=left}{o1,v3}
\fmf{dbl_plain_arrow,label=$p_4$,label.side=right}{o2,v3}
\fmf{scalar,tension=0.4,label=$k$,label.side=left}{v2,v1}
\fmf{scalar,tension=0.4,label=$k+p_1$,label.side=left}{v1,v3}
\fmf{scalar,tension=0.4,label=$k-p_2$,label.side=left}{v3,v2}
\end{fmfgraph*}
\end{fmffile}
 }\vspace{1cm}
 \subfigure[Bubble]{
 \centering\label{subfig:Scalar4ptBubble}
\begin{fmffile}{Scalar4ptBubble}
\begin{fmfgraph*}(50,35)
\fmfleft{i2,i1}
\fmfright{o2,o1}
\fmflabel{$h^{\mu_1\nu_1}$}{i1}
\fmflabel{$h^{\mu_2\nu_2}$}{i2}
\fmflabel{$h^{\mu_3\nu_3}$}{o1}
\fmflabel{$h^{\mu_4\nu_4}$}{o2}
\fmfv{decor.shape=circle,decor.filled=full,decor.size=3thick}{v1}
\fmfv{decor.shape=circle,decor.filled=full,decor.size=3thick}{v3}
\fmf{dbl_plain_arrow,label=$p_1$,label.side=right}{i1,v1}
\fmf{dbl_plain_arrow,label=$p_2$,label.side=left}{i2,v1}
\fmf{dbl_plain_arrow,label=$p_3$,label.side=left}{o1,v3}
\fmf{dbl_plain_arrow,label=$p_4$,label.side=right}{o2,v3}
\fmf{scalar,left=0.6,tension=0.4,label=$k$,label.side=left}{v3,v1}
\fmf{scalar,left=0.6,tension=0.4,label=$k+p_1+p_2$,label.side=left}{v1,v3}
\end{fmfgraph*}
\end{fmffile}
 }\hspace{1.2cm}
 \subfigure[3-graviton]{
 \centering\label{subfig:3h}
\begin{fmffile}{3hDiagram}
\begin{fmfgraph*}(50,35)
\fmfleft{i}
\fmfright{o3,o2,o1}
\fmflabel{$h^{\mu\nu}$}{i}
\fmfv{decor.shape=circle,decor.filled=full,decor.size=3thick}{v1}
\fmfv{decor.shape=circle,decor.filled=full,decor.size=3thick}{v3}
\fmf{dbl_plain_arrow,label=$p$,label.side=right}{i,v1}
\fmf{dbl_plain_arrow,tension=0.5}{o3,v3}
\fmf{dbl_plain_arrow,tension=0.5}{o1,v3}
\fmf{dbl_plain_arrow,tension=0.5}{o2,v3}
\fmf{scalar,left=0.6,tension=0.4,label=$k$,label.side=left}{v3,v1}
\fmf{scalar,left=0.6,tension=0.4,label=$k+p$,label.side=left}{v1,v3}
\end{fmfgraph*}
\end{fmffile}
}
 \begin{minipage}{5.5in}
 \caption{\small
 Feynman diagrams for the pseudo-amplitude in the free scalar theory.}
 \label{fig:ScalarLoopDiagrams}
 \end{minipage}
\end{figure}
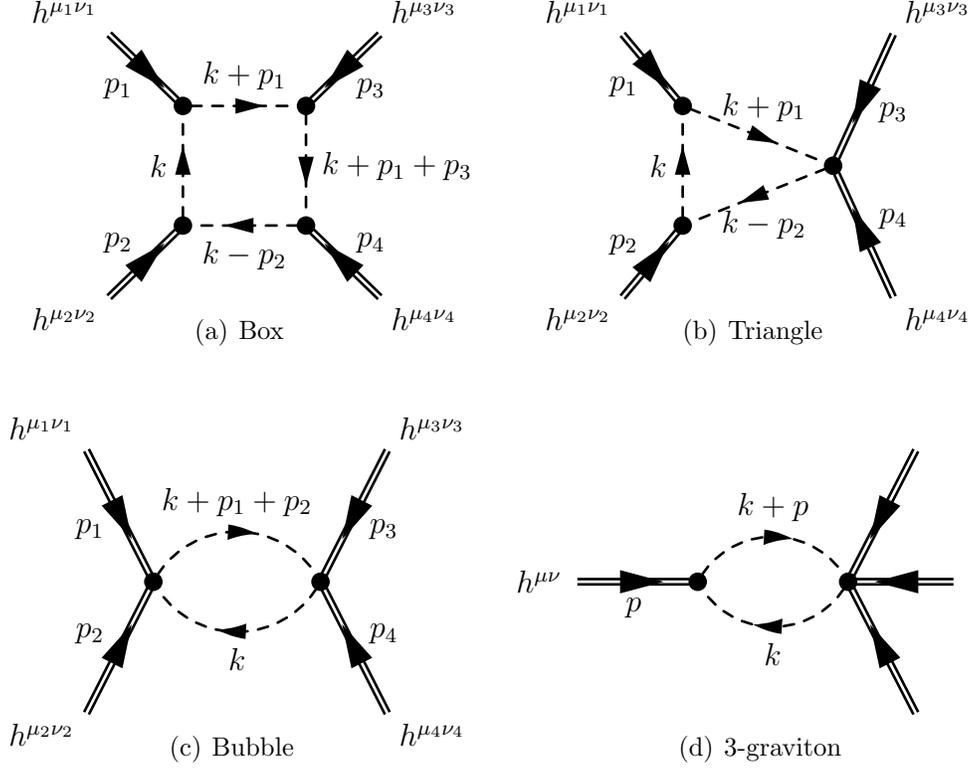

The pseudo-amplitude is given by the 1-loop Feynman diagrams shown
in Fig.~\ref{fig:ScalarLoopDiagrams}.
To examine the IR finiteness, we use the method of
``expansion by regions''~\cite{Smirnov:2002pj}.
This can be used to obtain the asymptotic behavior of a Feynman diagram in
various kinematic limits by summing over the contributions from
relevant regions of the loop momenta that contribute to the dimensionally
regularized integral.
To establish IR finiteness of the diagram it is sufficient to examine all
the dangerous (potentially IR divergent) regions and establish that
they are finite.
This amounts to a power-counting argument, and is simpler and more
transparent than the direct evaluation of diagrams, where the IR
finiteness appears to result from miraculous cancellations.
For 1-loop diagrams, the only dangerous regions are known to be
the soft, ultra-soft, and collinear limits.
We will consider first the massless limit $-p_i^2 = m^2 \to 0$
with $s, t$ fixed, and then the limit $t \to 0$ with
$m^2 = 0$ and $s$ fixed.

We will discuss the box diagram in Fig.~\ref{subfig:Scalar4ptBox} in detail,
and make only brief remarks about the other diagrams.
Before taking either the massless or the forward limit, the box diagram
contributes
\begin{equation}
i{\cal M} \propto \left\{ \begin{array}{l}
\int {\frac{{{d^4}k}}{{{{\left( {2\pi } \right)}^4}}}\frac{1}{{{k^2}{{\left( {k + {p_1}} \right)}^2}{{\left( {k - {p_2}} \right)}^2}{{\left( {k + {p_1} + {p_3}} \right)}^2}}} \, {\epsilon_{{\mu _1}{\nu _1}}}\left( {{p_1}} \right){\epsilon_{{\mu _2}{\nu _2}}}\left( {{p_2}} \right)\epsilon_{{\mu _3}{\nu _3}}\left( {{p_3}} \right)\epsilon_{{\mu _4}{\nu _4}}\left( {{p_4}} \right)} \\
 \times \left[ {{k^{{\mu _1}}}{{\left( {k + {p_1}} \right)}^{{\nu _1}}} - \frac{1}{2}{\eta ^{{\mu _1}{\nu _1}}}k\cdot\left( {k + {p_1}} \right) - \frac{1}{6}{\eta ^{{\mu _1}{\nu _1}}}{m^2}} \right]\\
 \times \left[ {{k^{{\mu _2}}}{{\left( {k - {p_2}} \right)}^{{\nu _2}}} - \frac{1}{2}{\eta ^{{\mu _2}{\nu _2}}}k\cdot\left( {k - {p_2}} \right) - \frac{1}{6}{\eta ^{{\mu _2}{\nu _2}}}{m^2}} \right]\\
 \times \left[ {{{\left( {k + {p_1}} \right)}^{{\mu _3}}}{{\left( {k + {p_1} + {p_3}} \right)}^{{\nu _3}}} - \frac{1}{2}{\eta ^{{\mu _3}{\nu _3}}}\left( {k + {p_1}} \right)\cdot\left( {k + {p_1} + {p_3}} \right) - \frac{1}{6}{\eta ^{{\mu _3}{\nu _3}}}{m^2}} \right]\\
 \times \left[ {{{\left( {k - {p_2}} \right)}^{{\mu _4}}}{{\left( {k + {p_1} + {p_3}} \right)}^{{\nu _4}}} - \frac{1}{2}{\eta ^{{\mu _4}{\nu _4}}}\left( {k - {p_2}} \right)\cdot\left( {k + {p_1} + {p_3}} \right) - \frac{1}{6}{\eta ^{{\mu _4}{\nu _4}}}{m^2}} \right]
\end{array} \right\} . \eql{MScalarbox0}
\end{equation}

\subsubsection*{Massless limit}
We begin with the massless limit $m^2 \ll s,t \sim Q^2$.
The soft region is defined as the regime in which all the components of $k$
are of order $m$, {\it i.e.}~$k^\mu \sim m$.
The ultra-soft region is defined by $k^\mu \sim \frac{m^2}{Q}$.
In these two regions, the propagator part of the loop integral can be reduced to
\begin{equation}
\frac{d^4 k}{k^2 \left(k+p_1\right)^2 \left(k-p_2\right)^2 \left(k+p_1+p_3\right)^2}\sim\frac{{{d^4}k}}{{{k^2}  (k \cdot Q) (k \cdot Q) {Q^2}}},
\end{equation}
which is at most logarithmically IR divergent.
Both vertices associated with $p_1$ and $p_2$ give a further suppression of at least one power of $k \sim m$,
\begin{subequations}
\begin{align}
\left[ {{k^{{\mu _1}}}{{\left( {k + {p_1}} \right)}^{{\nu _1}}} - \frac{1}{2}{\eta ^{{\mu _1}{\nu _1}}}k \cdot \left( {k + {p_1}} \right) - \frac{1}{6}{\eta ^{{\mu _1}{\nu _1}}}{m^2}} \right] &\lesssim O(k) , \\
\left[ {{k^{{\mu _2}}}{{\left( {k - {p_2}} \right)}^{{\nu _2}}} - \frac{1}{2}{\eta ^{{\mu _2}{\nu _2}}}k \cdot \left( {k - {p_2}} \right) - \frac{1}{6}{\eta ^{{\mu _2}{\nu _2}}}{m^2}} \right] &\lesssim O(k) .
\end{align}
\end{subequations}
Therefore, contributions from the soft and ultra-soft regions to the box diagram are finite.

The collinear regions are slightly more complicated.
They correspond to the case where the internal momentum at a vertex
becomes collinear with the external momentum.
For example, the collinear region for the vertex associated with
$p_1^\mu = \frac{\sqrt s}{2}\left(1,0,0,\sqrt{1-4m^2/s}\right)$ is defined as
\begin{equation}
\left\{ \begin{array}{l}
{k_ + } \equiv {k^0} + {k^3}\sim Q\\[3pt]
\displaystyle  {k_ - } \equiv {k^0} - {k^3}\sim \frac{{{m^2}}}{Q}\\[3pt]
k_ \bot ^i \equiv \left( {{k^1},{k^2}} \right)\sim m
\end{array} \right. , \eql{p1collinear}
\end{equation}
which gives $k^2\sim m^2$, $p_1 \cdot k \sim m^2$, and $k^\mu \propto p_1^\mu + O(m)$.
Other collinear regions are defined similarly, and can be obtained by
permutations of momenta.
In the collinear region defined by \Eq{p1collinear}, the propagator part goes as
\begin{equation}
\frac{{{d^4}k}}{{{k^2}{{\left( {k + {p_1}} \right)}^2}{{\left( {k - {p_2}} \right)}^2}{{\left( {k + {p_1} + {p_3}} \right)}^2}}}\sim \frac{{{m^4}}}{{{m^2} {m^2} {Q^2} {Q^2}}} .
\end{equation}
This is again logarithmic divergent by itself, but the vertex associated with $p_1$ contributes an additional suppression factor
\begin{equation}
\left[ {{k^{{\mu _1}}}{{\left( {k + {p_1}} \right)}^{{\nu _1}}} - \frac{1}{2}{\eta ^{{\mu _1}{\nu _1}}}k \cdot \left( {k + {p_1}} \right) - \frac{1}{6}{\eta ^{{\mu _1}{\nu _1}}}{m^2}} \right]\lesssim O( {{m^2}} ) .
\end{equation}
Here we have used $p_1 \cdot k \sim m^2$ as well as
the transversality of the graviton polarization tensor $p_1^{\mu_1}\epsilon_{\mu_1\nu_1}(p_1)=0$.
We see that the contribution from the collinear region is also finite.

\subsubsection*{Forward limit}
We now check the forward scattering limit $t \ll s \sim Q^2$,
setting $m^2 = 0$.
In this case, the pseudo-amplitude in \Eq{MScalarbox0} becomes
\begin{equation}
i{\cal M} \propto \left\{ \begin{array}{l}
\int {\frac{{{d^4}k}}{{{{\left( {2\pi } \right)}^4}}}\frac{1}{{{k^2}{{\left( {k + {p_1}} \right)}^2}{{\left( {k - {p_2}} \right)}^2}{{\left( {k + {p_1} + {p_3}} \right)}^2}}}{\epsilon_{{\mu _1}{\nu _1}}}\left( {{p_1}} \right){\epsilon_{{\mu _2}{\nu _2}}}\left( {{p_2}} \right)\epsilon_{{\mu _3}{\nu _3}}\left( {{p_3}} \right)\epsilon_{{\mu _4}{\nu _4}}\left( {{p_4}} \right)} \\
 \times \left[ {{k^{{\mu _1}}}{{\left( {k + {p_1}} \right)}^{{\nu _1}}} - \frac{1}{2}{\eta ^{{\mu _1}{\nu _1}}}k\cdot\left( {k + {p_1}} \right)} \right]\left[ {{k^{{\mu _2}}}{{\left( {k - {p_2}} \right)}^{{\nu _2}}} - \frac{1}{2}{\eta ^{{\mu _2}{\nu _2}}}k\cdot\left( {k - {p_2}} \right)} \right]\\
 \times \left[ {{{\left( {k + {p_1}} \right)}^{{\mu _3}}}{{\left( {k + {p_1} + {p_3}} \right)}^{{\nu _3}}} - \frac{1}{2}{\eta ^{{\mu _3}{\nu _3}}}\left( {k + {p_1}} \right)\cdot\left( {k + {p_1} + {p_3}} \right)} \right]\\
 \times \left[ {{{\left( {k - {p_2}} \right)}^{{\mu _4}}}{{\left( {k + {p_1} + {p_3}} \right)}^{{\nu _4}}} - \frac{1}{2}{\eta ^{{\mu _4}{\nu _4}}}\left( {k - {p_2}} \right)\cdot\left( {k + {p_1} + {p_3}} \right)} \right]
\end{array} \right\} . \eql{MScalarbox1}
\end{equation}
In the soft region $k^\mu\sim \sqrt{t}$ and the ultra-soft region $k^\mu\sim t/Q^2$, the propagator part goes like
\begin{equation}
\frac{{{d^4}k}}{{{k^2}{{\left( {k + {p_1}} \right)}^2}{{\left( {k - {p_2}} \right)}^2}{{\left( {k + {p_1} + {p_3}} \right)}^2}}}\sim \frac{{{d^4}k}}{{{k^2} (k \cdot Q) (k \cdot Q) t}} ,
\end{equation}
which has a $1/t$ divergence.
However, for transverse and traceless graviton polarizations, both vertices associated with $p_1$ and $p_2$ give a suppression factor of at least two powers of $k$:
\begin{subequations}
\begin{align}
\left[ {{k^{{\mu _1}}}{{\left( {k + {p_1}} \right)}^{{\nu _1}}} - \frac{1}{2}{\eta ^{{\mu _1}{\nu _1}}}k\cdot\left( {k + {p_1}} \right)} \right] &\to k^{\mu_1}k^{\nu_1} \lesssim O(k^2) ,  \\
\left[ {{k^{{\mu _2}}}{{\left( {k - {p_2}} \right)}^{{\nu _2}}} - \frac{1}{2}{\eta ^{{\mu _2}{\nu _2}}}k\cdot\left( {k - {p_2}} \right)} \right] &\to k^{\mu_2} k^{\nu_2} \lesssim O(k^2) .
\end{align}
\end{subequations}
Together, they make the contribution from these regions IR finite.

In the $p_1$ collinear region defined by \Eq{p1collinear}
(with $m$ replaced by $\sqrt{t}$), we have
$k \propto p_1 + O\bigl(\sqrt{t}\bigr) \propto p_3 + O\bigl(\sqrt{t}\bigr)$.
The propagator part is $1/t$ IR divergent
\begin{equation}
\frac{{{d^4}k}}{{{k^2}{{\left( {k + {p_1}} \right)}^2}{{\left( {k - {p_2}} \right)}^2}{{\left( {k + {p_1} + {p_3}} \right)}^2}}}\sim \frac{{{t^2}}}{{t \cdot t \cdot {Q^2} \cdot t}} ,
\end{equation}
but both the $p_1$ and $p_3$ vertices give a suppression factor proportional to $t$:
\begin{subequations}
\begin{align}
\left[ {{k^{{\mu _1}}}{{\left( {k + {p_1}} \right)}^{{\nu _1}}} - \frac{1}{2}{\eta ^{{\mu _1}{\nu _1}}}k\cdot\left( {k + {p_1}} \right)} \right] &\lesssim O(t) , \\
\left[ {{{\left( {k + {p_1}} \right)}^{{\mu _3}}}{{\left( {k + {p_1} + {p_3}} \right)}^{{\nu _3}}} - \frac{1}{2}{\eta ^{{\mu _3}{\nu _3}}}\left( {k + {p_1}} \right)\cdot\left( {k + {p_1} + {p_3}} \right)} \right] &\lesssim O(t) ,
\end{align}
\end{subequations}
where we have again used the transversality of the gravitons polarizations.

\subsubsection*{Summary}

We have shown that the box diagram in Fig.~\ref{subfig:Scalar4ptBox} is IR finite under both the massless limit and the forward limit, assuming transverse and traceless polarizations for the gravitons.
One can repeat the above procedure for the triangle diagram in Fig.~\ref{subfig:Scalar4ptTriangle} and the bubble diagram in Fig.~\ref{subfig:Scalar4ptBubble}.
They also turn out to be IR finite under both limits using
similar reasoning.

In addition, there is a special kind of diagram with the
3-graviton vertex shown in Fig.~\ref{subfig:3h}. This diagram depends on a single external momentum $p=p_i$.
We hence expect it to vanish under the massless limit
$-p_i^2=m^2\to 0$.
However, before we can claim this, the IR finiteness under this limit has to be checked.
This actually can be shown regardless of the details of the 3-graviton vertex. The diagram has the form:
\[
i\mathcal{M}\propto\int\frac{d^4 k}{(2\pi)^4} \frac{1}{k^2(k+p)^2} \epsilon_{\mu\nu}(p)V_{h\phi\phi}^{\mu\nu}(k+p,-k) V_3 , \eql{3hM}
\]
where we have denoted the the 3-graviton vertex by $V_3$.
Because $V_3$ is a polynomial of $p,k$ contracted with the three external polarization tensors, it is finite in the limit
$m^2 \to 0$, and we only need to check for IR divergences
in the rest of the diagram.
The only dangerous region for \Eq{3hM} is the collinear
region defined in \Eq{p1collinear}.
The propagator part is logrithmically divergent in this region
\[
\frac{d^4 k}{k^2(k+p)^2} \sim \frac{m^4}{m^2 m^2} .
\]
But the 1-graviton vertex provides additional suppression
factors
\[
\epsilon_{\mu\nu}(p)V_{h\phi\phi}^{\mu\nu}(k+p,-k) \lsim O(m),
\]
since $k^\mu\propto p^\mu+O(m)$ in the collinear region.

For free theories of fermions and vectors, we have checked that IR finiteness
holds by a similar analysis.
For fermions, we must consider perturbations of the vierbein
field ${\varphi^a}_\mu \equiv {e^a}_\mu-{\delta^a}_\mu$,
which corresponds to a metric perturbation
$h_{\mu\nu} = \varphi_{\mu\nu} + \varphi_{\nu\mu} + \eta_{ab} \gap {\varphi^a}_\mu \gap {\varphi^b}_\nu$.
We express the vierbein in terms of the metric perturbation by making the
gauge choice $\varphi_{\mu\nu}=\varphi_{\nu\mu}$ to define the energy-momentum
tensor.
In the free vector case, we follow the usual gauge-fixing procedure, including
ghost fields to subtract the unphysical vector intermediate states in
the $TTTT$ correlator.

\subsection{Imaginary Part of the Pseudo-amplitude}\label{appsec:ImaginaryPart}

In free CFTs, the full imaginary part of the pseudo-amplitude in the forward limit $\Im\mathcal{M}_{T_1 T_2\to T_3 T_4}(s) \propto \sigma \left(h_1 h_2\to \text{CFT} \right)$ can be directly evaluated, due to the asymptotic particle interpretation of the CFT states. This calculation is detailed below for the free scalar theory.
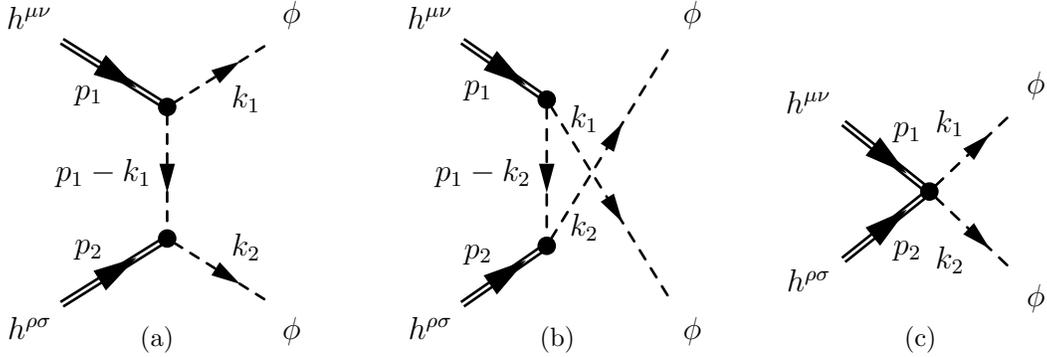
\begin{figure}
\centering
 \subfigure[]{
 \centering\label{subfig:ScalarM1}
\begin{fmffile}{ScalarM1}
\begin{fmfgraph*}(35,35)
\fmfleft{i2,i1}
\fmfright{o2,o1}
\fmflabel{$h^{\mu\nu}$}{i1}
\fmflabel{$h^{\rho\sigma}$}{i2}
\fmflabel{$\phi$}{o1}
\fmflabel{$\phi$}{o2}
\fmfv{decor.shape=circle,decor.filled=full,decor.size=3thick}{v1}
\fmfv{decor.shape=circle,decor.filled=full,decor.size=3thick}{v2}
\fmf{dbl_plain_arrow,label=$p_1$,label.side=right}{i1,v1}
\fmf{dbl_plain_arrow,label=$p_2$,label.side=left}{i2,v2}
\fmf{scalar,label=$k_1$,label.side=right}{v1,o1}
\fmf{scalar,label=$k_2$,label.side=left}{v2,o2}
\fmf{scalar,label=$p_1-k_1$,label.side=right}{v1,v2}
\end{fmfgraph*}
\end{fmffile}
 }\hspace{1cm}
 \subfigure[]{
 \centering\label{subfig:ScalarM2}
\begin{fmffile}{ScalarM2}
\begin{fmfgraph*}(35,35)
\fmfleft{i2,i1}
\fmfright{o2,o1}
\fmflabel{$h^{\mu\nu}$}{i1}
\fmflabel{$h^{\rho\sigma}$}{i2}
\fmflabel{$\phi$}{o1}
\fmflabel{$\phi$}{o2}
\fmfv{decor.shape=circle,decor.filled=full,decor.size=3thick}{v1}
\fmfv{decor.shape=circle,decor.filled=full,decor.size=3thick}{v2}
\fmf{dbl_plain_arrow,tension=1.5,label=$p_1$,label.side=right}{i1,v1}
\fmf{dbl_plain_arrow,tension=1.5,label=$p_2$,label.side=left}{i2,v2}
\fmf{scalar,tension=0,label=$k_1$,label.side=left}{v2,o1}
\fmf{scalar,tension=0,label=$k_2$,label.side=right}{v1,o2}
\fmf{scalar,label=$p_1-k_2$,label.side=right}{v1,v2}
\fmf{phantom}{v1,o1}
\fmf{phantom}{v2,o2}
\end{fmfgraph*}
\end{fmffile}
 }\hspace{1cm}
 \subfigure[]{
 \centering\label{subfig:ScalarM3}
\begin{fmffile}{ScalarM3}
\begin{fmfgraph*}(25,30)
\fmfleft{e1,i2,e2,e3,i1,e4}
\fmfright{d6,o2,d5,d4,d3,d2,o1,d1}
\fmflabel{$h^{\mu\nu}$}{i1}
\fmflabel{$h^{\rho\sigma}$}{i2}
\fmflabel{$\phi$}{o1}
\fmflabel{$\phi$}{o2}
\fmfv{decor.shape=circle,decor.filled=full,decor.size=3thick}{v}
\fmf{dbl_plain_arrow,label=$p_1$,label.side=left}{i1,v}
\fmf{dbl_plain_arrow,label=$p_2$,label.side=right}{i2,v}
\fmf{scalar,label=$k_1$,label.side=left}{v,o1}
\fmf{scalar,label=$k_2$,label.side=right}{v,o2}
\end{fmfgraph*}
\end{fmffile}
 }
 \begin{minipage}{5.5in}
 \caption{\small
 Tree-level diagrams relevant to the $hh\to\phi\phi$ scattering process in the free scalar theory.}
 \label{fig:ScalarTreeDiagrams}
 \end{minipage}
\end{figure}
In this case, the concrete form of the above relation is
\begin{equation}
2\Im \mathcal{M}_{T_iT_j\to T_iT_j}(s) = (2i)^4 \int { d\Pi_{\phi\phi} \left(k_1, k_2\right) {\left| {\mathcal{M}_{h_ih_j\to\phi\phi} \left(k_1, k_2\right)} \right|}^2 } , \eql{ScalarOPT}
\end{equation}
where the two-body phase space can be written in terms of the scattering angles as
\begin{equation}
\int {d\Pi_{\phi\phi} \left(k_1, k_2\right)} = \frac{1}{2}\frac{1}{{8\pi }}\int_{ - 1}^1 {\frac{{d\left( {\cos \theta } \right)}}{2}} . \eql{ScalarPhasespace}
\end{equation}
Note the presence of the symmetry factor $1/2$ due to identical final-state particles. There are three diagrams for the scattering amplitude $\mathcal{M}_{h_1h_2\to\phi\phi}\left(k_1, k_2\right)$, as shown in Fig.~\ref{fig:ScalarTreeDiagrams}. They add up to the total amplitude
\begin{equation}
{{\cal M}_{h_1h_2\to\phi \phi }} = {\epsilon_{\mu\nu}^1}\left( {{p_1}} \right){\epsilon_{\rho\sigma}^2}\left( {{p_2}} \right)\left\{ \begin{array}{l}
\frac{{k_1^\mu k_1^\nu k_2^\rho k_2^\sigma }}{{{{\left( {{p_1} - {k_1}} \right)}^2}}} + \frac{{k_2^\mu k_2^\nu k_1^\rho k_1^\sigma }}{{{{\left( {{p_1} - {k_2}} \right)}^2}}}\\
 + {\eta ^{\nu \sigma }}\big[ - \frac{1}{2}{\eta ^{\mu \rho }}\left( {{k_1}\cdot{k_2}} \right) + \left( {k_1^\mu k_2^\rho  + k_2^\mu k_1^\rho } \right) \\
\qquad\quad - \frac{1}{6}\left( {p_2^\mu p_1^\rho  - \frac{3}{2}{\eta ^{\mu \rho }}{p_1}\cdot{p_2}} \right) \big]
\end{array} \right\} .
\end{equation}
Here, we have readily taken the massless limit $-p_1^2=-p_2^2=m^2\to 0$. Splitting the result into the two helicity structures (same or opposite),
and expressing it in terms of the scattering angle $\theta$ and the azimuthal angle $\varphi$, we get
\begin{align}
{\cal M}_{h^+ h^+\to\phi\phi } (\theta,\varphi) &= \epsilon_{\mu \nu }^+ \left( {{p_1}} \right)\epsilon_{\rho \sigma }^+ \left( {{p_2}} \right)\left[ {\frac{{k_1^\mu k_1^\nu k_2^\rho k_2^\sigma }}{{{{\left( {{p_1} - {k_1}} \right)}^2}}} + \frac{{k_2^\mu k_2^\nu k_1^\rho k_1^\sigma }}{{{{\left( {{p_1} - {k_2}} \right)}^2}}} + {\eta ^{\nu \sigma }}\left( {\frac{s}{8}{\eta ^{\mu \rho }} + 2k_1^\mu k_2^\rho } \right)} \right]
\nonumber \\
&= \frac{s}{16}\left( {3{{\cos }^2}\theta  - 1} \right) , \\
{\cal M}_{h^+ h^-\to\phi \phi } (\theta,\varphi) &= \epsilon_{\mu \nu }^+ \left( {{p_1}} \right)\epsilon_{\rho \sigma }^- \left( {{p_2}} \right)\left[ {\frac{{k_1^\mu k_1^\nu k_2^\rho k_2^\sigma }}{{{{\left( {{p_1} - {k_1}} \right)}^2}}} + \frac{{k_2^\mu k_2^\nu k_1^\rho k_1^\sigma }}{{{{\left( {{p_1} - {k_2}} \right)}^2}}}} \right]
\nonumber \\
&= \frac{{{s}}}{16}{e^{ 4i\varphi }}\left( {1 - {{\cos }^2}\theta } \right) .
\end{align}
Making use of Eqs.~\eqref{eq:ScalarOPT} and~\eqref{eq:ScalarPhasespace}, we get the finite results
\begin{align}
2\Im\mathcal{M}_{++}^\text{scalar}(s) &= \frac{1}{{320\pi }}{s^2} , \\
2\Im\mathcal{M}_{+-}^\text{scalar}(s) &= \frac{1}{{480\pi }}{s^2} .
\end{align}
The imaginary part of the pseudo-amplitude can be calculated in the same way in free fermion and vector theories.

\subsection[Free Scalar Contribution in General d]{$\phi^2$ Contribution in General $d$}\label{appsec:Delta2}

We now show how to use the contribution of the $\phi^2$ operator
in free scalar theory away from
$d = 4$ to check the normalization of the coefficients appearing in our
sum rule.
Although the sum rule is only valid for $d = 4$, the optical theorem \Eq{opticaltheorem}
is valid for any $d$, and the contribution of $\phi^2$ to the imaginary
part of the amplitude is given by%
\footnote{The operator that satisfies the normalization \Eq{operatornormalization}
is really $\frac 1{\sqrt{2}} \phi^2$.}
\[
\Im \scr{M}_{++--}(s, d) = 5 \pi s^{d/2} \la_{TT\phi^2}^2 f^{(+)}(\phi^2, d)
+ \cdots.
\]
Defining this requires the analytic continuation of the polarization tensors
to general $d$, but in practice the contribution of $\phi^2$ can be computed using
only the $d$-independent relations
\[
	p_{i\mu} \epsilon_\pm^{\mu\nu}(p_i) = 0,
    \qquad
    \eta_{\mu\nu} \epsilon_\pm^{\mu\nu}(p_i) = 0,
    \qquad
    \eta_{\mu\rho} \eta_{\nu\sigma} \epsilon_+^{\mu\nu}(p_1) \epsilon_+^{\rho\sigma}(p_2) = 1.
\]
The calculation in the free scalar theory then gives
\[
5\pi \lambda_{TT\phi^2}^2 f^{(+)}(\phi^2, d)
	= \frac{(d - 2)^2 (d - 4)^2}
	{2^{2d+3} \pi^{(d-3)/2} \ggap d^2
    (d - 1) \gap \Gamma\bigl(\frac{d+1}{2}\bigr)} .
\]
This can be thought of as the contribution of an operator of dimension
$\De = d-2$ in general $d$.
It has a double zero as $d \to 4$, consistent with the fact that
there is a double zero in the contribution of a scalar operator $\O$
with dimension $\De$ in $d = 4$ in the limit $\De \to 2$.
In fact, if we assume that the contribution of a general scalar operator of
dimension $\De$ in $d$ spacetime dimensions is an analytic function of
$\De$ and $d$, then the coefficient of the double zero at the point
$d = 4$, $\De = 2$ must agree.
Using our convention \Eq{lambdaTTO}, the
OPE coefficient in the free scalar theory is given by
\[
	\lambda_{TT\phi^2} = \frac{d (d-2)^2}{2  \sqrt{2} (d-1)} C_T
    \xrightarrow{d \to 4} \frac{4 \sqrt{2}}{9 \pi^4},
\]
which gives
\[
\sfrac{5}{2} f^{(+)}(\phi^2, d)
\xrightarrow{d \to 4}
\frac{3^2 \pi^6}{2^{17}} ( \Delta - 2)^2,
\]
where $\De = d -2$.
This precisely agrees with \Eq{fscalar}, giving an
independent check of the normalization of $f(\O)$.


\bibliographystyle{utphys}
\bibliography{mycites}

\end{document}